\documentclass[10pt,a4paper,twoside]{article}
\usepackage[utf8]{inputenc}
\usepackage[left=4.4cm,right=4.4cm]{geometry}
\usepackage[authoryear,round]{natbib}
\usepackage{changepage}
\usepackage{threeparttablex}
\usepackage{longtable}
\usepackage{booktabs}
\usepackage{amsmath}
\usepackage{amsfonts}
\usepackage{amssymb}
\usepackage{bm}
\usepackage{graphicx}
\usepackage{subcaption}  
\usepackage[font=footnotesize]{caption}
\usepackage{nameref} 
\newcommand{\degree}{\ensuremath{^\circ}}
\newcommand{\mi}{\mathrm{i}} 

\newcommand\frontmatter{
    \cleardoublepage
    \pagenumbering{roman}}
\newcommand\mainmatter{
    \cleardoublepage
    \pagenumbering{arabic}}
\newcommand{\drop}{1truecm}
\newcommand*{\titlenew}{\begingroup
\parindent=0pt
\vspace*{\drop}
\rule{12.2cm}{0.5mm}
\vskip 0.3truecm
{\Huge\bfseries Planetary Ring Dynamics}\\ [\baselineskip]
\vspace*{0.4truecm}
{\huge {The Streamline Formalism\\ 
{\LARGE {\itshape 2. Theory of Narrow Rings and Sharp Edges}}
} 
}\par
\rule{12.2cm}{0.5mm}
\vfill
\begin{figure}[h]
    \centering
    \includegraphics[width=\textwidth]{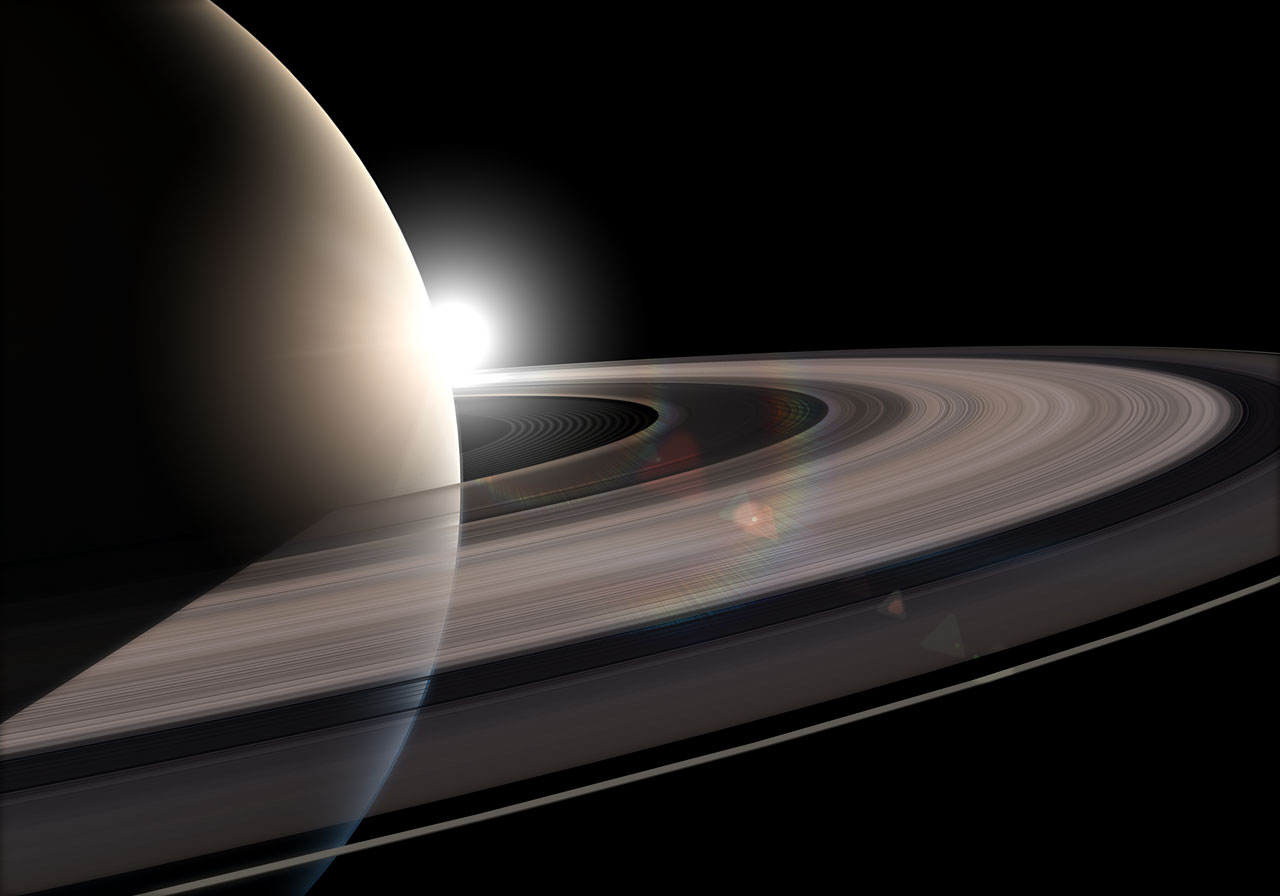}
\end{figure}
\vfill
{\Large{\itshape Pierre-Yves Longaretti}}\\ [\baselineskip] 
{\large IPAG, CNRS \& UGA\\}
\texttt{pierre-yves.longaretti@univ-grenoble-alpes.fr}\\
\rule{12.2cm}{0.3mm}
\begin{figure}[h]
    \centering
    \begin{subfigure}[b]{0.15\textwidth}
        \includegraphics[width=\textwidth]{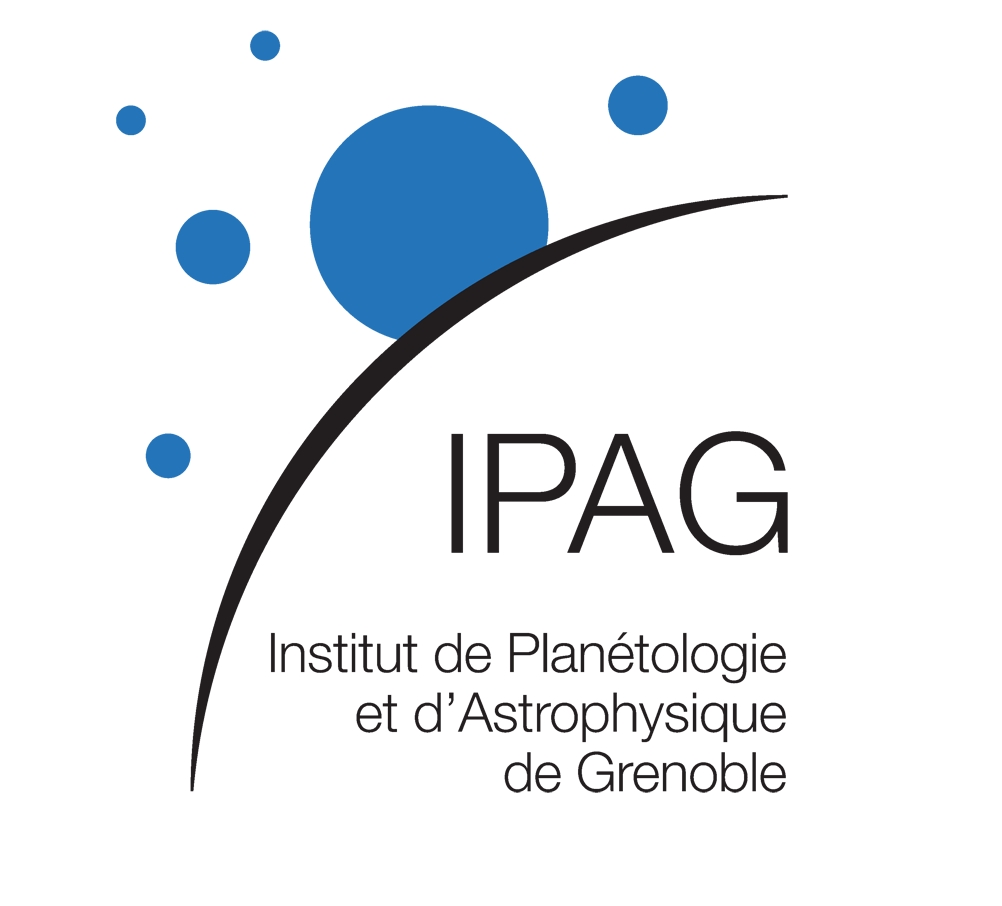}
    \end{subfigure}
    \hfill
    \begin{subfigure}[b]{0.13\textwidth}
        \includegraphics[width=\textwidth]{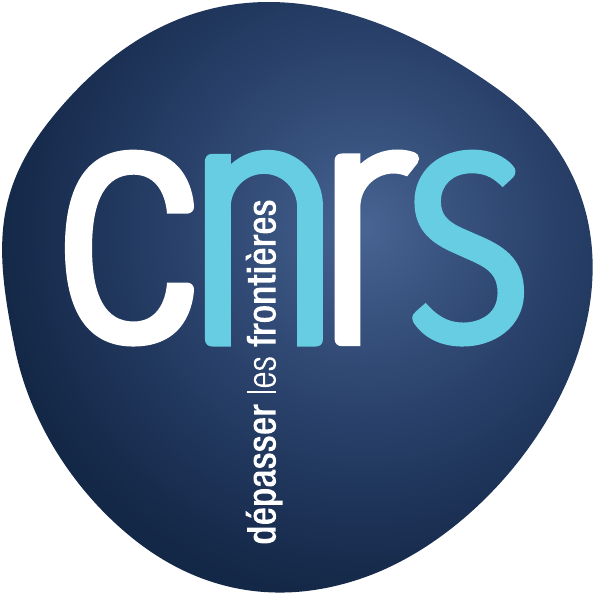}
    \end{subfigure}
    \hfill
    \begin{subfigure}[b]{0.2\textwidth}
        \includegraphics[width=\textwidth]{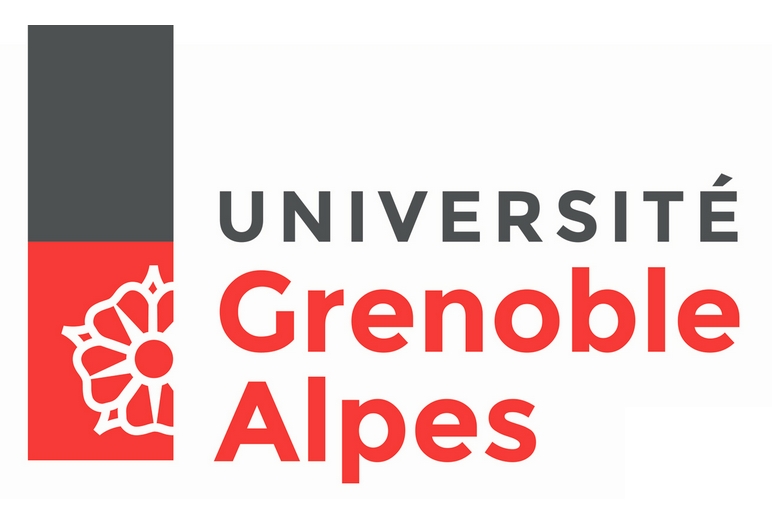}
    \end{subfigure}
\end{figure}
\endgroup}
\begin{document}
%
%
\titlenew
\thispagestyle{empty}
\clearpage
%
%
\newpage\null\thispagestyle{empty}\newpage
\frontmatter
\section*{}
\markboth{Contents}{}
\vspace*{-1cm}
\tableofcontents
\markboth{Contents}{}
\clearpage
\section*{Foreword}
\label{sec:foreword}
\addcontentsline{toc}{section}{\nameref{sec:foreword}}
\markboth{Foreword}{}

The present material presents a long overdue complement to my initial notes on the streamline formalism initially published in the \textit{Goutelas} series of volumes \citep{L92} and recently corrected and posted on the \textit{arXiv}. It will appear later on this year in the upcoming book (\textit{Planetary Ring Systems}), edited by Matt Tiscareno and Carl Murray and to be published at Cambridge University Press. These two pieces of work are brought together as part 1 and part 2 of a mini-series on planetary ring dynamics, posted on the \textit{arXiv}; part 1 is quoted here as \cite{L92} and can also be accessed directly on the \textit{arXiv}  (\textit{astro-ph/1606.00759}).

The present material covers the features of large scale ring dynamics in perturbed flows that were not addressed in the initial lecture notes; this includes an extensive coverage of all kinds of ring modes dynamics (except density waves which have been covered in part 1), the origin of ring  eccentricities and mode amplitudes, and the issue of ring/gap confinement. This still leaves aside a number of important dynamical issues relating to the ring small scale structure, most notably the dynamics of self-gravitational wakes, of local viscous overstabilities and of ballistic transport processes. These topics will be covered in the book mentioned above, among others.

As this material is designed to be self-contained, there is some $30\%$ overlap with part 1 of this two-part series: the basics of the streamline formalism are covered in both, albeit in more details in part 1, and most probably in a more pedagogical way for students and newcomers. This corresponds to the two sections devoted here to theoretical background (sections \ref{sec:kin}, \ref{sec:pert}); the discussion of the standard self-gravity model of rigid precession is also common to both parts (addressed here in section \ref{ssg}). The discussion of the stress-tensor treatment is much more extensive in part 1. In any case, the literature coverage on all these questions has been updated here wherever relevant with respect to part 1.

\bigskip

Please feel free to send me any feedback on these notes, if only to point out the unavoidable remaining mistakes and typos

\vskip 1truecm
\hfill \textit{Pierre-Yves Longaretti, January 5, 2017}

\vfill
\noindent
Photo credits:\\
\noindent Cover: NASA\\
\noindent Backpage:
\begin{itemize}
\item Collection of early drawings of Saturn by various observers, from Huyghens' \textit{Systema Saturnium} (1659).
\item Guido Bonatti, from his \textit{De Astronomia Tractatus X Universum quod ad iudiciariam rationem nativitatum} (Basel, 1550)
\end{itemize}

\clearpage
\newpage\null\thispagestyle{empty}\newpage
%
%
\mainmatter
\setcounter{figure}{0}    
\section{Introduction}\label{sec:int}

Narrow rings and sharp edges are typical features of what are usually referred to as perturbed dense rings: perturbed because they display small deviations from circularity, and dense because these are vertically thin rings with optical depth of order unity. The last two decades have seen a large focus on small scale phenomena in such rings (in particular on self-gravitational wakes and viscous overstabilities; see the chapters by Cuzzi et al., Stewart et al., and Salo et al.), but a substantial part of our current theoretical understanding of the large scale structure of such perturbed rings stems from a formal framework put forward by Borderies, Goldreich and Tremaine (hereafter BGT) in the late 1970s and throughout the 1980s,  known as the \textit{streamline formalism} (\citealt{GT79b,BGT82,BGT83b,BGT83a,BGT85,BGT86,BGT89}). Furthermore, this groundwork has spurred a number of other analytical and numerical studies relating to the same issues. As a consequence, our discussion draws heavily on the GT/BGT work, with other relevant analyses discussed along the way.

The streamline formalism is essentially a fluid treatment of ring systems, based on the use of the Boltzmann equation for describing microphysical phenomena (but semi-heuristic prescriptions are quite useful as well on this front) and borrowing perturbation techniques from celestial mechanics to tackle the large scale dynamics. It relies on orderings relevant for ring systems. 

For rings with optical depth of order unity, the shortest timescales are defined by the collision frequency $\omega_c$ and the mean motion $n$, with $\omega_c \gtrsim n$ and $n^{-1}\sim \mathrm{ a\ few\ hours}$ in Saturn's rings. The basic circular motion is externally perturbed by satellites and possibly the planet's internal oscillation modes, and internally by the ring's self-gravity and by internal stress (pressure or stress tensor) due to interparticle collisions. Electromagnetic forces have very little effect on the dynamics of major rings, due to the large particle sizes that constitute most of their mass; atmospheric drag may induce an extra torque in the Uranian rings, but this point is ignored in this review unless otherwise stated. Similarly, ballistic transport processes (see the chapter by Estrada et al.) have limited relevance to the problems discussed in this chapter, although they might play some role in some narrow rings characterized by broad edges. 

All perturbations are much weaker than the planet's attraction, so that their characteristic time-scales are much larger than the orbital time-scale, typically of the order of years for the shortest ones. From the kinematical point of view, such perturbations induce only weak deviations from circularity. Eccentricities range typically from $\sim 10^{-5}$ (e.g.\ for density waves) to $\sim 10^{-2}$ (e.g.\ in narrow eccentric rings), with similar figures for inclinations. However, typical dimensionless radial gradients of eccentricity or phase are usually of order unity and can produce large density contrasts. 

The two orderings just described (small dynamical perturbations and small kinematical deviations from circularity) underlie the general philosophy of the formalism. Deviations from circularity are treated to first order in eccentricity, but radial gradients are not linearized; due to the short collisional time-scale, the ring stress-tensor reaches quasi-equilibrium very fast and evolves quasi-statically with the macroscopic dynamics; for the same reason, a vertical quasi-static equilibrium is often assumed and the very small ring thickness makes the use of vertically integrated dynamical equations a very common procedure. All forces but the planet's are treated as perturbations. As in any other fluid disk problem, the dynamics is investigated for specific forms of the deviations from circularity. 

A specificity of the formalism, related to its reliance on celestial mechanics techniques, is the (semi-)Lagrangian fluid description adopted instead of the more common Eulerian one. This review conforms with a widespread usage in ring dynamics in which the Lagrangian derivative $D/Dt$ is denoted $d/dt$, but the reader should not forget that one deals with a fluid problem and not with individual particle motions. It is also customary to describe first the kinematics underlying the specific forms of motions under scrutiny and then address the related dynamical problem; this fosters physical insight, and we will follow this traditional pattern here.

An important piece of understanding that has emerged from this framework over the years is the realization that all narrow ring modes and edge modes stem from the same fundamental physics. Let us introduce some definitions here. An edge mode is an $m$-lobe pattern seen on a ring edge (e.g., the Saturn's A and B ring outer edge modes); its amplitude is non-negligible on a limited radial extent from the edge. Such edge modes may also be seen at narrow ring edges. In this case, an edge mode by definition does not encompass the whole width of the ring, in contradistinction with a global narrow ring mode. Eccentric rings are only a particular form of global modes and are characterized by $m=1$; $m\neq 1$ global modes differ in quantitative details from $m=1$ modes but not in their most essential characteristics. Edge modes may seem less immediately connected to global modes, but still emerge from the same theoretical picture; in particular, both edge modes and global narrow ring modes can be seen as trapped standing density waves. Making this underlying unity apparent is one of the main themes of this chapter. 

Only planar motions are described, firstly because the bulk of the theoretical literature focuses on motions confined to the equatorial plane of the planet, and secondly because the theory of inclined rings and bending waves is blueprinted \textit{mutatis mutandis} from their eccentric counterparts. 

The first two sections below draw mostly on \cite{BGT83a,BGT85}, except on specific points where relevant references will be given as needed. A somewhat more systematic form of exposition has been adopted here with respect to these original references. A more detailed description of this material can be found in \cite{L92}, as well as the streamline formalism theory of linear and nonlinear density waves, which is not reviewed in this chapter (see the chapter by Stewart et al.). There is some overlap between this earlier work and the present one. However, this earlier presentation is more adapted as an introduction to the topics discussed here, while the present exposition is rather aimed at researchers already familiar with ring issues and looking for an in-depth coverage; it is also substantially updated. 

The formal and physical aspects of the streamline formalism have evolved in the course of the dozen papers published by the GT/BGT team, later papers not only improving but sometimes superseding previous ones. This review is intended to present these results in a systematic and self-consistent way along with contributions by other researchers; it is also intended to bridge the gap between simplified expositions and the existing (and sometimes daunting) literature, although in a somewhat compact way in the two first sections. As a consequence, the material assembled here has two somewhat antithetic flavors: formal and mathematical in the first part, and more physical and heuristic in the second. Sections \ref{sec:kin} (kinematics), \ref{sec:pert} (perturbation equations) and \ref{sec:2str} (two-streamline model) are more abstract in content, except for the heuristic discussion of section \ref{sec:press} (stress tensor), as they focus on the formalism itself. The reader interested only in physical applications can largely ignore section \ref{sec:pert} (except for the definitions of relevant quantities) and \ref{sec:2str} and jump directly from section \ref{sec:kin} to \ref{sec:times}, where the time-scale hierarchy and related simplifications are discussed in more detail. From section \ref{sec:freeeigen} onwards, dynamical equations are used in a reduced form with appropriate references to earlier sections when needed, and the discussion focuses on the physics. The sections devoted to specific issues (\ref{sec:freeeigen} to \ref{sec:shep}) can be read independently from one another; they deal respectively with the origin of narrow rings uniform precession, the description of edge and narrow ring modes as trapped waves, the origin of narrow ring eccentricities and edge mode amplitudes, and the shepherding mechanism of narrow rings and sharp edges. A table of symbols is provided at the end of the main text. 

The systematization effort undertaken here has led to corrections and extensions of existing results, and occasionally to the derivation of new ones. The most important are the following, by order of appearance in the text:

\begin{itemize}
\item The form of the stress tensor perturbation Eqs.~\eqref{t1} and \eqref{t2} is corrected.
\item The two-streamline model of \cite{BGT83a} is formulated for all types of modes (section \ref{sec:2str}).
\item The expression of the pattern speed of free modes (section \ref{sec:freepatspeed}) does not appear to have been derived previously.
\item The physics of the trapped wave picture of these modes (section \ref{trapped}) is presented in some detail, following an unpublished preliminary analysis by Peter Goldreich.
\item The \cite{GT81} analysis of the excitation of narrow ring eccentricities by external satellites turns out to be erroneous on an important aspect (see section \ref{sec:satamp} and the Appendix) and has been accordingly corrected. This analysis is also reformulated for isolated resonances and not only overlapping ones, as the former setting is more relevant to actual rings; systematic torque expressions are given in this regime (section \ref{sec:narrow}).
\item The \cite{GT79a} initial picture of narrow ring confinement (section \ref{sec:wakeconfnarrow}) is systematically revisited. Relatedly, the present discussion of the satellite mass limits (section \ref{sec:satmass}) extends and corrects earlier ones \citep{BGT84,GP87}.
\end{itemize}

It is hoped that providing a comprehensive and systematic exposition of the formalism and the physical issues to which it can be applied may motivate researchers in the field to produce the detailed analyses of key processes that are still lacking to this date.

\section{Theoretical background: kinematics}\label{sec:kin}

On the one hand, ring fluid particles must follow eccentric orbits, as the Navier-Stokes equation in Lagrangian variables with all forces neglected but the planet's  (zeroth order approximation) reduces to a test particle equation of motion ($d\mathbf{r}/dt$ $= - \nabla \Phi_p$ where $\Phi_p$ is the potential of the planet). To lowest order in deviation from circularity, this gives
\begin{eqnarray}
r & = & a_e[1-\epsilon\cos M_e],\label{epir}\\
\theta & = & \varpi_e+ M_e+2 \frac{\Omega}{\kappa}\epsilon\sin M_e,\label{epiphi}
\end{eqnarray}
\noindent where $a_e$, $\epsilon$, $\varpi_e$ and $M_e$ are the epicyclic semi-major axis, epicyclic eccentricity, epicyclic periapsis angle and epicyclic mean anomaly, respectively. The epicyclic mean longitude is $\varphi_e=\varpi_e+M_e$, and $\Omega$ and $\kappa$ are the epicyclic angular velocity and horizontal frequency at $a_e$ (see expressions below). For unperturbed motions, $dM_e/dt=\kappa$ and $d\varphi/dt=\Omega=\dot{\varpi}_e + \kappa$. 

The related radial and angular velocities read
\begin{eqnarray}
\frac{dr}{dt} & = & a_e\epsilon\kappa\sin M_e,\label{ur}\\
\frac{d\theta}{dt} & = & \Omega +2\Omega\epsilon\cos M_e.\label{utheta}
\end{eqnarray}
Following common celestial mechanics practice transposed to fluid flows, perturbed motions are represented by the same relations, except that the epicyclic elements must depend on time\footnote{For readers unfamiliar with this procedure, this is just a special application of the method of variation of constants. The procedure is well-known in the theory of osculating elliptic motions, where Gauss or Laplace equations constitute the most used form of the related evolution equations \citep{MD99,R88}. The equivalent equations of perturbed epicyclic motions are given in section \ref{sec:pert} (see  \citealt{L92} for more details).}.

Epicyclic elements \citep{BL87,LB91} are used because they directly stand for the observed quantities, unlike their elliptic counterparts (also known as the osculating or Keplerian elements). The subscript ``\textit{e}" is used as a reminder of this fact in this kinematics section; it will be dropped from section \ref{sec:proc} (dynamical processes) onwards to alleviate notations. The epicyclic elements used here are not the usual ones, but have been defined in order for all relevant expressions to bear a closer analogy to the more familiar elliptic ones.

The usual specific energy $E$ and angular momentum $H$ have the following expressions in epicyclic variables
\begin{eqnarray}
E & = &  \Phi_p(a_e)+\frac{\Omega^2a_e^2}{2}+\mathcal{O}(\epsilon^4),\label{specen}\\
H  & = & \Omega a_e^2\left[1-\frac{1}{ 2}\left(\frac{\kappa}{\Omega}\right)^2
\epsilon^2\right]+\mathcal{O}(\epsilon^4),\label{specang}
\end{eqnarray}
\noindent where $\Phi_p$ is the potential of the planet.  

For completeness, in terms of the leading planetary oblateness parameter $J_2$, the potential, the epicyclic angular velocity and epicyclic frequency read
\begin{eqnarray}
\Phi_p(a_e) & = &\frac{G M_p}{a_e} \times  \left[-1+\frac{1}{2}\left(\frac{R_p}{a_e}\right)^2 J_2 \right],\label{pot}\\
\Omega(a_e) & = & n(a_e)\ \times\! \left[\ \  1+\frac{3}{4}\left(\frac{R_p}{a_e}\right)^2 J_2\right],\label{Omega}\\
\kappa(a_e) & = & n(a_e)\ \times\! \left[\ \ 1-\frac{3}{4}\left(\frac{R_p}{a_e}\right)^2 J_2\right],\label{kappa}
\end{eqnarray}
\noindent where $M_p$ and $R_p$ are the planet mass and radius. More complete expressions can be found in \cite{BL87} and \cite{BL94}. For future use, we have also introduced an effective elliptic mean motion defined by
\begin{equation}\label{meanmotion}
n(a_e)=\left(\frac{GM_p}{a_e^3}\right)^{1/2}.
\end{equation}

On the other hand, the perturbed collective motion of ring fluid particles is also \textit{assumed} to constitute an $m$-lobed \textit{mode} in a frame rotating with an angular velocity denoted the pattern speed $\Omega_p$; this assumption clearly restricts the types of motion under investigation, but is adapted for single mode dynamics\footnote{Multiple mode motions and issues related to mode coupling are briefly discussed in the conclusion. Ignoring mode coupling, Eq.~\eqref{rmode} can be generalized to a superposition of various $m$ modes; the reader is referred to \cite{L89b} for a formulation of this problem.}. This choice is clearly motivated by observations, and the dynamical equations self-consistently specify the conditions of existence of such motions. The related kinematical equations read:
\begin{eqnarray}
r(a_e,\varphi_e,t) & = & a_e \{ 1-\epsilon(a_e,t) \times\cos\left[ m(\varphi_e-\Omega_p t)+ m\Delta(a_e,t) \right]\},\label{rmode}\\
\theta(a_e,\varphi_e,t) & = & \varphi_e+2\frac{\Omega}{\kappa}
\epsilon(a_e,t)\times
 \sin\left[ m(\varphi_e-\Omega_p t)+ m\Delta(a_e,t) \right],\label{phimode}
\end{eqnarray}
\noindent In these relations, ($a_e,\varphi_e$) represent the circular motion the fluid particle would have in the absence of perturbation, and are used as (semi-)Lagrangian labels; $m\Delta$ is referred to as the \textit{streamline apsidal shift} in the remainder of this review. Note however that for observational fits, one usually defines $\delta_m=-\Delta$. Finally, following \cite{BGT83a} and \cite{SYL85}, let us define the complex eccentricity: 
\begin{equation}\label{compe}
Z \equiv \epsilon \exp\mi m\Delta.
\end{equation}
The complex eccentricities of the discrete streamlines (in a discrete formulation) are collected in a finite array, called the \textit{complex eccentricity vector} in the remainder of the text. 

Quite often, the $m$-lobe shape can be considered stationary in the rotating frame, in which case $\epsilon$ and $\Delta$ are independent of $t$; however, a time-dependence may occur, e.g., under the action of viscous overstabilities, or in the relaxation phase to equilibrium in which case the time dependence is transient (see section \ref{sec:2str} for more details). The dynamics of satellite-induced wakes can be incorporated in this framework, provided that $m$ assumes continuous real values instead of discrete integer ones, and by allowing the eccentricity to depend on the azimuthal coordinate $\varphi$ (see section \ref{shflux}).

\begin{figure*}[ht]
\centering
\includegraphics[width=\textwidth]{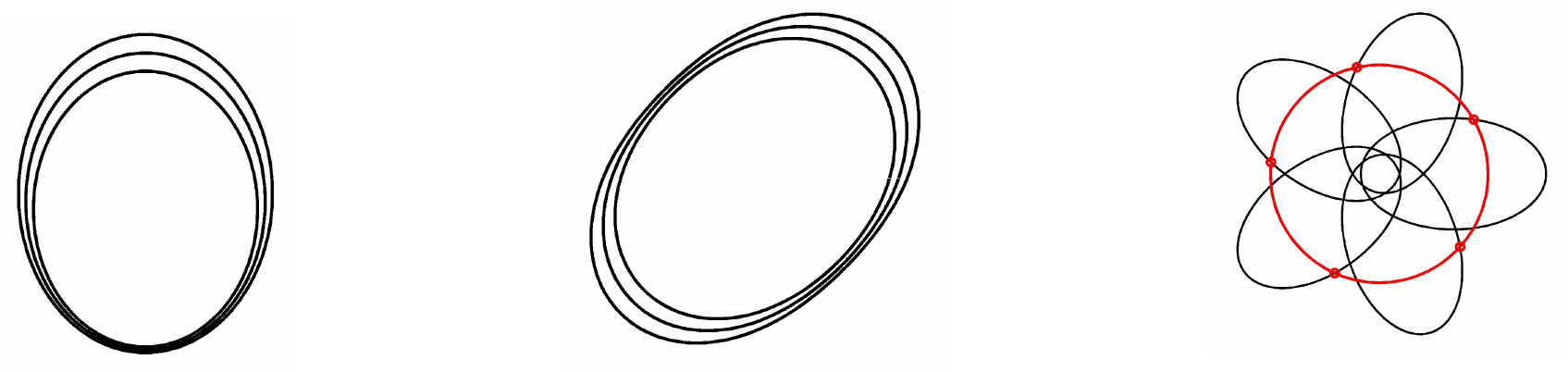}
%
\caption{Sketch of mode patterns as observed in a narrow ring. Eccentricities have been greatly amplified. Left: $m$=1 mode; middle: $m$=2 mode; right: infinitely narrow $m=0$ mode. For this last mode, all fluid particles belong to the same circle oscillating radially (in red), while occupying synchronized azimuthal locations along their individual orbits (in black), thus producing a radially pulsating circle. This mode is kinematically different from $m\neq 0$ modes, for which orbits and streamlines are identical in the rotating frame.}
\label{fig:rings}
\end{figure*}

For free modes, the pattern speed $\Omega_p$ is defined by the entire perturbed region (see section \ref{sec:freepatspeed}); for resonantly forced modes, the pattern speed is defined by the satellite\footnote{Or by the forcing agent if not a satellite, e.g.\ planetary oscillation modes, or possibly, other ring modes. The resonance condition quoted here applies to circular background motions. More complex resonances are discussed in section \ref{sec:satamp}.}: $\Omega_p = \Omega_s + k/|m|\ \kappa_s$ [$k$ and $m$ are integers specified by the satellite potential Fourier expansion and $\Omega_s$ and $\kappa_s$ are the satellite rotation and epicyclic frequency, respectively; see, e.g.\ \citealt{GT82} or \citealt{Sh84} for a discussion of this expression, and right below Eq.~\eqref{syncons}]. In this case, an implicit choice of origins of time and angles has been made in the previous relations: at $t=0$, the satellite is at periapse and its mean epicyclic longitude is zero. As a consequence, for $k=0$ resonances, the origin of angles in the rotating frame is the satellite's mean epicyclic longitude. Finally, for $m=0$ modes, one needs to replace the phase $m(\varphi_e-\Omega_p t)+ m\Delta$ by $\Omega_pt + \Delta$:
\begin{eqnarray}
r(a_e,\varphi_e,t) & = & a_e \{ 1-\epsilon(a_e,t)\cos\left[ \Omega_p t + \Delta(a_e,t) \right]\},\label{rmode0}\\
\theta(a_e,\varphi_e,t) & = & \varphi_e+2\frac{\Omega}{\kappa}
\epsilon(a_e,t)\sin\left[ \Omega_p t+ \Delta(a_e,t) \right].\label{phimode0}
\end{eqnarray}
Note that for $m\neq 0$, when the $m$-lobe pattern appears stationary in the frame rotating at the pattern speed (i.e., when the eccentricity and apsidal shift are time-independent), Eq.~\eqref{rmode} describes both the fluid particle orbits and streamlines, hence the name of the formalism; this property is not true for the $m=0$ mode. The shape of various modes is sketched on Figure~\ref{fig:rings} to illustrate these considerations.

Eqs.~\eqref{epir}, \eqref{epiphi} and \eqref{rmode}, \eqref{phimode} can be satisfied simultaneously only if
\begin{equation}\label{anom}
M_e = m(\varphi_e-\Omega_p t)+m\Delta,
\end{equation}
\noindent i.e., if the following relations hold\footnote{The contribution of the perturbations to $d\varphi_e/dt$ is comparatively negligible; see section \ref{sec:pert}.} \citep{BGT85}
\begin{eqnarray}
\frac{dm\Delta}{dt} & = & -m(\Omega-\Omega_p) + \Omega-\dot\varpi_e,\label{phasetime}\\
\varpi_0 & = & \varphi_0(1-m)-m\Delta_0,\label{phaseinit}
\end{eqnarray}
\noindent where $\varpi_0$, $\varphi_0$ are the periapse angle and mean longitude of the fluid particle at $t=0$ and $\Delta_0$ the streamline apsidal shift at the same time. The second relation expresses the constraint that all particles with identical $a_e$ must satisfy in order to belong to a common $m$-lobe streamline in the rotating frame, while the first is a necessary condition for the pattern to be maintained at all times (similar relations exist for an $m=0$ mode). From Eqs.~\eqref{Omega}, \eqref{kappa} and \eqref{meanmotion} and for stationary modes, Eq.~\eqref{phasetime} implies that:
\begin{eqnarray}
\dot\varpi_{pert} &=& -m(\Omega-\Omega_p)+(\Omega-\dot\varpi_{plan})\nonumber\\
& \simeq & \left[\frac{3}{2}(m-1)+\frac{21}{4}\left(1+\frac{m-1}{2}\right)\left(\frac{R_p}{a_r}\right)^2 J_2\right] \times n(a_r)\frac{a_e-a_r}{a_r},\label{syncons}
\end{eqnarray}
\noindent where $a_r$ is the resonance radius (defined next), and where 
$\dot\varpi_{pert}$ and $\dot\varpi_{plan} \equiv \Omega-\kappa$ are the
contributions of the perturbing acceleration and of the planet to the precession rate, respectively. The resonance radius is traditionally defined by $m[\Omega(a_r)-\Omega_p]=\pm[\Omega(a_r)-\dot\varpi_{plan}(a_r)]$; here, equivalently and consistently with Eqs.~\eqref{phasetime} and \eqref{syncons}, it is defined by $m[\Omega(a_r)-\Omega_p]=[\Omega(a_r)-\dot\varpi_{plan.}(a_r)]$ with $m > 0$ or $m<0$; the $+$ sign corresponds to an inner Lindblad resonance (ILR) and the $-$ sign to an outer Lindblad resonance (OLR); corotation resonances are largely ignored in this work, except in section \ref{sec:ecc}. Finally, the ratio $\Omega(a_r)/\Omega_p\simeq (|m|+k):(|m|\mp 1)$ is often used to label a Lindblad resonance; the near equality is an equality if one ignores the contribution of the planet oblateness (as $\Omega_p=\Omega_s + k\kappa_s/|m|$) and the minus (plus) sign in the denominator corresponds to an ILR (OLR).

One sees that the perturbing accelerations must produce a secular variation of
the line of the apses on the synodic time-scale in order to maintain the mode shape throughout the perturbed region and prevent orbit crossing\footnote{Perturbed regions are always very limited in azimuthal extent, so that $(a_e-a_r)/a_r \ll 1$; this ensures that the required dynamical agent is indeed a perturbation.}. This precession rate is enforced by collective effects, the ring self-gravity being often invoked to play this r\^ole. Note also that for an elliptic ring ($m=1$), the required contribution of the perturbations to the precession rate is reduced by a factor $\mathcal{O}(J_2)$. Therefore, the $m=1$ mode is easier to maintain than other modes over a given span in semi-major axis (it requires a lower level of perturbation), or instead it can exist over more extended regions in semi-major axis for a given magnitude of the perturbing acceleration.

We conclude this section on kinematics with the behavior of the ring surface density. This is derived through standard Lagrangian reasoning by noting that the conservation of a mass element $\delta m$ between its unperturbed ($\delta m = \sigma_0 a_e\delta a_e \delta \varphi_e$) and perturbed ($\delta m= \sigma r\delta r\delta\theta$) locations leads to the following relation between the unperturbed and perturbed surface density
\begin{equation}
\sigma(a_e,\varphi_e)=\frac{\sigma_0(a_e)}{J}=\frac{\sigma_0(a_e)}{1-q\cos[m(\varphi_e-\Omega_p t) + m\Delta
+\gamma]},\label{sig}
\end{equation}
\noindent where
\begin{eqnarray}
q\cos\gamma & = & a_e\frac{d \epsilon}{d a_e},\label{qcos}\\
q\sin\gamma & = & ma_e\epsilon\frac{d\Delta}{da_e}.\label{qsin}
\end{eqnarray}
\noindent $J$ is the Jacobian of the change of variables $r,\theta \rightarrow a_e,\varphi_e$, to lowest order in eccentricity, and is dominated by the contribution of $\partial \epsilon/\partial a_e$ in narrow rings modes, and $\partial m\Delta/\partial a_e$ in wakes (far enough downstream) and density waves; in general $q\gg \epsilon$, justifying the neglect of terms of order $\epsilon$. Note that the definition of the complex eccentricity Eq.~\eqref{compe} implies that 
\begin{equation}\label{qcomplex}
q e^{\mi\gamma}=a\frac{dZ}{da}e^{-\mi m\Delta}.
\end{equation}

Note also that Eq.~\eqref{sig} gives the surface density at the perturbed location as $\sigma(r,\theta)$ once ($a_e,\varphi_e$) are expressed in terms of $r,\theta$. Finally, $q <1$ is enforced as $q > 1$ would imply streamline crossing. Collective effects (i.e., self-gravity and/or collisions) prevent this from happening.

\section{Theoretical background: perturbation equations}\label{sec:pert}

The perturbation equations of the epicyclic osculating elements of ring fluid particles can be obtained from standard variation of the constants techniques, with the help of the energy and angular momentum motion integrals. Noting by $R$ and $S$ the radial and tangential perturbing accelerations, they read \citep{LB91}:
\begin{eqnarray}
\frac{da_e}{dt} & = & \frac{2}{\kappa}\left[R\epsilon\sin M_e+\frac{\Omega}{\kappa}S\left(1+\epsilon\cos M_e\right)\right] +\mathcal{O}(\epsilon^2),\label{aepi}\\
\frac{d\epsilon}{dt} & = & \frac{1}{\kappa a_e}\left[R\sin  M_e + 2\frac{\Omega}{\kappa}S\cos M_e\right] +\mathcal{O}(\epsilon),\label{eepi}\\
\frac{d\varpi_e}{dt} & = & \dot\varpi_{plan.} 
+\frac{1}{\kappa a_e\epsilon}\left[-R\cos M_e
+2\frac{\Omega}{\kappa}S\sin M_e\right] +\mathcal{O}(\epsilon^0),\label{pomepi}\\
\frac{d M_e}{dt} & = & \kappa+\frac{1}{\kappa a_e\epsilon}
\left[R\cos M_e - 2\frac{\Omega}{\kappa}S\sin M_e\right] +\mathcal{O}(\epsilon^0),\label{meanepi}
\end{eqnarray}
\noindent where  $\dot\varpi_{plan.} = \Omega-\kappa$  represents the effect of the planet oblateness on the precession of the apses [see Eq.\eqref{syncons}]. As $d\varphi_e/dt - \Omega=d(\varpi_e+M_e)/dt - \Omega$ is negligible compared to $d\varpi_e/dt$, only the first three equations are needed. 

Also, $\Omega/\kappa=1 + \mathcal{O}(J_2)$. Because $R$ and $S$ are small (compared to the planet's acceleration) and $\epsilon\ll 1$, one can replace this ratio with $1$ in Eqs.~\eqref{rmode}, \eqref{rmode0}, \eqref{aepi}, \eqref{eepi}, \eqref{pomepi} and \eqref{meanepi}; one can also replace everywhere $\kappa$ and $\Omega$ by the effective mean motion $n$ except in $\Omega - \kappa$, to the same level of precision. In this limit, these relations become formally identical to their elliptic counterparts, except for the notable fact that $a_e$ and $\epsilon$ differ from the osculating elliptic $a$ and $e$ by terms of order $J_2$ (in particular, for circular motion, this makes $a_e =r$ and $\epsilon=0$ hold exactly for the epicyclic osculating elements). This simplification is made from now on.

The perturbation equations are given below both in the continuum limit and in their discretized, $N$ streamline form, which is more appropriate for numerical work and for some heuristic discussions. As usual, the perturbation equations are averaged to remove the shortest (orbital) time-scale from the problem. It is customary in celestial mechanics to average over time, but in fact, the averaging variable can be freely chosen; it turns out that averaging over $\varphi_e$ instead of time is more appropriate for the fluid problem discussed here where $a_e$ and $\varphi_e$ must be treated as independent variables (see \citealt{L92} for more details on this question).

The relations of the following subsections are presented in a full and rather compact way for generality and ease of reference, but are mostly used to discuss simple models and gain physical insight in the remainder of the theoretical analysis presented in this chapter. For the epicyclic periapsis angle, only the perturbation contribution to Eq.~\eqref{pomepi} is given. From now on, and in order to alleviate notations, we drop the subscript ``\textit{e}" on the epicyclic osculating elements.

\subsection{Self-gravity}\label{sec:sg}

Let us consider two $m$-lobed streamlines (indexed by $i$ and $j$) with orbital elements $a_i,\epsilon_i, \Delta_i$ and $a_j,\epsilon_j, \Delta_j$ respectively. We wish to compute the gravitational perturbation of streamline $j$ on streamline $i$. In fact, it is sufficient to compute the gravitational perturbation on a fluid particle of streamline $i$, because this perturbation is identical for all fluid
particles, once averaged over the orbital phase $\varphi$. For $\Delta a_{ij}\equiv a_i-a_j$ small enough, one can locally identify the streamline with a straight line and find the perturbing acceleration with the help of Gauss's theorem. The gravitational acceleration $\boldsymbol{g}_{sg}$ on a fluid particle is then given by \citep{GT79b}
\begin{equation}\label{sgacc}
\boldsymbol{g}_{sg}=\frac{2G\lambda_j}{\Delta_c}\boldsymbol{u},
\end{equation}
\noindent where $\boldsymbol{u}$ is the unit vector perpendicular to streamline $j$ and directed towards this streamline from the considered fluid particle of streamline $i$, $\lambda_j$ the linear mass density of streamline $j$, and $\Delta_c$ the distance of the fluid particle to streamline $j$ along ${\bf u}$. Note that this asymptotic approximation has not only been used in papers based on the streamline formalism, but actually in most if not all theoretical studies of the r\^ole of self-gravity in rings (e.g., \citealt{SYL85,SDLYC85,PM05})

Projecting on the radial and azimuthal directions gives the relevant acceleration to lowest non-vanishing order in eccentricity; averaging over $\varphi$ yields the required perturbation equations, with $m_j=2\pi a_j\sigma_0^j\delta^{\pm}(a_j)$ being the mass of streamline $j$ [$\delta^{\pm}(a_j)$ is the streamline width\footnote{The rationale of this notation will become apparent later on.}]:
\begin{eqnarray}
\left(\frac{da_i}{dt}\right)_{sg} & = & -\frac{2(m-1)n}{\pi}\frac{m_j}{M_p}
\frac{a}{\Delta a_{ij}}a\epsilon_i \times H(q_{ij}^2)q_{ij}\sin\gamma_{ij},\label{asg}\\
\left(\frac{d\epsilon_i}{dt}\right)_{sg} & = & \frac{n}{\pi}\frac{m_j}{M_p}
\frac{a}{\Delta a_{ij}}H(q_{ij}^2) q_{ij}\sin\gamma_{ij},\label{esg}\\
\left(\frac{d\varpi_i}{dt}\right)_{sg} & = & \frac{n}{\pi}\frac{m_j}{M_p}
\frac{a}{\epsilon_i\Delta a_{ij}}H(q_{ij}^2)
q_{ij}\cos\gamma_{ij},\label{pomsg}
\end{eqnarray}
\noindent where $H(q^2)$ is defined by
\begin{equation}\label{hq2}
H(q^2)\equiv \frac{1}{2\pi q}\int_{-\pi}^{\pi}\frac{\cos u}{1-q\cos u}du
= \frac{1-(1-q^2)^{1/2}}{q^2(1-q^2)^{1/2}}.
\end{equation}
\noindent and\footnote{Do not confuse the imaginary number $\mi$ with the index $i$.}
\begin{equation}
q_{ij}{\exp}(\mi\gamma_{ij})=\frac{a_i\epsilon_i - a_j\epsilon_j{\exp}
[\mi m(\Delta_j-\Delta_i)]}{\Delta a_{ij}}.\label{qij}
\end{equation}
The action of the ring self-gravity results from the sum of these contributions over all streamlines $j\neq i$. In these expressions, as well as in the following ones for the pressure tensor and satellite perturbations, all slowly varying spatial dependences on $a$ can be ignored, except in the difference $a_i - a_j$. 

Eqs.~\eqref{phasetime}, \eqref{esg} and \eqref{pomsg} lead to the following expression for the self-gravity contribution to the time evolution of the complex eccentricity $Z_i$:
\begin{eqnarray}
\left(\frac{dZ_i}{dt}\right)_{sg} & = &\! \!\!\!\!  -\mi \sum_{j\neq i}\frac{n}{\pi}\frac{m_j}{M_p} a^2
H(q_{ij}^2)\frac{Z_i-Z_j}{(\Delta a_{ij})^2}\nonumber\\
& = &\! \!\!\!\!  -\mi \int d a_j \frac{2 n_i a_i^3 \sigma^j_0}{M_p} H(q_{ij}^2)\frac{Z_i-Z_j}{(\Delta a_{ij})^2},\label{zsg}
\end{eqnarray}
with a similar expression for $da_i/dt$. 

Self-gravity conserves both energy and angular momentum, globally. It just redistributes it among streamlines. Eqs.~\eqref{specen}, \eqref{specang}, \eqref{asg} and \eqref{esg} allow us to compute the related changes; the task is simplified by noting that the self-gravitational potential is stationary in the rotating frame for stationary streamlines\footnote{Deviations from stationarity are slow and therefore negligible.}, so that changes in specific energy and angular momentum are related by Jacobi's integral and by introducing the self-gravity energy and angular momentum luminosities $L_E^{sg}(a_i) = \Omega_p L_H^{sg}(a_i)$ (i.e. the fluxes of energy and momentum across the entire streamline $i$). These quantities are more compactly expressed in continuous form
\begin{eqnarray}
L_E^{sg} & = & \Omega_p L_H^{sg} = -4\pi Gm\Omega_p\int_0^a da_1\ \sigma_0(a_1)a_1\epsilon_1\times\nonumber\\ 
& & \quad \int_a^{+\infty}da_2\ \sigma_0(a_2)a_2
\frac{H(q_{12}^2)q_{12}\sin \gamma_{12}}{a_1-a_2},\label{sgflux}
\end{eqnarray}
but the equivalent discrete sum is readily obtained. With this definition,
\begin{eqnarray}
\left(\frac{dE_i}{dt}\right)_{sg} & = & -\frac{1}{2\pi a_i \sigma^i_0}\frac{\partial L_E^{sg}}{\partial a} = -\frac{\Delta^{\pm} (L_E^{sg})}{m_i},\label{specesg}\\
\left(\frac{dH_i}{dt}\right)_{sg} & = & -\frac{1}{2\pi a_i \sigma^i_0}\frac{\partial L_H^{sg}}{\partial a} = -\frac{\Delta^{\pm} (L_H^{sg})}{m_i},\label{specangmomsg}
\end{eqnarray}
with
\begin{equation}
\Delta^{\pm} (X) \equiv X^{i,i+1}-X^{i-1,i},\label{deltaop}
\end{equation}
where $X^{i,j}$ is evaluated at the boundary between streamlines $i$ and $j$ [i.e. in $(a_i+a_j)/2$ for evenly spaced streamlines]. Note that at the ring edge, $\Delta^{\pm}$ involves only one side; this is required because the luminosities are conserved quantities: whatever flux from inner streamlines crosses a streamline boundary must be deposited on the outer streamlines\footnote{This property is more transparent if one writes an equation for the energy and angular momentum density rather than for their specific counterpart; see \citealt{BGT85}.}. Note also that if streamlines are aligned [i.e., $m\Delta(a)$ is constant], the luminosities vanish.

\subsection{Resonant satellite forcing}\label{sec:sat}

We only consider here the effect of a single satellite resonance on a circular ring. The question of the effect of multiple resonances on the overall evolution of a narrow eccentric ring is more complex and will be discussed later on (section \ref{sec:ecc}).

It is customary in fluid descriptions of ring systems to expand the perturbing potential with respect to the satellite motion to obtain a function of ($r,\theta$); this leads to a double Fourier expansion of the satellite potential, one in azimuth $\theta$ and one in time, as the satellite orbit is closed in its precessing frame. The terms of the resulting Fourier series are of the form (see, e.g., \citealt{GT80,GT82,Sh84})
%
\begin{equation}\label{satpot}
\phi_s(r,\theta,t)=\Phi_{mk}(r/a_s)\cos m(\theta-\Omega_p t),
\end{equation}
with $m \ge 0$ and $k\in \mathbb{Z}$; here, one conventionally assumes that $m< 0$ at outer Lindblad resonances (OLR), but this does not introduce new terms in this Fourier expansion. The Fourier coefficients $\Phi_{mk}$ can be expressed in terms of Laplace coefficients (\citealt{MD99}; a particularly synthetic and convenient derivation can be found in Appendix A of \citealt{Sh84}); they are of order $e_s^{|k|} \ll 1$, where $e_s$ is the satellite eccentricity. The pattern speed $\Omega_p = \Omega_s + k/|m|\ \kappa_s$ has been discussed in the kinematics section (\ref{sec:kin}), as has the implicit choice in origins of time and angle.

This form of the potential leads to the following expressions of the phase-averaged perturbation equations:

\begin{eqnarray}
\left(\frac{da_i}{dt}\right)_s & = & n a(m-1)\epsilon _i\frac{a\Psi_{mk}}{G
M_p}\sin m\Delta_i,\label{asat}\\
\left(\frac{d\epsilon_i}{dt}\right)_s & = & -n \frac{a\Psi_{mk}}{2 G
M_p}\sin m\Delta_i,\label{esat}\\
\left(\frac{d\varpi_i}{dt}\right)_s & = & \frac{n}{\epsilon_i} 
\frac{a\Psi_{mk}}{2 G M_p}\cos m\Delta_i,\label{pomsat}
\end{eqnarray}
\noindent where one has defined
\begin{equation}\label{psimk}
\Psi_{mk}(a) \equiv a\frac{d\Phi_{mk}}{da} + 2m\Phi_{mk}.
\end{equation}
This quantity is an effective potential energy characterizing the strength of the satellite ($m,k$) Lindblad resonance.

For completeness, let us express Eqs.~\eqref{esat} and \eqref{pomsat} in terms of $Z_i$ (Eq.~\eqref{compe}):
\begin{equation}
\left(\frac{dZ_i}{dt}\right)_s = -\mi n 
\frac{a\Psi_{mk}}{2 G M_p}.\label{zsat}
\end{equation}
Eq.~\eqref{asat} gives the rate of resonant energy exchange with the satellite for streamline $i$
\begin{equation}\label{specsat}
\left(\frac{dE_i}{dt}\right)_{s}= \Omega_p\frac{T^s_i}{m_i}= \frac{\Omega_p}{2} \Psi_{mk}\epsilon_i\sin m\Delta_i
\end{equation}
where $T^s_i$ is the satellite torque on streamline $i$ (the rate of energy and angular momentum exchange are again related through Jacobi's constant). In integral form, the total torque
reads
\begin{equation}\label{torque}
T_s=\int da\ \pi ma\sigma_0\epsilon\Psi_{mk}\sin
m\Delta=\int da\ \mathfrak{T}_s,
\end{equation}
where $\mathfrak{T}_s$ is the torque density and the integral extends over the whole perturbed region. The satellite torque is important only at resonances, where it leads to secular exchanges of angular momentum with the ring. Note that the torque density is negative for an outer satellite (inner Lindblad resonance) and positive for an inner satellite (outer Lindblad resonance), either due to density wave physics (see e.g.\ \citealt{Sh84}) or because of collisional dissipation \citep{GT81}. 

\subsection{Stress tensor}\label{sec:press}

\subsubsection{Perturbation equations}\label{sec:stresseq}

The pressure tensor is the most difficult of all perturbing agents to quantify from a theoretical point of view. Because the collision time-scale of dense rings is comparable to the orbital time-scale, none of the usual asymptotic regimes of fluid dynamics or physical kinetics applies; furthermore, the ring particles' collisional properties are not well-known.

Although many articles have been devoted to analyzing ring kinetics in unperturbed (circular) background motion, very few theoretical studies have focused on the behavior of the pressure tensor in perturbed flows, either in the small filling-factor ---dilute --- \citep{BGT83b,SDLYC85} or large filling-factor ---compact --- \citep{BGT85} limits. Furthermore the available studies rely on a number of simplifying assumptions (identical hard spheres, etc). However, they give some guidelines on appropriate modifications of and deviations from the common (simplest) pressure-viscosity description with constant kinematic viscosity (see the chapter by Colwell et al.). These theories are not presented here in any detail; the interested reader will find an \textit{ab initio} construction and discussion in \cite{L92}, besides the original publications mentioned above.

Probably the most important piece of physics that is lacking from available characterizations of the ring effective pressure tensor in perturbed ring regions relates to the role of small-scale self-gravitational wake two-dimensional weak turbulence, as quantified in circular Keplerian flows by, e.g., \cite{DI01} and \cite{YOD12}; this process has also been studied in the context of accretion disk theory (see, e.g., \citealt{G01} and the chapter by Latter et al. for more details). Some numerical work on this front would be welcome; in the absence of such studies, the results of these authors will be used as a first approximation to quantify wake-induced transport in eccentric background motion. 

The pressure tensor is defined by\footnote{The pressure tensor is the opposite of the stress tensor, and the two designations may indifferently be used.}
\begin{equation}
P_{\alpha\beta}= \int dz \int (v_\alpha-u_\alpha)(v_\beta-u_\beta)fd^3\boldsymbol{v},\label{presseq}
\end{equation}
\noindent where $f(\boldsymbol{r},\boldsymbol{v},t)$ is the distribution function of ring particles; $\boldsymbol{v}$ designates the velocity of individual ring particles, while $\boldsymbol{u}$ refers to their local average (i.e., it is a fluid particle velocity); $\alpha$ and $\beta$ stand for components along orthonormal vector directions $\bm{e}_\alpha, \bm{e}_\beta$ (such as $\bm{e}_r$, $\bm{e}_\theta$, $\bm{e}_z$). This mean velocity is computed from the standard osculating velocity except that the mean anomaly is expressed through Eq.~\eqref{anom}; this follows because one wishes to understand the stress tensor response to the assumed modal motion.

The pressure tensor captures the effect of particle collisions from one streamline on its neighbors. It is often estimated in the earlier BGT papers from the vertically integrated force per unit length of streamline exerted by the outside material, 
$-(\boldsymbol{n} \cdot \boldsymbol{e}_{\alpha})P_{\alpha\beta}\boldsymbol{e}_{\beta}$ ($\alpha,\beta=r,\theta$, $\boldsymbol{n}$ being the normal to the streamline boundary). However, this approach is not precise enough to capture the effect of collisional energy loss \citep{BGT85,BGT89}, which can only be obtained from the exact form of the acceleration:
\begin{eqnarray}
R & = & -\frac{1}{\sigma}\left[\frac{1}{r}
\frac{\partial(rP_{rr})}{\partial
r}+\frac{1}{r}\frac{\partial P_{r\theta}}{\partial\theta}-
\frac{P_{\theta\theta}}{r}\right],\label{radialstress}\\
S & = & -\frac{1}{\sigma}
\left[\frac{1}{r^2}\frac{\partial(r^2 P_{r\theta})}{\partial r} +
\frac{1}{r}\frac{\partial
P_{\theta\theta}}{\partial\theta}\right].\label{azimstress}
\end{eqnarray}
The resulting phase-averaged perturbation equations on streamline $i$ from the neighboring ring material read

\begin{eqnarray}
\left(\frac{da_i}{dt}\right)_{vis} & = & 
-\frac{1}{\pi a^2_i n^2_i \sigma^i_0}\frac{\partial (L_E^{vis})}{\partial a} + \frac{1}{a_i n_i\sigma^i_0}(2q_it^i_1 - 3a^i_{r\theta})\nonumber\\
& = & -\frac{2}{a_i n^2_i m_i}\Delta^{\pm} (L_E^{vis}) + \frac{2\pi\delta^{\pm}(a)}{n_i m_i}(2q_it^i_1 - 3a^i_{r\theta}),\label{arphi}\\
\left(\frac{d\epsilon_i}{dt}\right)_{vis} & = & \frac{1}{a_i n_i \sigma_0^i}\left[-\frac{\partial}{\partial a}(t_1\cos\gamma+t_2\sin\gamma + (t^i_1\sin\gamma_i -t^i_2\cos\gamma_i)\frac{\partial m\Delta}{\partial a} \right]\nonumber\\
& = & \frac{2\pi}{n_i m_i}\left[
-\Delta^{\pm}(t_1\cos\gamma+t_2\sin\gamma) + (t^i_1\sin\gamma_i -t^i_2\cos\gamma_i)\delta^{\pm} (m\Delta)\right],\nonumber\\ \label{t1}\\
\left(\frac{d\varpi_i}{dt }\right)_{vis} 
& = & \frac{1}{\epsilon_i a_i n_i \sigma_0^i}\left[\frac{\partial}{\partial a}(t_1\sin\gamma -t_2\cos\gamma) + (t^i_1\cos\gamma_i+t^i_2\sin\gamma_i)\frac{\partial m\Delta}{\partial a}
\right]\nonumber\\
& = & \frac{2\pi}{\epsilon_i n_i m_i}\left[\Delta^{\pm}(t_1\sin\gamma -t_2\cos\gamma) + (t^i_1\cos\gamma_i+t^i_2\sin\gamma_i)\delta^{\pm}(m\Delta)
\right],\nonumber\\ \label{t2}
 \end{eqnarray}
\noindent where $m_i$ is the total mass of the considered streamline, $\gamma$ is defined through Eqs.~\eqref{qcos} and \eqref{qsin}, and $\delta^{\pm}(a)$ the streamline width. $\Delta^{\pm}$ is the operator introduced in Eq.~\eqref{deltaop}; as before, at the boundary $\Delta^{\pm}$ involves only one side as the finite difference stands for a conserved quantity; $\delta^{\pm}(m\Delta)$ on the other hand stands for the variation across the considered streamline, including for the two boundary streamlines so that the difference of definition between $\Delta^\pm$ and $\delta^\pm$ is important only for these boundary streamlines. $t_1,t_2$ and $a_{r\theta}$ are defined by\footnote{This definition as well as the results Eqs.~\eqref{arphi}, \eqref{t1} and \eqref{t2} differs from \cite{BGT83a,BGT85,BGT86} in several respects, but is consistent with \cite{LR95}; see \cite{L92} for details. The major difference comes from the new terms proportional to $\delta^{\pm}(m\Delta)$. The presence of these terms is unavoidable for two reasons: firstly, the perturbation equations must be expressed in principle in terms of the exact solution, which involves the identity $M=m(\varphi-\Omega_p t) + m\Delta$; secondly, the averaging over $\varphi$ relies on the fact that $a$ and $\varphi$ are independent semi-Lagrangian labels.}
\begin{eqnarray}
t_1 & = & s_{rr}+2c_{r\theta},\label{t1def}\\
t_2 & = & 2s_{r\theta}-c_{rr},\label{t2def}\\
c_{\alpha\beta}+\mi s_{\alpha\beta} & = & \langle\exp(\mi M')P_{\alpha\beta}(M')\rangle,\label{cijsijdef}\\
a_{\alpha\beta} & = & \langle P_{\alpha\beta}(M')\rangle,\label{aijdef}
\end{eqnarray}
\noindent where the bracket notation stands for the azimuthal average\footnote{The pressure depends on azimuth only through $M'$; see e.g.\ \cite{BGT83b,SDLYC85}.} over $M'=M_i+\gamma_i$. In Eq.~\eqref{arphi}, $L_E^{vis}$ is the viscous energy luminosity (the flux of energy through an entire streamline boundary). To leading order in $\epsilon$,
\begin{equation}\label{visflux}
L_E^{vis} = n L_H^{vis} + \mathcal{O}(\epsilon)= 2\pi n a^2a_{r\theta} + \mathcal{O}(\epsilon),
\end{equation}
where the viscous angular momentum luminosity $L_H^{vis}$ has also been introduced for later use. The expression for $L_H^{vis}$ makes direct physical sense by noting that, to leading order in eccentricity, the torque at the streamline boundary exerted by the outer material is $a P_{r\theta}$, per unit length. As a consequence, $a_{r\theta}$ is the effective stress tensor component that characterizes radial viscous energy and angular momentum fluxes. The connection between the energy and angular momentum fluxes results directly from Eqs.~\eqref{specen} and \eqref{specang}. Finally, $t_1$ and $t_2$ are viscous-like and pressure-like quantities, an interpretation justified below.
 
Consequently the first term on the right-hand side of  Eq.~\eqref{arphi} represents the viscous energy redistribution across streamlines, while the second term represents the loss of orbital energy due to dissipation during collisions as shown explicitly by \cite{BGT85}, so that $2q t_1 - 3a_{r\theta} < 0$. Note also that in perturbed flows, this dissipation term is smaller than the luminosity contribution by a factor $\sim w/a$ where $w$ is the width of the perturbed region; however, in circular flows (i.e., in the standard viscous diffusion regime), the dissipation and flux terms are of the same order of magnitude.

The pressure tensor contribution to the evolution of $Z_i$ follows directly from Eqs.~\eqref{t1} and \eqref{t2}:
 \begin{eqnarray}
\left(\frac{dZ_i}{dt}\right)_{vis}  & = & \mi \frac{Z_i}{|Z_i|} \frac{1}{a_i n_i \sigma_0^i}\left(
\frac{\partial}{\partial a}[(t_2 +\mi t_1)e^{\mi \gamma}] + (-t^i_1 +\mi t^i_2)e^{\mi \gamma^i}\frac{\partial m\Delta}{\partial a}\right)\nonumber\\
& = & \mi \frac{Z_i}{|Z_i|} \frac{2\pi}{n m_i}\left(
\Delta^\pm[(t_2 +\mi t_1)e^{\mi \gamma}] + (-t^i_1 +\mi t^i_2)e^{\mi \gamma}\delta^\pm(m\Delta)\right).\label{zvis}
\end{eqnarray}

The remainder of this section is devoted to the questions of angular momentum flux reversal and energy dissipation, as they play a crucial r\^ole in the theory of sharp boundaries (section \ref{sec:shep}). The properties of four different physical models will be summarized to this purpose: the standard, constant kinematic viscosity hydrodynamic limit, the low-filling factor limit, the dense granular flow limit, and finally the transport induced by self-gravitational wakes.

\subsubsection{Constant viscosity hydrodynamic limit}\label{sec:hydro}

This limit has at least one important limitation, but nonetheless possesses a number of qualitative features that are common to all models and are more easily brought to light than in more sophisticated approaches; this makes it a convenient starting point to grasp the relevant features of the stress tensor behavior in perturbed regions. In this limit, $P_{\alpha\beta}=P\delta_{\alpha\beta} -\nu(\sigma_0/J)(\partial u_\alpha/\partial x_\beta + \partial u_\beta/\partial x_\alpha)$ ($P$ being the pressure and $\delta_{\alpha\beta}$ the Kronecker symbol, with cartesian coordinates used), so that the pressure tensor coefficients read\footnote{The bulk viscosity term is neglected for simplicity, as these relations are used here in a qualitative way.}

\begin{eqnarray}
t_2 & = & -q H(q^2) \sigma_0 c^2 ,\label{t2hydro}\\
t_1 & = & - \frac{ q}{(1-q^2)^{3/2}}\nu \sigma_0 n,\label{t1hydro}\\
a_{r\theta} & = & \frac{3 - 4q^2}{2(1-q^2)^{3/2}}\nu \sigma_0 n.\label{arphihydro}
\end{eqnarray}
\noindent A simple form has been assumed here for the pressure, for illustrative purposes ($P=\sigma c^2$ where $c$ is the velocity dispersion, assumed independent of azimuth for simplicity). This shows that $t_2\propto P_0$ is a ``pressure-like" coefficient, while $t_1\propto\nu$ and $a_{r\theta}\propto \nu$ are ``viscous-like" (a generic feature). Note that $t_1,t_2 \le 0$ while $a_{r\theta} \ge 0$ for small $q$ and changes sign for $q\equiv q_a=\sqrt{3}/2\approx 0.87$ ($a_{r\theta} \le 0$ for $q \ge q_a$). This change of sign is common to all theories explored so far, although the quantitative details vary (in particular, $q_a$ is smaller in the more sophisticated theories summarized next). This reflects the fact that for sufficiently strong perturbation (high enough $q$) the shear of the flow changes sign not only locally, but on average along any given streamline; this is the origin of the angular momentum flux reversal needed for the standard BGT model of sharp edge confinement (section \ref{sec:shep}).

The main limitation of the hydrodynamic limit with constant viscosity is that it cannot capture properly the energy dissipation required to ensure the confinement of sharp ring boundaries (section \ref{shflux}) and overestimates the value of $q_a$. In particular, for this dissipation to occur, the effective kinematic viscosity \textit{must} depend either on the velocity dispersion (for dilute rings) or the surface density (for compact rings); adding an internal energy equation on the basis of the constant kinematic viscosity prescription cannot resolve the issue, as this will not affect the rate of energy injection from the fluid mean motion to internal motions (velocity dispersion). More elaborate physics is required for this, that is now described.

\subsubsection{Hot dilute limit (local transport)}\label{sec:dilute}

Quite a few analytic and numerical studies have been made in this regime for unperturbed (circular) Keplerian flows, but very few for perturbed (eccentric) ones. The two available analyses were performed with the help of the Boltzmann equation, with two different forms of the collision term --- Boltzmann \citep{BGT83b} and Krook \citep{SDLYC85}. These two analyses neglect a number of important processes or conditions, such as the coupling between the translational and rotational degrees of freedom, mutual gravitational stirring of ring particles, or the existence of a distribution of particle sizes, and make use of highly simplified particle collisional properties --- although to some extent the Krook model can be parametrized to mimic some of these features.

In this regime, the pressure tensor coefficients $t_1,t_2, a_{r\theta}$ naturally scale as $\sigma_0 c_0^2$ where $c_0$ is the unperturbed velocity dispersion. The reversal of the angular momentum luminosity coefficient $a_{r\theta}$ can only be characterized numerically; \cite{BGT83b} have shown that this reversal occurs for $q(\tau)=q_a\simeq 0.6$ --- $0.8$ for an optical depth $\tau\sim 1$, noticeably lower than in the constant viscosity hydrodynamic limit.

There are only two ways to produce substantial enhancements of energy dissipation in this limit with respect to unperturbed regions: either through the effect of streamline compression and/or through increases of the velocity dispersion. This is now discussed with the help of the following heuristic argument adapted from \cite{GT78a}. 

From elementary kinetic theory reasoning, one can see that $\nu_l\sim \omega_c l^2$, where $l$ and $\omega_c$ stand for the mean free path and collision frequency respectively. The collision frequency can be expressed as \citep{SS85}
\begin{equation}\label{colfreq}
\omega_c\sim F n \tau, 
\end{equation}
where $F=(1+4\pi G \rho/ n^2)^{1/2}$ is the increase in vertical epicyclic frequency due to the ring's vertical self-gravity. The enhancement factor $F$ is typically of order 3 or 4 in Saturn's rings with respect to the usual estimate $n\tau$. The optical depth, as defined by \cite{SS85} is $\tau\simeq \sigma d^2/m_p$ within a factor of order unity ($d$ is the particle size and $m_p$ the particle mass). This expression for the collision frequency is valid as long as particles are not close-packed.

The mean free path $l\sim c/ n\sim a\epsilon$ when $\tau\ll 1$ (due to infrequent collisions during epicyclic oscillations) and $l\sim c/\omega_c$ when $\tau\gg 1$. The two regimes can be approximately captured by 
\begin{equation}\label{kinvis}
\nu_l\sim \frac{\omega_c c^2}{n^2(1+\omega_c^2/n^2)} \sim \frac{Fc^2\tau}{n(1+F^2\tau^2)}.
\end{equation}
The velocity dispersion is determined in turn by the balance of collisional dissipation of random motions per unit mass\footnote{The coefficient of restitution $\varepsilon_c$ is not to be confused with the eccentricity $\epsilon$.}, $\dot{E_c}\sim \omega_c(1-\varepsilon^2_c)c^2$, and viscous energy injection into random motions from the velocity shear per unit mass, $\dot{E}_c=2\nu n^2 G(q)$ [$G(q)$ is the dimensionless velocity shear\footnote{The velocity shear is $S=(u_{\alpha\beta}u_{\alpha\beta})^{1/2}$ where $u_{\alpha\beta}=(\partial u_\alpha/\partial x_\beta + \partial u_\beta/\partial x_\alpha)/2$ in cartesian coordinates (implicit summation over repeated indices); $G(q)=S/\Omega$.}]. The balance of the two yields \citep{BGT85}
\begin{align}\label{dispvel}
\varepsilon^2_c(c)\sim 1 - \beta\frac{G(q)}{1+F^2\tau^2},
\end{align}
where $\beta$ is a constant of order unity ($\beta\sim 0.5$, somewhat model dependent). This indirectly sets the magnitude of the velocity dispersion through the dependence of $\varepsilon_c$ on $c$.

Note that this argument ignores many actual complications (velocity anisotro-py, particle size distribution, particle spin, non-zero filling factors, etc) but does capture an essential qualitative feature of dilute rings: as the shear increases with the level of perturbation $q$, the velocity dispersion $c$ must increase due to energy balance requirements, at given ring optical depth $\tau$. Note however that Eq.~\eqref{dispvel} implicitly limits the dimensionless shear: $G(q)\lesssim 1+F^2\tau^2$. Consequently, large energy injection rates into random motion do not come from the dimensionless shear factor, but must arise from increases in the kinematic viscosity from large velocity dispersion (i.e., $\varepsilon_c\approx 0$). This heuristic conclusion is confirmed by numerical calculations. Such large enhancements of the velocity dispersion are beyond the scope of the hydrodynamic limit with constant kinematic viscosity, although more sophisticated versions may capture the relevant physics.

However, the local-transport limit applies rarely if at all in major rings, as this would require the disk to be hot enough (i.e., thick enough), which in turn requires sufficiently inelastic collisions for a ring of ice hard spheres, as seems likely . In practice, such a condition seems unlikely to occur for realistic ring particle collisional behavior, which are most likely essentially ice hard spheres\footnote{A more quantitative heuristic version of this argument was made by \cite{BGT85} in perturbed regions.}

As a consequence, let us now turn to the more relevant opposite limit, which can be captured in a granular flow model described next. 

\subsubsection{Cold dense granular flow limit (nonlocal transport) limit}\label{sec:compact}

Another scaling of the pressure tensor applies in this opposite limit, where the ring volume mass density $\rho$ is comparable to the particle density due to close-packing so that $t_1,t_2, a_{r\theta} \propto \rho n^2 H_0^3 \propto \sigma_0^3$ ($H_0$ is the unperturbed ring thickness) and the ring behaves more like a (non-Newtonian) liquid than a gas. 

Perturbed rings in this regime have been explored from a granular flow model by \cite{BGT85}. This theory also predicts that angular momentum flux reversal occurs for large enough perturbations and that $t_1$ can be positive for small $q$, in which case viscous overstabilities can occur; this is the only known theory having this property. In any case, $t_1$ must be negative for large $q$ because energy dissipation during collisions [the first term on the right-hand side of  Eq.~\eqref{arphi}] requires $3 a_{r\theta} > 2 q t_1$ while $a_{r\theta} < 0$ for $q > q_a$.

The energetics associated with granular flows differs from the other two limiting cases. The microphysics needs to be adapted to the regime where the ring particle size $d$ is larger or much larger than the particle distance\footnote{The distance $s$ is the distance between the particle's surfaces, not between the particle's centers.} $s$ ($d\gg s$); in particular, the dynamic viscosity reads\footnote{For an elementary justification of this form of the viscosity, see \cite{H83} and \cite{L92}.} 
\begin{equation}
\eta_{nl}\sim \omega_c \rho d^2 \sim \rho d^2 c/s,\label{kinvis2}
\end{equation}
where $\omega_c$ is the collision frequency and $\rho$ is the particle density, nearly equal to the ring density, while $\eta_{nl} n G(q) \sim  \omega_c  \rho c^2 \sim \rho c^3/s$ from the balance of energy dissipation [cf \citealt{BGT85}, Eqs.~(47) and (52), respectively; alternatively, see \citealt{L92}, section 5.2]. This implies
\begin{equation}\label{dispvel2}
c\sim n d G(q). 
\end{equation}
Note that the close packing condition ($s\ll d$) prevents the velocity dispersion from canceling out of the internal energy balance, so that the dependence of $\varepsilon$ on $c$ becomes secondary in setting the magnitude of $c$; incidentally, this also decouples the velocity dispersion from the optical depth (a characteristic of dilute systems pointed out above) and the two can be chosen independently, as long as the large filling factor condition is maintained. 

The collision frequency in the granular flow model can be substantially enhanced with respect to Eq.~\eqref{colfreq}: particles move at velocity $\sim c$ over a distance $\sim s$ between two collisions so that the collision frequency reads\footnote{For the second approximation, see Eq.~\eqref{dispvel2} above. For the last approximation, Eqs.~(47) and (56) of \cite{BGT85} imply that $d^3\sim H_o^2 s$; see also \citealt{L92}, section 5.2. The first approximation assumes $\omega_c > n$, as does the last expression in Eq.~\eqref{kinvis2}, whereas the first is always valid.}:
\begin{equation}\label{colfreq2}
\omega_c\sim \frac{c}{s}\sim \frac{d}{s} n\sim F \left(\frac{H_0}{d}\right)^2 n, 
\end{equation}
The enhancement with respect to the usual estimate ($\Omega\tau$) can easily be of an order of magnitude or even much larger. 

The magnitude of the related energy dissipation can be quantified in the granular flow model in a rather straightforward manner, as explicit equations have been derived by \cite{BGT85} for the various pressure tensor coefficients. Performing this task for parameters appropriate to dense systems, the granular flow model predicts\footnote{Calculations by the author.} that $a_{r\theta}$ changes sign for $q=q_a\approx 0.78$ and that $2q_a |t_1(q_a)|\gtrsim 10^2 a_{r\theta}(0)$ so that the increase in dissipation in perturbed flows with respect to circular ones can actually be quite substantial, in particular in the regime that is relevant for maintaining sharp boundaries (section \ref{sec:shep}).

Note that, in contrast with the previous case, most of this enhancement in energy dissipation comes from the rather steep scaling with the surface density ($\propto \sigma^3$), i.e.\ from streamline compression, and not from the velocity dispersion, which always remains of order $n d$.

\subsubsection{Transport regimes and stress tensor models}\label{sec:pressmod}

It is of some interest to assess more precisely the conditions in which one of these two regimes dominates, as the two modes of transport (local and nonlocal) are always present. Note that the first expressions of Eqs.~\eqref{kinvis} and \eqref{kinvis2} are always valid, whereas the subsequent ones make assumptions about the physical conditions. In particular, note that for $\tau\gg 1$, Eqs.~\eqref{colfreq} ($\omega_c\propto H_0/d$) and \eqref{colfreq2} ($\omega_c\propto (H_0/d)^2$) are incompatible as the first ignores the enhancement in collision frequency due to close-packing; the first one requires low filling factors ($FF\ll 1$ so that $s\gg d$) while the second one assumes large filling factors ($FF\sim 1$ with $s\ll d$).

Nevertheless, keeping the collision frequency unspecified, one has
\begin{equation}
\frac{\nu_{nl}}{\nu_l} \sim \left(\frac{n d}{c}\right)^2 (1+\omega_c^2/n^2).\label{transpratio}
\end{equation}
In the close-packing regime, ($d \gg s$) nonlinear transport dominates due to Eqs.~\eqref{dispvel2} and \eqref{colfreq2}, as expected. In the dilute regime, $d\ll s$, small filling factor), it is often assumed that local transport dominates but the above relation makes this assumption not sufficient and the ring must be sufficiently hot as well\footnote{I thank Henrik Latter for pointing this out to me.}. Eq.~\eqref{dispvel} implies that the coefficient of restitution should always be significantly different from unity, which in turn implies for hard spheres that $c \gg n d \sim 1$ mm/s in Saturn's rings (see, e.g., \citealt{BHL84}). On the other hand the results of \cite{SDLYC85} imply that for optical depth of order unity and strong enough perturbation, the particles may be elastic enough, making the ratio Eq.~\eqref{transpratio} of order unity. 

In any case, for rings $\tau\gtrsim 1$, if particles are too inelastic, a close-packed configuration where nonlocal transport dominates is unavoidable. Conversely in quasi-monolayer rings with $\tau \lesssim 1$ both modes of transport are expected to be of comparable magnitude, although in this case, using the granular flow model heuristics seems more appropriate at least qualitatively. In fact, it is unclear if (or even unlikely that) the local transport dominates anywhere in the major ring systems of Saturn and Uranus.

This discussion nevertheless indicates that it would be very useful to have a stress tensor model that can span the two regimes for relevant choices of the model parameters. Such a model has been devised by
\cite{BGT86}, based on the following generic physical constraints that have been discussed in this section:
\begin{enumerate}
\item Streamlines cross at $q=1$, so that the pressure tensor is
likely to diverge as $q\rightarrow 1$, or even for smaller values of $q$ from Eq.~\eqref{dispvel}.
\item $t_1$ and $t_2$ vanish as $q\rightarrow 0$, because the flow,
and therefore the pressure tensor components, become axisymmetric in
this limit. Hence, it is reasonable to assume that $t_{1,2}\propto q$ 
for small $q$, as this dependence is characteristic of both the local and nonlocal transport models. Notice that $t_1$ is negative in the first, whereas it is positive for small values of $q$ in
the dense model; $t_2$ is negative in both models.
\item We have shown that the energy dissipation due to inelastic
collisions implies $2qt_1<3a_{r\theta}$. This is a general result, which
does not depend on the choice of the collisional model.
\item We have also seen that $a_{r\theta}$ is in general positive
for small $q$, so that angular momentum flows outwards. However, as
$q\rightarrow 1$, the direction of the angular momentum flow is
reversed, so that there is some value $q=q_a(\sigma_0)$ for which
$a_{r\theta}=0$ and the transition between the two regimes occur. For
dense systems, $q_2$ is independent of $\sigma_0$.
\end{enumerate}

From these general considerations, \cite{BGT86} were motivated to devise simple empirical formul\ae\ for the three
coefficients $a_{r\theta}, t_1, t_2$, of the form
\begin{eqnarray}
a_{r\theta} & = & B_a\sigma_0^b\frac{q_a-q}{(q_c-q)^c},\\
t_1 & = & B_1\sigma_0^b q\frac{q_1-q}{(q_c-q)^c},\\
t_2 & = & B_2\sigma_0^b q\frac{q_2-q}{(q_c-q)^c}.
\end{eqnarray}
In these equations, $B_a, B_1$ and $B_2$ are positive quantities,
$0<q_a<1$, $q_1<q_a$ is either positive or negative, and $q_2$ is
negative. In the model for dense systems, $b=3$, $q_c=1$ and the 
rapid divergence of the viscous coefficients is well represented by $c\simeq 3$. In principle, $q_a$, $q_1$, $q_2$ and $q_c$ are functions of $\sigma_0$, except for dense systems. Orders of magnitude for $B_a$, $B_1$ and $B_2$ can be obtained from the comparison of these relations with specific models.

This model gives good semi-quantitative agreement with observed properties, in particular in the analysis of density wave damping, although a detailed quantitative comparison probably calls for more sophisticated models \citep{BGT86,SDLYC85,LSS16}. Another related issue is to investigate to which extent it is possible to connect this type of modelling with phenomenological physical constraints on ring rheology. It must also be noted that most hydrodynamical models used in the literature (e.g., \citealt{PL88}, \citealt{ST99} or \citealt{LO09}) make use of a simple viscosity prescriptions of the form $\nu=\nu_0(\sigma/\sigma_0)^\beta$ which allows for the existence of overstabilities in a simple and efficient manner for large enough $\beta$; notice in this respect that all stress tensor coefficients scale like $\sigma^3$ in the dense limit. However, such models fail to capture some of the constraints mentioned above; this limits their ability to properly describe the libration dynamics of narrow rings (discussed in section \ref{sec:2str}), for example, or the self-limiting amplitude of viscous overstabilities. 

\subsubsection{$N$-body simulation results}\label{sec:sim}

There are many analytical and numerical studies of the stress tensor behavior in \textit{unperturbed} (circular flows). These can't be surveyed here, but it might be worth pointing out at least the analysis of \cite{AT86}, which bears some relation to the granular flow model discussed above, and displays furthermore a possible gas-liquid transition that is not captured in this simpler model.

However, there does not appear to be more recent theoretical analyses characterizing the behavior of the pressure tensor from first principles in perturbed (non-axisymmetric) flows. Numerical $N$-body simulations might constitute a tool of choice here, as they are based on \textit{ab initio} --- albeit idealized --- physics instead of \textit{ad hoc} models, but have mostly focused in the past two decades on the study of viscous overstabilities and of wakes (in locally gravitationally unstable rings) in a background circular Keplerian shear flow\footnote{This limitation is a constraint of the form of the shearing box model that has been widely used in such studies.}. The number of numerical simulations of wakes is by now rather large (e.g., \citealt{S92a,S92b,R94,S95,DI99,RSLCS10,MFKS15}), including semi-analytically quantified results on the associated ``viscous" transport \citep{DI01,TI01,TOD03,YOD12}. The dynamical properties of wakes will be reviewed in the chapter by Salo et al., and are therefore only briefly summarized right below.

At the same time, an important shift of focus from viscous instabilities (i.e. non-oscillating instabilities) to viscous overstabilities (i.e., oscillating instabilities) has occurred in the theoretical and numerical literature \citep{PL88,LR95,ST95,ST99,SSPS00,SSS01,SSSP01,SS03,LO06,LO08,LO09,LO10,RL13}; this followed from the double realization that the conditions for viscous instabilities were not realized in dense rings due to finite particle sizes and related excluded volumes, both from analytic studies and $N$-body simulations \citep{AT86,WT88,A91,S91,R94}, while the analytical theory of \cite{BGT85} predicted from first principles the existence of overstabilities. The chapter by Stewart et al. will provide the reader with more information on the dynamics of small-scale viscous overstabilites.

There are two notable exceptions to the focus on unperturbed shear flows in $N$-body simulation: \cite{HS92} and \cite{M96}/. The first simulated a whole ring while the second implemented an extension of the shearing box $N$ body technique of \cite{WT88} to modal perturbations relevant for a narrow ring. Both studies found viscous angular momentum reversal at large enough $q$, but the second one, by construction, is much less noisy than the first. \cite{M96} also confirmed the presence of a viscous overstability for large enough filling factors; on this last question, the simulated behavior is in qualitative agreement with the theoretical results of \cite{BGT85}, while \citeauthor{M96}'s extension of BGT's granular formalism to smaller filling factors seems to account for the quantitative differences between BGT's analytical results and the numerical ones. 

\subsubsection{Self-gravitational wakes and transport}\label{sec:sgw}

The occurrence of self-gravitational wakes in rings can dramatically amplify the ring effective transport, altering substantially the traditional view summarized so far. The topic is briefly overviewed here for circular background flow, in the absence of relevant information in perturbed flows; for more details, the reader is referred to Heikki Salo's chapter.

Self-gravitational wakes are the primary if not sole explanation of the A ring brightness asymmetry, as realized very early on \citep{CGH76}. These structures are known to exist both in the A and B rings (see, e.g., \citealt{CESSC07,HNSWBBBC07,NH10}). \cite{RSLCS10} have performed a very detailed study of the mass and surface density of the rings, under the self-consistent assumption that self-gravitational wakes are present in the A and B ring and comparing observed optical depths with simulation averages; they find $\sigma_0\sim 50$ g/cm$^2$ in the A ring, and $\sigma_0\sim$ 250 --- 480 g/cm$^2$ in the B ring; however, this estimate for the B ring surface density is contradicted by substantially smaller values deduced more recently from the analysis of the dispersion relation of density waves (about a factor of 5; see \citealt{HN16}). 

Self-gravitational wakes are particular instances of self-gravitational instabilities, similar to self-gravitationally unstable density waves. The occurrence of such instabilities is controlled by Toomre's $Q$ parameter \citep{T64}:
\begin{equation}
Q\equiv \frac{c\kappa}{\pi G\sigma_0},\label{toomreq}
\end{equation}
where $c$ is the sound speed (gas disk) or the velocity dispersion (particle disk). Axisymmetric instabilities occur for $Q < Q_c\approx 1$, but non-axisymmetric ones up to $Q \lesssim 2$ ; the characteristic length $\lambda$ of instability is $\lambda \lesssim \lambda_c = 4\pi^2 G\sigma_0/\kappa^2$ where $\lambda_c$ is the critical wavelength of axisymmetric Jeans' instabilities \citep{JT66,S95}; $\lambda_c\sim 100$~m in Saturn's rings.

It is of some interest to compute the minimal critical surface density $\sigma_c$ obtained for the minimal velocity dispersion $c^*\simeq n d$:
\begin{equation}
\sigma_c \simeq \frac{M_p d}{2\pi a^3}.\label{sigc}
\end{equation}
Identifying the effective particle diameter in this expression with the cutoff of the size distribution (an upper limit for this quantity), one finds that $\sigma_c \lesssim 100$ g/cm$^2$ in Saturn's rings, which is consistent with the existence of wakes in the A and B ring, taking into account the crudeness of the estimate.

\cite{LK72} were the first to recognize that self-gravity produces a stress in the fluid and a related angular momentum luminosity $L^{sg}_H =\int dz\ r^2 g_r g_\theta/4\pi G$, where $g_\alpha$ is the self-gravitational acceleration in direction $\alpha=r,\theta$. As wakes are trailing (non-axisymmetric) features, the azimuthal acceleration does not vanish and a net angular momentum luminosity is carried by self-gravity. In a background circular motion, this luminosity can be described as an effective viscosity $\nu_{sg}$ implicitly defined by\footnote{See Eqs.\eqref{visflux} and \eqref{arphihydro} in the $q\rightarrow 0$ limit.}
\begin{equation}
L^{sg}_H = 3\pi\sigma_0\nu_{sg} n a^2.\label{sglum}
\end{equation}
\cite{DI01} and \cite{YOD12} have computed the self-gravitational stress from $N$-body simulations of wakes in a shearing box setting, using Hill's equations of motion; the equivalent viscosity can be computed either from a direct numerical calculation of the self-gravitational flux or from the dissipation of energy \citep{TOD03}. The results can be captured by the following simple approximate expression of $\nu_{sg}$:
\begin{equation}
\nu_{sg}\simeq C(r_h^*)\frac{G^2\sigma_0^2}{n^3},\label{nusg}
\end{equation} 
where $C(r_h^*)\simeq 26{r_h^*}^5$ for non-rotating particles \citep{DI01} and $C(r_h^*)\simeq 53{r_h^*}^5$ otherwise \citep{YOD12}; $r_h^*$ is the dimensionless Hill radius:
\begin{equation}
r_h^*= \left(\frac{2m_p}{3M_p}\right)^{1/3}\frac{a}{d}.\label{rh}
\end{equation}
For non-rotating particles, $C(r_h^*) \sim 6$ --- 20 in the B ring and 26 --- 40 in the A ring \citep{DI01}. The dimensional scaling is easily understood from the critical length of the instability $\lambda_c$, which implies that $\nu_{sg}\propto n\lambda_c^2 \propto G^2\sigma_0^2/n^3$, but the dimensionless steep scaling with the Hill radius is less easily captured by a heuristic argument; however, a positive correlation with the dimensionless Hill radius is expected, as particles fill more and more of their Hill sphere as one moves closer to the planet, inhibiting the formation of self-gravity wakes.

The resulting amplification of the ring viscosity and related angular momentum transport is quite substantial. In the close-packing limit discussed above, and ignoring wakes for the time being, the granular flow model gives 
\begin{equation}
\nu = \nu_m\simeq n\left(\frac{\sigma_0}{\rho}\right)^2,\label{numin}
\end{equation}
where $\rho$ is the ring particle internal density; this is the minimum viscosity that can be achieved in a ring. For $\sigma_0 = 100$ g/cm$^2$ and $\rho\simeq 1$ g/cm$^3$, $\nu_m\sim 1$ cm$^2$/s, whereas $\nu_{sg}\sim 50$ --- 100 cm$^2$/s for the B ring ($\sigma_0 = 100$ g/cm$^2$) and $\sim 100$ --- 200 cm$^2$/s for the A ring ($\sigma_0 = 50$ g/cm$^2$). Typically, the self-gravitational viscosity exceeds the minimum one by two orders of magnitude.

In practice, the minimum viscosity may not be reached in Saturn's rings, as close-packing there leads to the onset of self-gravitational instabilities and wakes. In dilute rings, the collisional viscosity is dominated by the \cite{GT78a} estimate, Eq.~\eqref{kinvis}; in such a context (possibly in the C ring) the surface density seems too low and the velocity dispersion too large for wakes to form.

Finally, note that perturbed regions are often characterized by an increased dissipation [see, e.g., section \ref{sec:resconf} on sharp edges confinement and Eq.~\eqref{dissexcess}]. If this translates into an increase of the ring velocity dispersion, the critical surface density Eq.~\eqref{sigc} will also increase by at least one order of magnitude. On the other hand, in close-packing configurations, the granular flow model predicts an increase of surface density and scale height without increase of the velocity dispersion to cope with an increased dissipation; but the underlying assumption (interparticle distances much smaller than the particles' size) may not apply in actual rings. The fate of wakes in perturbed regions is therefore uncertain and more detailed analyses are required before conclusions can be drawn on this topic. On this front, note that \cite{LS05} simulations of the Encke gap shows that self-gravitational wakes may be periodically created and destroyed by the moonlet wake, but the overall effect of the wake on the issue of transport and confinement is not quantitatively characterized by these authors. It would also be of some interest to characterize the critical value of $q$ for angular momentum flux reversal in perturbed flows where the transport is dominated by self-gravitational wakes.

\subsection{Mean eccentricity}\label{sec:meanecc}

One can also define a mean complex eccentricity 
\begin{equation}\label{meanevec}
\bar{Z} \equiv \frac{\sum_{i=1}^N m_i Z_i}{\sum_{i=1}^N m_i}.
\end{equation}
Because $q_{ij}=|Z_i - Z_j|/|a_i - a_j|=q_{ji}$ this mean complex eccentricity is conserved under the action of the ring self-gravity 
\begin{equation}\label{meanz}
\left(\frac{d\bar{Z}}{dt}\right)_{sg}=0.
\end{equation}
The radial action\footnote{This expression assumes $\epsilon \ll 1$.} $\sum m_i a^{1/2}|Z_i|^2$ (where $a$ is taken constant here consistently with earlier approximations) is also conserved under the action of the ring self-gravity \citep{PM05}.

The conservation of the mean complex eccentricity under the action of the ring self-gravity (and near conservation under the stress tensor) motivates us to define the mean eccentricity as
\begin{equation}\label{meanecc}
\bar{\epsilon} \equiv |\bar{Z}|.
\end{equation}
This definition makes physical sense; for example, in an hypothetical situation where one has two streamlines with equal eccentricity and opposite apsidal phase, one would intuitively assume that the two streamlines' mean eccentricity is vanishing. 

When the streamlines are aligned or nearly aligned, this definition reduces to a mass average of the eccentricities:
\begin{equation}\label{meane}
\bar{\epsilon} = \frac{\sum_{i=1}^N m_i \epsilon_i}{\sum_{i=1}^N m_i}.
\end{equation}
Nevertheless, the definition of a mean eccentricity is somewhat arbitrary. For example \cite{BGT83a} use two different definitions, first $\sum |Z_i|$ and secondly the magnitude of the mean complex eccentricity $|\bar{Z}|$; their results indicate that the difference in the eccentricity damping rate amounts to at most a factor of order unity (there is no difference for a two-streamline model). One might also define a mean eccentricity from the square root of the radial action. 

Our definition is motivated both by the intuitive idea that the mean eccentricity should be controlled by where most of the mass is located in the perturbed region, and by analytic simplifications that follow from this choice in the evaluation of the effects of the ring self-gravity and stress tensor on this mean eccentricity. However, it is probably less easily accessible from observations than a simple algebraic mean. Note that for $\bar{\epsilon} \gg \delta \epsilon$ (as observed in the major narrow rings), the numerical difference between the various possible definitions is negligible. 

\section{Dynamics: general discussion}\label{sec:proc}

The constitutive pieces of the streamline formalism have been presented in the two previous sections. We are now in position to review our current state of understanding of a number of important issues.

Eq.~\eqref{syncons} embodies what constitutes probably the most fruitful aspect of ring dynamics, along with constraints derived from energy and angular momentum budgets. This relation expresses that weak forces, most notably the ring self-gravity and, to a lesser but non-negligible extent, the stress tensor, must adjust the ring fluid particle epicyclic frequency in order to maintain the streamline $m$-lobe pattern throughout the perturbed region. As such, this constraint yields rather directly the nonlinear dispersion relation of density waves \citep{BGT85}, the standard self-gravity model for uniform precession of narrow rings \citep{GT79b,BGT83a} and the pressure-modified uniform precession model \citep{CG00}; it also underlies the trapped-wave picture of narrow ring and edge modes and consequently specifies the pattern speed of free modes.

Before analyzing these problems, it is convenient to collect here all the terms leading to the evolution of the complex eccentricity vector $Z_i$ and semi-major axis $a_i$
\begin{eqnarray}
\frac{dZ_i}{dt}& = & -\mi \sum_{j\neq i} \frac{n}{\pi}\frac{m_j}{M_p} a^2
H(q_{ij}^2)\frac{Z_i-Z_j}{(\Delta a_{ij})^2} +\mi \frac{Z_i}{|Z_i|} \frac{2\pi}{n m_i}
\left[\Delta^\pm(t_2 +\mi t_1)e^{\mi \gamma}\right.\nonumber\\
& & \left. + (-t_1 +\mi t_2)e^{\mi \gamma}\Delta^\pm(m\Delta)\right] -\mi n 
\frac{a\Psi_{mk}}{2 G M_p} +\mi (m\Omega_i - \kappa_i)Z_i -\mi m\Omega_p Z_i,\nonumber\\ \label{completez}\\
\frac{da_i}{dt} & = & -\sum_i\frac{2(m-1)n}{\pi}\frac{m_j}{M_p}
\frac{a}{\Delta a_{ij}}a\epsilon_i -\frac{2}{a_i n^2_i m_i}\Delta^{\pm} (L_E^{vis}) +\nonumber\\
& & \frac{2\pi\delta^{\pm}(a)}{n_i m_i}(2q_it^i_1 - 3a^i_{r\theta}) + n a(m-1)\epsilon _i\frac{a\Psi_{mk}}{G
M_p}\sin m\Delta_i,\label{completea}
\end{eqnarray}
where $\kappa_i=\kappa(a_i)$ and $\Omega_i=\Omega(a_i)$. One recognizes successively the contributions of the ring self-gravity, stress tensor, a satellite resonant forcing term and  --- for $Z_i$ only --- the planet-induced differential motion with respect to the pattern speed\footnote{See the dynamical perturbations section (\ref{sec:pert}) for the definition of the various quantities. Recall that $H$ is a quantity of order unity, that $t_1$ and $t_2$ are the viscous- and pressure-like parts of the pressure tensor while $a_{r\theta}$ characterizes its energy and angular momentum radial fluxes, and that $\Psi_{mk}$ has the dimension of a potential energy and specifies the strength of the satellite resonant component}. The satellite driving is described through a single resonance in a circular background motion in these equations; the effect of multiple and/or more complex resonances is more subtle and will be discussed in section \ref{sec:satamp} (eccentricity excitation by external satellites).

Eqs.~\eqref{completez} and \eqref{completea} involve a number of time-scales and a wealth of dynamical effects that can hardly be grasped in a single all-encompassing analysis. In fact, to date, no published analysis makes use of this full set of equations, although some are very close to this objective (e.g., \citealt{BGT86}). Instead, we first discuss some important features through a simple two-streamline model ignoring for the time being the effect of external satellites. This is the object of the next section. 

This will bring to light a number of relevant time-scales, whose hierarchy and meaning will be discussed in section \ref{sec:times}. This time-scale analysis will show in turn that it is not only simpler but to some extent legitimate to look first into the more detailed dynamics of stationary free modes and examine later the origin of free and forced mode mean amplitudes, which are generally driven on a longer time scale, ignoring in a first step changes in semi-major axis under the hypothesis that these structures are shepherded by satellites or other dynamical agents. This leads to the formulation of a non-linear eigenvalue problem for the complex eccentricity which plays a prominent role in the analysis of all kinds of modes. 


\subsection{Two-streamline model}\label{sec:2str}

This type of model was first introduced by \cite{BGT83a} for the study of narrow elliptic rings. It is generalized here in a straightforward manner to apply to all narrow ring modes and edge modes, in order to make more transparent the close analogy between the two contexts. We incorporate successively the ring self-gravity and stress tensor in this discussion; the action of satellites is briefly discussed in section \ref{sec:times} (time scales) but a more extensive discussion is deferred to sections \ref{sec:satamp} (satellite forcing of ring eccentricities) and \ref{sec:shep} (shepherding). The first three subsections deal with the dynamics of $Z$ only; the last one considers the radial diffusion driven by the viscous contribution to $da/dt$.

This material is presented in rather abstract form. Its physical implications will be discussed in a less compact way in later sections. 

\subsubsection{Self-gravity}\label{sec:2strsg}

For now, let us keep only the self-gravity and planet contributions in Eq.~\eqref{completez}. This allows us right below to define a self-gravitational characteristic time-scale $\Omega_{sg}^{-1}$ \citep{BGT83a}. We introduce two auxiliary variables $\delta Z = Z_2 - Z_1$ and $\bar{Z}=(Z_1 + Z_2)/2$. Conversely, $Z_1=\bar{Z}+\delta Z/2$ and $Z_2=\bar{Z}-\delta Z/2$; this specifies the entire problem once ($\bar{Z},\delta Z$) are known. The total perturbed region mass is $M_r$; for simplicity, the two streamline masses are assumed equal; it is possible to formulate this model for unequal streamline masses, but this increases the complexity of the analysis without obvious benefits in either insight or precision.

Defining $\omega_i=m\Omega_i - \kappa_i$, one has
\begin{eqnarray}
\frac{d\bar{Z}}{dt} & = & -\mi m\Omega_p \bar{Z} +\mi \frac{\omega_1 Z_1+\omega_2 Z_2}{2},\label{meansg}\\
\frac{d\delta Z}{dt} & = & -\mi\Omega_{sg}\delta Z +\mi (\omega_2 Z_2 - \omega_1 Z_1) - \mi m \Omega_p \delta Z,\label{libsg}\\
\Omega_{sg} & = & \frac{n}{\pi}\frac{M_r}{M_p}\left(\frac{a}{\delta a}\right)^2 H(q^2),\label{sgfreq}
\end{eqnarray}
where Eq.~\eqref{meanevec} has been used. 

The value of $q=a|\delta Z|/|\delta a|$ is determined later on. In this definition, $\delta a = a_2 - a_1$ is the characteristic width of the perturbed region; by construction, this is half the full extent between the inner and outer boundary of the perturbed region, with similar definitions for $\delta\epsilon$ and $\delta m\Delta$ (as $a_i$ corresponds to the center of streamline $i$). For narrow ring modes, this is specified by the inner and outer edges of the ring but no such simple constraint exists for edge modes. 

\paragraph{Equilibrium.}

The two-streamline model has a fixed point. Imposing $d\bar{Z}/dt=0$ yields
\begin{equation}
m\Omega_p=\frac{\omega_1 \epsilon_1+\omega_2 \epsilon_2}{\epsilon_1+  \epsilon_2}.\label{2strpat}
\end{equation}
%
%
%
Defining as usual the resonance radius $a_r$ such that $\omega(a_r)=m\Omega_p$, and noting $\omega'=d\omega/da$ at $a=a_r$ [see Eq.~\eqref{syncons}], $d\Delta Z/dt=0$ leads to
%
%
%
%
\begin{equation}
\Omega_{sg}\delta\epsilon = \omega'\bar{\epsilon}\delta a.\label{2streq}
\end{equation}
In the self-gravitational frequency, $q=a\delta\epsilon/\delta a$ with $\delta\epsilon=|\epsilon_2-\epsilon_1|$. This determines the equilibrium configuration of the ring once a mean eccentricity $\bar{\epsilon}=(\epsilon_1 + \epsilon_2)/2$ is specified. From a physical point of view, this second relation expresses that the self-gravity contribution to the precession rate balances the differential apsidal shift that the planet tends to impose to the two streamlines, a central concept in the standard model of uniform precession of narrow rings (section \ref{ssg}). 

Narrow ring modes are usually characterized by $\delta \epsilon \ll \bar{\epsilon}$ while $\delta \epsilon = \bar{\epsilon}$ holds for an edge mode, to the level of precision adopted here. Consequently, Eq.~\eqref{2streq} provides a relation between the perturbed region width $\delta a$ and the mass $M_r$ of an edge mode. Note also that in an edge mode, the resonance location is closer to the edge than to the middle of the perturbed region as a consequence\footnote{For edge modes, the larger streamline eccentricity is 3 times the smaller one in the two-streamline approximation, due to the vanishing of the mode amplitude far from the edge. This implies that the resonance location does not provide a precise enough estimate for the radial extent of the mode from the edge, a somewhat counterintuitive conclusion that is however mitigated if the surface density is not uniform. Nevertheless, both from Eq.~\eqref{2strpat} and from the results displayed on Fig.~\ref{fig:trapped} for the nodeless mode, it appears that the width of the perturbed region is $\gtrsim 2|a_e - a_r|$ where $a_e$ and $a_r$ are the edge and resonance locations, respectively.} of Eq.~\eqref{2strpat}. 

\paragraph{Internal oscillation (``libration").}

It is interesting to first look at the dynamics with self-gravity alone, i.e., ignoring the planet and pattern speed contributions in Eqs.~\eqref{meansg} and \eqref{libsg}. This makes the above equilibrium solution trivial, $\delta Z^{eq} = 0$; the mean amplitude $\bar{Z}^{eq}$ is arbitrary and set to zero for simplicity\footnote{In practice, this means that both $\epsilon^{eq}$ and $\delta\epsilon^{eq}$ are small enough so that $q$ is dominated by the libration/circulation motion and not by the equilibrium contribution. This may be relevant for a narrow ring, but for an edge mode, one can at best have $\epsilon^{eq}\simeq \delta\epsilon^{eq}\simeq |\delta Z_0|$.}.

Due to the $\mi$ factor ($\sqrt{-1}$) on the right-hand side and assuming for the time-being that $q$ is constant, the self-gravitational frequency Eq.~\eqref{sgfreq} describes oscillations (``librations") of narrow rings. Indeed, this equation is immediately integrated into 
\begin{equation}\label{libsg2}
\delta Z=\delta Z_0 e^{-\mi\Omega_{sg}(t-t_0)},
\end{equation}
i.e., a rotation of complex eccentricity difference in the complex plane in the clockwise direction\footnote{For our definition of the phase $m\Delta$; see Eq.\eqref{phasetime}.}; this result self-consistently justifies the assumption of constant $q= a |d\delta Z_0/da|$. Such motions can be either viscously damped \citep{BGT83a} or viscously excited \citep{LR95} --- see next subsection. An $N$-streamline model will display $N$ oscillation modes that differ by the number of radial nodes in their oscillation amplitude\footnote{An $N$-streamline model will also describe modes differing by their number of radial nodes in $\epsilon$; see section \ref{trapped} on the trapped wave picture of ring modes.}. 

This type of motion may be relevant, e.g., to explain the dynamical origin of the beating frequency observed between the free and forced $m=2$ B ring edge modes (see chapter by Nicholson et al.), but a detailed modelling would be required to quantify the merits of this suggestion; an alternative explanation is discussed in section \ref{trapped} (trapped wave picture of ring modes).

The term libration is somewhat of a misnomer to describe these oscillations as the streamlines are clearly circulating with respect to one another in the context examined here, but we will show right next that small amplitude oscillations around the equilibrium point are indeed librations, a context more relevant for narrow rings.

\paragraph{Generic solution.}

Qualitatively, the generic solution to the motion is a superposition of a libration as described by Eq.~\eqref{libsg} and of the forced equilibrium Eqs.~\eqref{2streq} and \eqref{2strpat}. Although this expectation is supported by the numerical solutions of \cite{LR95}, this generic solution is cumbersome for arbitrary libration amplitudes due to the $q$ dependence of $\Omega_{sg}$, which makes the pattern speed non stationary and the motion non harmonic. It is advisable to \textit{define} the pattern speed as the constant quantity satisfying Eq.~\eqref{meansg} once the streamline oscillations are averaged out, because this averaging\footnote{This time-averaged pattern speed will be accessible to observations only if data are obtained on a long enough time frame, a rather demanding condition due to the large libration time scales in rings --- years to tens of years.} implies that $\langle d\bar{Z}/dt\rangle =0$. 

A simple solution can nevertheless be found for small amplitude librations; in this case, the instantaneous pattern speed is nearly identical to the time averaged one. To this effect, let us define $Z_i=\epsilon^{eq}_i +z_i$ with $|z_i|\ll \bar{\epsilon}^{eq}$ where the superscript $eq$ refers to the equilibrium solution just described. We can assume that $\Delta^{eq}_1=\Delta^{eq}_2=0$ without loss of generality; therefore,
\begin{eqnarray}
Z_i 
& = & (\epsilon^{eq}_i + \delta\epsilon_i) +  \mi\epsilon^{eq}_i\delta(m\Delta_i) + \mathcal{O}(\delta\epsilon_i^2,\delta(m\Delta_i)^2)
\end{eqnarray}
i.e., $z_i=\delta\epsilon_i+\mi\epsilon^{eq}\delta(m\Delta_i)$. This assumption implies that $q=q^{eq}$ to leading order in $|z_i|$. To the same level of precision, one assumes that $\Omega_p=\Omega_p^{eq}$. Finally, neglecting terms of relative order\footnote{Such terms are negligible for narrow rings but the small parameter is equal to one for edge modes. However, even in this case, the neglected terms represent only 25\% of the leading ones. As the two-streamline model is used only semi-quantitatively, this approximation is not important. Note however that for edge modes, $\bar{z}$ must also display a similar oscillation of similar amplitude.} $\delta\epsilon^{eq}/\bar{\epsilon}^{eq}$, Eqs.~\eqref{meansg} implies that $\bar{z}\simeq 0$ and \eqref{libsg} becomes
\begin{equation}\label{libeq}
\frac{d\delta z}{dt} = -\mi\Omega_{sg}\delta z.
\end{equation}
Consequently, the libration motion is again given by Eq.~\eqref{libsg2}, except for the substitution of the libration amplitude $\delta z$ to the total difference $\delta Z$. 

Such motions might contribute to the sometimes substantial residuals observed in the analysis of narrow ring modes and edge modes (see the observational chapter by Philip Nicholson, Richard French and Joseph Spitale in this volume). Here again, specific data analyses and theoretical modelling will be needed to ascertain or disprove this possibility.

\subsubsection{Stress tensor}\label{sec:2strstress}

From Eq.~\eqref{zvis}, two relevant time scales $\lambda_1^{-1}$ (viscous) and $\lambda_2^{-1}$ (pressure) can be identified \citep{BGT83a} when the apsidal shift derivative term is not dominant, a situation appropriate to narrow ring and edge modes:
\begin{eqnarray}
\lambda_1 & = & \ \ \frac{8\pi a t_1}{q n M_r\delta a},\label{t1freq}\\
\lambda_2 & = & -\frac{8\pi a t_2}{q n M_r\delta a},\label{t2freq}
\end{eqnarray}
The choice of sign in the second equation makes $\lambda_2 >0$. These definitions lead to the following contributions to Eq.~\eqref{libsg}:
\begin{equation}\label{libvis}
\left(\frac{d\delta Z}{dt}\right)_{vis} =\left[-\mi\lambda_2 +\lambda_1 \right] \delta Z,
\end{equation}
to leading order (small apsidal shift across the perturbed region). The contribution to $d\bar{Z}/dt$ is not given, as it results in a correction to the pattern speed related to the average apsidal shift across the perturbing region, which we show right below to be negligible.

It is clear that $\lambda_2$ contributes to the libration motion, changing the rotation frequency of the complex eccentricity difference into $\Omega_{sg}-\lambda_2$, while $\lambda_1$ represents either collisional damping ($\lambda_1 < 0$) or amplification ($\lambda_1 > 0$). Generally speaking, the time-scales $\lambda_1^{-1}$ and $\lambda_2^{-1}$ are about an order of magnitude larger than the self-gravitational time-scale (see also section \ref{ssg} on the standard self-gravity model).

More precisely, the stress tensor corrects the self-gravity equilibrium solution into
\begin{eqnarray}
\frac{\delta\epsilon^{eq}}{\bar{\epsilon}^{eq}} = \frac{(\Omega_{sg}-\lambda_2)\omega'\delta a}{\lambda_1^2+(\Omega_{sg}-\lambda_2)^2} & \simeq & \frac{\omega'\delta a}{\Omega_{sg}},\label{2streps}\\
\delta(m\Delta)^{eq} =  -\frac{\lambda_1}{\Omega_{sg}-\lambda_2}\frac{\delta\epsilon^{eq}}{\bar{\epsilon}^{eq}} & \simeq & -\frac{\lambda_1}{\Omega_{sg}} \frac{\delta\epsilon^{eq}}{\bar{\epsilon}^{eq}},\label{2straps}
\end{eqnarray}
i.e., the main effect is to produce a small apsidal shift across the perturbed region. Let us also note here that external satellites influence these relations indirectly through their role on the mean eccentricity rather than directly through their action on the ring precession (for complements or qualifications, see sections \ref{sec:times} on time scales, \ref{sec:otherprec} on alternative rigid precession models, \ref{trapped} on the trapped wave picture of ring modes and \ref{sec:ecc} on the excitation of ring eccentricities).

Similarly, the amplitude of small librations becomes
\begin{equation}\label{libsg3}
\delta z=\delta z_0 e^{-\mi(\Omega_{sg}-\lambda_2)(t-t_0)}\exp\left(\int_{t_0}^t\lambda_1 dt\right),
\end{equation}
where $\delta z = \delta\epsilon + i \bar{\epsilon}^{eq}\delta(m\Delta)$ as before. The amplitude $|\delta z_0|\exp\int\lambda_1 dt$ stops evolving when $\lambda_1(q)=0$; for viscously stable rings, librations are fully damped. For overstable rings, and in the absence of external perturbing agents such as satellites, finite amplitude librations can be maintained at the value of $q=q_1$ where $\lambda_1$ changes sign\footnote{A sign change is an unavoidable consequence of the dissipation constraint $2q t_1 - 3a_{r\theta} < 0$; see section \ref{sec:press}.}. Because of this, the presence of a libration of constant amplitude is either the signature of a viscous overstability or a transient phenomenon. This will be further discussed in section \ref{sec:overvis} (viscous overstability driving of ring mode amplitudes).

\subsubsection{Mean eccentricity}\label{sec:2strmeane}

It is also interesting to look at the evolution of the mean eccentricity [Eq.~\eqref{meane}]
\begin{equation}\label{edamp}
\frac{1}{\bar{\epsilon}}\left(\frac{d\bar{\epsilon}}{ dt}\right)_{sg}=-\frac{\Omega_{sg}+\lambda_2}{4}\frac{\delta\epsilon}{\bar{\epsilon}}\delta(m\Delta).
\end{equation}
The $\lambda_2$ term is the dominant stress-tensor contribution to the evolution of the mean eccentricity, but it has not been captured in \cite{BGT83a} because of their approximation in the calculation of the stress tensor. The rate of change of the mean eccentricity due to the stress tensor is in any case substantially smaller\footnote{This result seems to imply that the mean eccentricity is conserved for a spherical planet, where $\delta\epsilon$ vanishes at equilibrium. In fact, sub-dominant terms --- neglected in the present analysis --- lead to a very slow decay. \cite{GT81} order of magnitude estimates give the minimum damping rate which implies that the mean eccentricity vector decays in all cases, albeit over a much longer time scale if the planet has no effect on the relative apsidal drift of streamlines.} than the one due to the ring self-gravity in Eq.~\eqref{edamp}. 

From the results of the preceding subsections, one notes here that usually $\delta\epsilon^{eq}/\bar{\epsilon^{eq}} \lesssim$ or $\ll 1$ and $|\delta(m\Delta)^{eq}| \ll 1$ so that the mean eccentricity evolution time scale is substantially longer than the self-gravitational one. Note that the presence of librations implies the existence of small amplitude oscillations of the mean eccentricity, not captured by Eq.~\eqref{edamp}. Note finally that depending on the sign of $\lambda_1$, Eq.~\eqref{2straps} implies either a growth or a damping of the mean eccentricity. This point is further discussed in section \ref{sec:overvis} (viscous overstability excitation of ring modes amplitudes).

\subsubsection{Radial spreading}

It is of interest to also consider the consequences of a two-streamline model on the radial viscous diffusion of perturbed ring regions. This analysis does not bear on the overall viscous diffusion over the entire rings of Saturn, a question recently revisited by \cite{SCCB10}, but only on the spreading of the perturbed region of width $\sim \delta a$. The main point of \cite{SCCB10} is worth recalling, though. Either from a simple random walk argument or from an order of magnitude estimate of the mass diffusion equation resulting from viscous transfers, a standard estimate of the spreading time of an unperturbed ring of width $\Delta r$ is $t_{sp}\sim (\Delta r)^2/\nu$ where the kinematic viscosity is often expressed as $\nu\sim c^2\tau/n$ for rings of optical depth $\tau\sim 1$. \cite{SCCB10} pointed out that in the later stages of evolution, the kinematic viscosity $\nu$ dependence on the ring physical conditions implied $\Delta r\propto t^{1/4}$ instead of the scaling $t^{1/2}$ derived with a constant viscosity; this point opens new perspectives on the age of the rings, which may be much older than previously estimated.

Our aim here is slightly different. We want to characterize how fast a narrow ring would spread or a narrow gap would close in the absence of confinement (``shepherding"). To answer a more focused question of this type, one can ignore the optical depth variation associated with spreading. It is also useful to distinguish narrow rings and sharp edges in this exercise.

\paragraph{Narrow rings.}

Noting by $\tau_{sp}$ the characteristic spreading time scale of ring and defining $\bar{a}= (a_1+a_2)/2$, one finds that Eq.~\eqref{completea}, to leading order, yields $d\bar{a}/dt=0$ and
\begin{equation}\label{tspnarrow}
\tau_{sp}^{-1}\equiv \frac{1}{\delta a}\frac{d\delta a}{dt} =  \frac{16\pi}{n M_r}\frac{a}{\delta a} a_{r\theta}(\bar{a}).
\end{equation}
From a physical point of view, this result expresses that the inner streamline transfers energy (and correlatively angular momentum) at a rate $L_E^{vis}$ to the outer one and this results mostly in spreading as the rate of energy dissipation is $\sim\delta a/a$ smaller than the viscous flux of energy in perturbed regions [as can be seen from Eq.~\eqref{arphi}].

The spreading time-scale is comparable to the viscous time-scale $\lambda_1^{-1}$. In other words, unconfined eccentric narrow rings will typically spread in a matter of a few tens to a few hundred years, based on either of the two scalings (dilute and compact) of the pressure tensor discussed in section \ref{sec:press}. 

The outwards angular momentum flux from the inner streamline to the outer one related to the energy flux is, to leading order in eccentricity, $L_H^{vis} = L_E^{vis}/n = 2\pi a^2 a_{r\theta}$ from Eq.~\eqref{visflux}. If narrow rings are shepherded by external satellites, the spreading time scale can be significantly longer. Time scale estimates can be obtained from $|T^s|\simeq L_H^{vis}$ (section \ref{sec:shep} on edge shepherding). This leads to a new spreading time scale, characterizing the time needed for the satellite to move away by $\delta a$
\begin{align}\label{tspsat}
\tau_{sat}^{-1}\simeq\frac{1}{\delta a}\frac{d a_s}{dt}\simeq \frac{4\pi a}{n M_s\delta a}a_{r\theta}.
\end{align}
This increases the spreading time scale by a factor
\begin{equation}\label{tspamp}
\frac{\tau_{sat}}{\tau_{sp}}\simeq \frac{4M_s}{M_r},
\end{equation}
assuming the shepherd satellites do not themselves receive or transfer energy and angular momentum through other resonances. As the shepherd satellites also transfer energy to the ring, the ring dissipation must adjust to accommodate both energy and angular momentum fluxes  once quasi-equilibrium is achieved. The shepherding dynamics is further discussed in section \ref{shflux}.

\paragraph{Sharp edges.} 

The main difference with a narrow ring is that $d\bar{a}/dt$ may not necessarily vanish for an unconfined ring edge. On the other hand, in the two-streamline model, the boundary streamline semi-major axis $a_b$ obeys 
\begin{equation}
\left|\frac{d a_b}{dt}\right| = \frac{8\pi a}{n M_r} a_{r\theta}
\end{equation}
The gap closing time will consequently be
\begin{equation}\label{taugap}
\tau_{gap}^{-1}\simeq \frac{16\pi}{n M_r}\frac{a}{\Delta_g}a_{r\theta}
\end{equation}
for a gap of width $\Delta_g$ (the extra factor of two comes from the fact that both edges of the gap move to close it). 

The concept of perturbed region width is somewhat ambiguous for an unshepherded edge. The idea here is that all edges are perturbed (as attested by the detection of various edge modes) and that this perturbation defines the perturbed region width, but that the dynamical source of perturbation may not confine the edge, especially if it is internal to the edge. 

There is no generic rule to quantify the ratio $\Delta_g/\delta a$, but $\delta a$ is nevertheless needed to estimate the gap closing time as it enters the perturbed region mass $M_r$. Another natural way to estimate the gap closing time is to identify the edge streamline width with the gap width ($\Delta_g/\delta a\sim 1$).

The Huygens gap inner edge (the B-ring edge) is shepherded by Mimas, and provides one example where the width of the perturbed region is constrained from the outside. For this gap, $\Delta_g\sim 300$ km. The resonance with Mimas 2:1 lies about 15 km inside the B ring edge, so that one has in this case $\Delta_g \sim 20\delta a$, if one assumes that the resonance defines the width of the perturbed region; this assumption is not warranted, as discussed at the end of section \ref{sec:2strsg} (self-gravity two-streamline model) and the effective two-streamline width may be a few times larger, reducing the ratio $\Delta_g/\delta a$ to a factor of a few. No such simple rule may be used for the other contexts where no resonance is known, such as the Russell or Jeffreys gaps, and a numerical analysis of edge mode structures along the lines described in section \ref{trapped} (trapped wave picture of ring modes) is probably needed to derive relevant rules relating the edge mode amplitudes (when detected) to the width of the perturbed region. In the absence of such a detailed analysis (the scope of which exceeds the bounds of the present review), one may simply assume that $\Delta_g/\delta a\sim 1$ --- 10.

Under this assumption, a simple upper limit can be placed on the gap closing time scale by setting $a_{r\theta} \sim \sigma_0 c^2$ (an order of magnitude appropriate for dilute rings) with $c\sim n d$, the minimal dispersion velocity that can be reached in rings. With these prescriptions, Eq~\eqref{taugap} gives, e.g., $\tau_{gap}\lesssim (8 n)^{-1}\delta a \Delta_g/d^2\sim 10^4$ --- $10^5$ yrs for the Jeffreys and Russell gaps.

If the gap edges are shepherded by a satellite orbiting within the gap, no closing will occur. If they are shepherded by some other mechanism, such an external satellite, the closing time scale is increased by a factor depending on the motion of the confining agent, which may depend on a number of factors. For example, the B-ring edge is shepherded by Mimas, whose motion is constrained on the one hand by its resonant interactions with the rings, and on the other hand by its resonance with Tethys. The minimum increase factor is again given by Eq.~\eqref{tspamp}.

\subsection{Time scale ordering and free modes nonlinear eigenvalue equation}\label{sec:times}

\subsubsection{Time scales}\label{sec:subtimes}

The previous two-streamline analysis has brought to light a number of time scales that give more substance to the time scale hierarchy mentioned in the introduction. Generally, the self-gravitational frequency is substantially larger than the stress tensor ones:
$\Omega_{sg} >$ or $\gg \lambda_1\sim \lambda_2$ [defined in Eqs.~\eqref{sgfreq}, \eqref{t1freq} and \eqref{t2freq}]. Typically, $\Omega_{sg}^{-1}$ is of the order of a few years to $10^3$ years while the stress-tensor time scales are typically an order of magnitude larger\footnote{This estimate refers to smooth gradients, of length scale comparable to the extent of the perturbed region; locally, e.g., close to an edge, the pressure and viscous time scales can be significantly reduced, but on a much smaller radial extent; see sections \ref{sec:cg} --- pressure-modified self-gravity model --- and \ref{sec:shep} --- shepherding --- for more details.}. $\Omega_{sg}^{-1}$ characterizes both the time-scale of self-gravity in the rigid precession equilibrium and the time scale of libration/circulation around the equilibrium; $\lambda_1^{-1}$ is the time-scale under which librations reach their steady-state amplitude at given mean eccentricity; in other words, the equilibrium of Eqs.~\eqref{2streps} and \eqref{2straps}  as well as the equilibrium amplitude of librations (vanishing or finite) of Eq.~\eqref{libsg3} are reached in a time-scale $\lambda_1^{-1}$, whatever the initial state. 

Also, the mean eccentricity evolves on a time-scale $\tau_\epsilon\simeq 4\lambda_1^{-1}( \bar{\epsilon}^{eq}/ \delta\epsilon^{eq})^2$, from Eqs.~\eqref{2straps} and \eqref{edamp} where $ \bar{\epsilon}^{eq}$ and $\delta\epsilon^{eq}$ are the equilibrium mean eccentricity and eccentricity difference across the ring (i.e., their values at vanishing libration about equilibrium). For narrow ring modes, this is usually substantially larger than the time required for the libration amplitude to reach its asymptotic value, vanishing or finite. The mean eccentricity reaches equilibrium on this time scale either by itself (change of sign or vanishing of $\lambda_1$) or under the added effect of an external agent, such as a satellite, the characteristic time scale of which must be comparable to the viscous one otherwise no equilibrium could be achieved. This matching occurs through the evolution of both $\bar{\epsilon}^{eq}$ and $\delta\epsilon^{eq}$, as $\lambda_1$ depends on $q$ and as the evolutions of $q$ and $\bar{\epsilon}^{eq}$ are coupled: the evolution of the mean eccentricity stops when $\lambda_1(q)$ reaches the right magnitude, i.e., when $q$ has the right magnitude, as $q$ increases with $\bar{\epsilon}^{eq}$ [cf Eq.~\eqref{2streps}] and as $\lambda_1 \rightarrow -\infty$ as $q \rightarrow 1$; in these conditions, a balance between the excitation by the satellite and viscous damping is unavoidable (the satellite eccentricity driving is discussed in more detail in section \ref{sec:ecc} on ring eccentricity excitation). 

In the absence of shepherding, both narrow rings and sharp edges would spread on a time-scale $\tau_{sp}\sim \lambda_1^{-1}$ from Eq.~\eqref{tspnarrow}; a narrow gap would close in a time scale $\tau_{sp}\Delta_g/\delta a$ where $\Delta_g$ and $\delta a$ are the gap and perturbed region size, respectively [cf Eq.~\eqref{taugap}]. These estimates must be increased if the viscous flux $\propto a_{r\theta}$ is reduced by angular momentum flux reversal throughout a narrow ring (section \ref{sec:shepnarrow}). Shepherding by external satellites may also increase this time scale by a factor of at least $M_s/M_r$ [cf Eq.\eqref{tspsat}]. However, if the presence of shepherd satellites is likely and compatible with the Voyager bounds on unseen small satellite sizes and masses in the Uranian ring system, the Saturnian system seems largely devoid of such objects except possibly for the outermost features, in agreement with Roche limit arguments. Considering the short time scales involved in radial diffusion, in the remainder of this review, gaps and ringlets are assumed to be shepherded (possibly by an as yet unidentified dynamics), although the shape of some edges (such as the Jeffreys gap inner edge) may suggest that they are actually not shepherded but instead are in the process of closing; this question will be briefly discussed again in the concluding section \ref{sec:misc}.

To conclude this discussion, we note that the viscous and pressure time scales $\lambda_1^{-1}$, $\lambda_2^{-1}$ and $\tau_{sp}$ are associated with smooth gradients across the whole perturbed regions. There can also be localized enhancements of the ring pressure and collisional dissipation in an actual ring that cannot be captured by a simple two-streamline analysis; these can lead to substantial deviations from the analysis of the two-streamline model in section \ref{sec:2str} (see the pressure-modified self-gravity model, section \ref{sec:cg} and the constraints related to the shepherding dynamics, section \ref{sec:shep}).

\subsubsection{Eigenvalue equation}\label{sec:subeigen}

The preceding considerations have interesting consequences. On time scales $\lambda_1 \ll \tau \ll \tau_\epsilon, \tau_{sat}$ (that is, $\tau$ slower than viscous damping but faster than eccentricity evolution and satellite migration), a narrow ring can be considered as effectively free if the effects of satellites on the time evolution of the complex eccentricity $dZ/dt$ can be neglected, an assumption discussed and justified in more detail in section \ref{sec:otherprec} (alternative rigid precession processes). For the same reason, the diffusion dynamics can be ignored on such time scales if narrow rings and sharp edges are shepherded. This motivates us to focus on the dynamics of $Z$ (ignoring $a$) for free-modes as a first step towards a more general theoretical analysis of the dynamics.

In such conditions, Eq.~\eqref{completez} contains all the relevant information and reduces to
\begin{eqnarray}\label{eigen}
& & \!\!\!\!\!\!\!\! -\sum_{j\neq i} \frac{n}{\pi}\frac{m_j}{M_p} a^2
H(q_{ij}^2)\frac{Z_i-Z_j}{(\Delta a_{ij})^2}\nonumber\\
& & \!\!\!\!\!\! + \frac{Z_i}{|Z_i|} \frac{2\pi}{n m_i}
\left[\Delta^\pm(t_2 +\mi t_1)e^{\mi \gamma} + (-t_1 +\mi t_2)e^{\mi \gamma}\delta^\pm(m\Delta)\right]\nonumber\\
& & \qquad + (m\Omega_i-\kappa_i)Z_i = m\Omega_p Z_i.
\end{eqnarray}
\noindent This relation is in fact a nonlinear eigenvalue problem\footnote{Once $q_{ij}$, $q$, $\gamma$, $t_1$ and $t_2$ are expressed in term of $Z$.} for the complex eccentricity $Z_i$ (eigenvector) and the pattern speed $\Omega_p$ (eigenvalue). It plays a major role in the understanding of the dynamics of both global narrow ring modes and edge modes and its properties will be discussed in some detail under various approximations in what follows.

\section{Uniform precession of narrow ring modes and edge modes}\label{sec:freeeigen}

The uniform precession of narrow rings, and more generally of all kinds of ring modes, arises either from collective effects or from an external forcing agent. Among theories relying on collective effects, one finds the standard self-gravity model \citep{GT79b,BGT83a}, the pressure-modified self-gravity model \citep{CG00}, the possible r\^ole of shocks (suggested by \citealt{GT79b} but never investigated) and the precessional pinch idea of \cite{DM80}. For external agents, the most often quoted suggestion is that a satellite may enforce rigid ring precession, but \cite{GT79b} showed that this was unpromising.

Of these various suggestions, the most robust still appears to be the pressure-modified self-gravity model (section \ref{sec:cg}), which draws largely on the standard self-gravity model (section \ref{ssg}). Both models can be analyzed within the context of the free mode eigenvalue problem formulated in section \ref{sec:subeigen}. The other suggestions will be discussed in section \ref{sec:otherprec}. 

We start with a generic relation for free mode pattern-speeds that is valid for both the standard and pressure-modified self-gravity model. 

\subsection{Free-mode pattern speed}\label{sec:freepatspeed}

Our purpose here is to capture some interesting consequences of Eq.~\eqref{eigen} for the pattern speed itself. 

First, assume that the pressure tensor term is negligible. In the linear limit, $H(q_{ij}^2) \rightarrow 1/2$, and Eq.~\eqref{eigen} (with $t_1=t_2=0$) is identical for both the real and imaginary part of $Z$. As a consequence, $m\Delta$ is a constant\footnote{Density wave theory, where $m\Delta$ grows steeply with radius, is largely ignored in this chapter, but is discussed in the chapter by Stewart et al.} that can be set to zero for an appropriate choice of the origin of angles. These constant-$m\Delta$ solutions can be straightforwardly generalized to the nonlinear regime. In this case, Eq.~\eqref{eigen} reduces to the constraint of vanishing apsidal shift rate $dm\Delta/dt=\mathrm{Im}(Z^{-1}dZ/dt)$.

Interestingly, a constant-$m\Delta$ solution also holds if one keeps the pressure-like coefficient $t_2$ while discarding the viscous-like one $t_1$. In this approximation, Eq.~\eqref{eigen} produces the following specific form of Eq.~\eqref{phasetime}:
\begin{eqnarray}\label{eigenbis}
\sum_{j\neq i} \frac{n}{\pi}\frac{m_j}{M_p} a^2
H(q_{ij}^2)\frac{\epsilon_i-\epsilon_j}{(\Delta a_{ij})^2} -  \frac{2\pi}{n m_i}\Delta^{\pm}(t_2) + (\kappa_i - m\Omega_i)\epsilon_i  - m\Omega_p \epsilon_i = 0,
\end{eqnarray}
\noindent with $q_{ij}=a\Delta\epsilon_{ij}/\Delta a_{ij}=q_{ji}$ where $\Delta x_{ij}=x_i - x_j$. Multiplying Eq.~\eqref{eigenbis} by $m_i$ and summing over $i$ leads to a mutual cancellation of all self-gravitational and pressure terms so that
\begin{equation}\label{patspeed}
m\Omega_p=\frac{\sum_i m_i\epsilon_i (m\Omega_i-\kappa_i)}{\sum_i m_i\epsilon_i}.
\end{equation}
\noindent This relation generalizes Eq.~\eqref{2strpat}. If furthermore the mean eccentricity dominates the eccentricity difference between the two edges, as in the major Uranian rings, $\epsilon_i$ is nearly constant across the ring so that this relation reduces to a simple mass average:
\begin{equation}\label{patspeed2}
m\Omega_p\simeq\frac{\sum_i m_i (m\Omega_i-\kappa_i)}{\sum_i m_i}.
\end{equation}
Finally, note that Eq.~\eqref{patspeed} applies to both the standard self-gravity model (section \ref{ssg}) and the pressure modified self-gravity model of \cite{CG00} (section \ref{sec:cg}). In particular, in this last model, $t_1\simeq 0$ and $\gamma\simeq 0$ but the pressure-like $t_2$ term is dominant on the edges, whereas it is negligible everywhere in the standard self-gravity model.

These considerations imply that Eq.~\eqref{patspeed} should provide a precise estimate of the pattern speed of edge modes and narrow ring modes as long as apsidal shifts as well as any non-stationarity can be neglected, due, e.g., to viscous overstabilities or transient phenomena, and assuming that the kinematics (eccentricity and surface density profile) can be sufficiently precisely constrained from observations. This useful relation does not seem to have been pointed out yet in the literature.

\subsection{Standard self-gravity model of narrow eccentric rings}\label{ssg}

To go further, one must actually solve the eigenvalue problem formulated above in section \ref{sec:subeigen}. Although the standard self-gravity model of \cite{GT79b} is now known to be inadequate at least in a number of instances, it is a useful first step for the more sophisticated model of section \ref{sec:cg} (pressure-modified self-gravity model) and is therefore discussed first\footnote{A similar model has been devised to enforce uniform precession in inclined rings; see \cite{BGT83d}}. This model was first devised by \cite{GT79b} for the $\epsilon$ ring of Uranus and soon recognized by these authors to apply to all narrow ring modes, although this idea appeared in print only much later \citep{L89b}. With appropriate changes in the boundary conditions, the same ideas are relevant to sharp edge modes.

\begin{figure*}[t!]
\centering
\includegraphics[width=\textwidth]{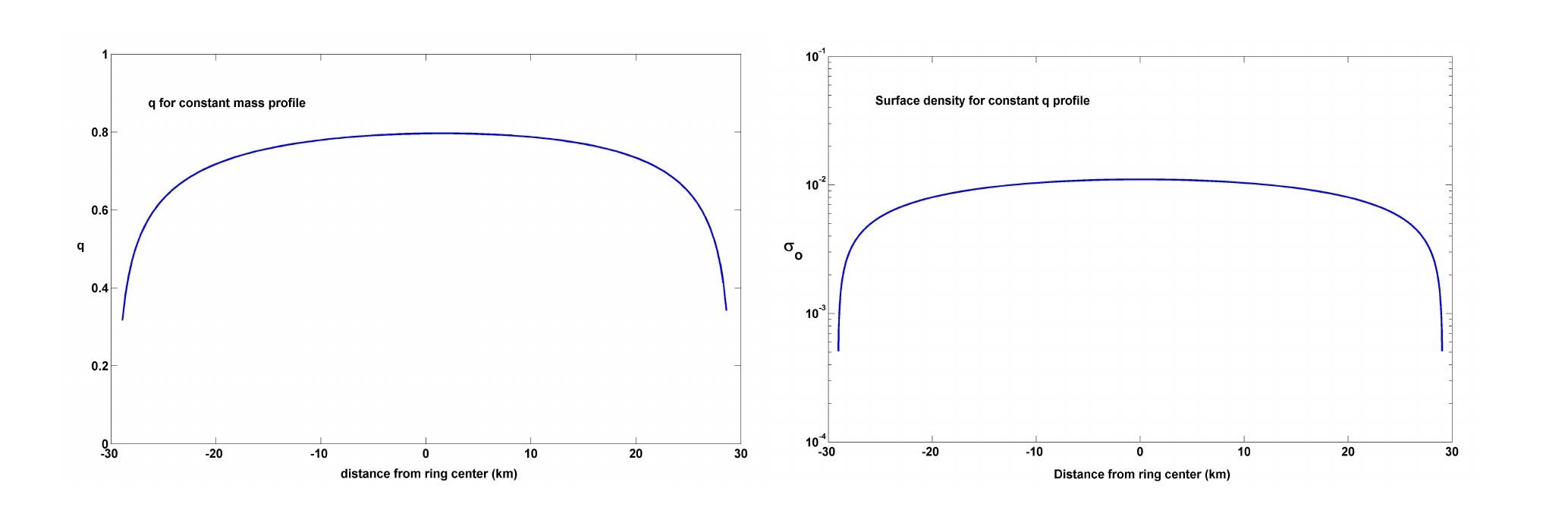}
\caption{Illustration of the standard self-gravity model, using parameters pertaining to the $\epsilon$ ring of Uranus. The profiles in the two instances (constant $q$ or constant $\sigma_0$) are very similar, a property reflecting the qualitatively similar role played by these two quantities in the nonlinear eigenvalue problem. The mass profile concavity is opposite to the characteristic double-peak profile observed in a number of narrow rings, including the $\epsilon$ ring.}
\label{fig:ssg}
\end{figure*}

This model of narrow ring uniform precession focuses on time-scales where the long term evolution of the ring due to the shepherding process is in quasi-static equilibrium while assuming that the ring self-gravity alone [first term on the left-hand side of Eq.~\eqref{eigenbis}] compensates the synodic shift of streamlines due to the planet-induced motion [third term on the left-hand side of Eq.~\eqref{eigenbis}] to produce a constant pattern speed [right-hand side of Eq.~\eqref{eigenbis}]. It is also assumed that the libration motion discussed in section \ref{sec:2str} (two-streamline model) is fully damped (i.e. $t_1 < 0$), and the small apsidal shift due to the ring dissipation is neglected.

This problem does not necessarily have a single solution\footnote{Actually, $\epsilon_i=0$ is a solution as well, which is precisely why this is an eigenvalue problem, although not a linear one.} (see section \ref{trapped} on the trapped wave picture of ring modes), so that one must introduce some constraint in order to lift this degeneracy. In their initial work, \cite{GT79b} introduced the ring mass profile $h_i=m_i/M_r$ ($M_r$ is the total mass of the ring), and solved the $N$ equations Eq.~\eqref{eigenbis} for $\epsilon_2,\dots,\epsilon_{N-1}$, $M_r$ and $\Omega_p$ for given edge eccentricities $\epsilon_1$, $\epsilon_N$ and mass profile $h_i$. Fig.~\ref{fig:ssg} exemplifies this type of solution for a constant mass profile and a constant $q$ profile.

This approach yields two remarkable results:

\begin{enumerate}
\item A relation between the ring geometry, as characterized by the eccentricities of its edges and its mass profile, the planet-induced differential precession and the ring mass.
\item A systematically positive eccentricity gradient between the inner and outer edge for all $m \ge 1$ modes. This follows from the sign of the gradient in Eq.~\eqref{syncons}.
\end{enumerate}

The main properties of the self-gravity model are captured by the simple two-streamline approximation first put forward by \cite{BGT83a} that has been described in section \ref{sec:2str}. Recall that in this model, the two streamlines have equal masses $m_1=m_2=M_r/2$ and that $q \simeq {a\delta\epsilon}/{\delta a}$. It is of some interest to express the results obtained in this earlier section in a less compact form appropriate for narrow rings.

As narrow rings generally have $\delta\epsilon\ll \epsilon$, the pattern speed Eq.~\eqref{2strpat} becomes
\begin{equation}\label{patspeed3}
\Omega_p=\frac{m-1}{m}n+\frac{3}{2m}J_2 n\left(\frac{R_p}{a}\right)^2\left[1+\frac{m-1}{2}\right]
\end{equation}
\noindent consistently with Eq.~\eqref{patspeed2}. 

Furthermore, for an $m=1$ mode, the equilibrium relation Eq.~\eqref{2streps} reads
\begin{equation}
\frac{\delta\epsilon}{\epsilon}=\frac{21\pi}{4}J_2\frac{M_p}{ M_r}\left(\frac{R_p}{
a}\right)^2\left(\frac{\delta a}{a}\right)^3\frac{1}{H(q^2)},\label{sgmass}
\end{equation}
\noindent while for $m\neq 1$,
\begin{equation}
\frac{\delta\epsilon}{\epsilon}=\frac{3\pi(m-1)}{2}
\frac{M_p}{M_r}\left(\frac{\delta a}{a}\right)^3\frac{1}{H(q^2)}.\label{sgmass2}
\end{equation}
where Eq.~\eqref{patspeed3} has been used; $H(q^2)$ is defined in Eq.~\eqref{hq2} with $q=a\delta \epsilon/\delta a$. This relates parameters describing the overall shape of the ring, $a$, $\delta a$, $\epsilon$, $\delta\epsilon$ to the mass of the ring $M_r$ as advertised above; therefore these relations have been used in practice to derive narrow ring masses. Recall also that the apsidal shift is given by Eq.~\eqref{2straps} as
\begin{equation}
\delta(m\Delta)\simeq -\frac{\lambda_1}{\Omega_{sg}}\frac{\delta\epsilon}{\epsilon},
\nonumber
\end{equation}
\noindent where $\Omega_{sg}$ and $\lambda_1$ are the self-gravitational and viscous-like frequencies defined in Eqs.~\eqref{sgfreq} and \eqref{t1freq}. For typical values of
the ring surface density ($\sigma_0\sim 50$ g/cm$^2$),
$\Omega_{sg}^{-1}$ is of the order of a few years to a few tens of years (e.g.\ $\sim 9$ years for the $\alpha$ ring) while $\lambda_1^{-1}$ is usually much longer than $\Omega_{sg}^{-1}$; e.g., with $c\sim 1$ mm/s and assuming $t_1\sim\sigma_0 c^2$, $\lambda_1^{-1}\sim 90$ years
for the $\alpha$ ring. This conforms with observations for most narrow ring modes; the large apsidal shift of some Saturnian ringlets is quite exceptional in this regard, which may point towards an interesting and unexpected behavior of the ring stress tensor or towards the existence of overstable libration/circulation motion in the ring. When the apsidal shift can be neglected, the width of the ring, which is given by $W=J\delta a$ as a function of azimuth, becomes proportional to the radius ($W\propto r$), as observed in the major narrow rings of Uranus. Note also that these two relations predict $\delta\epsilon>0$ for $m>0$, as systematically observed.

Because of these successes, the standard self-gravity model was widely regarded for a while as the correct explanation of the rigid precession of narrow elliptic rings. The model was however strongly questioned after the Voyager 2 encounter with Uranus, which revealed that the masses of the Uranian rings predicted by Eq.~\eqref{sgmass} were underestimated by a factor of $\sim 10$; the problem is particularly acute for the $\alpha$ and $\beta$ rings
\citep{GP87}. The first piece of evidence comes from the ring surface densities derived from the radio occultation data \citep{T86,E91,F91}. Another piece of evidence derives from the existence of an unexpectedly extended exosphere around Uranus. This is the source of an extra torque acting on the ring, which shepherd satellites  --- or other confining agents --- must balance in addition to the viscous torque in order for narrow rings to exist. It turns out that this constraint also requires ring surface densities about ten times larger than the ones predicted by the model (for more details about this point, see \citealt{GP87}). Finally, a detailed analysis of the variations of the optical depth profile of the $\epsilon$ ring of Uranus reveals that the detailed predictions of the model are incompatible with the data \citep{GSLK95}. This is not surprising considering the double peak structure of the $\epsilon$ ring, whereas the standard self-gravity model tends to produce an inverse concavity (see Fig.\ref{fig:ssg}).

In order to address these difficulties \cite{CG00} have devised a more sophisticated model that is presented in the next section.

\subsection{Pressure-modified self-gravity model}\label{sec:cg}

\cite{BGT82} were the first to note that the shepherding requirement implies a substantial increase in dissipation at a (narrow or broad) ring edge [see Eq.~\eqref{dissexcess} below]. \cite{CG00} pointed out in turn that this enhanced dissipation induces a localized but substantial apsidal shift rate due to the related increased pressure, which the ring-self gravity needs to counterbalance in addition to the planet-induced one. It turns out that for usual narrow eccentric rings ($m=1$), this additional apsidal shift rate is so large that it actually dominates the rigid precession dynamics throughout the whole ring, with the ring self-gravity propagating the constraint from the ring edges to the interior.

\subsubsection{Heuristic discussion} 

\cite{CG00} provide an insightful heuristic discussion whose essential point is reproduced here. In this argument, it is useful to distinguish three zones in the ring, with respect to the role of the increased pressure just mentioned, denoting by $c_b$ the velocity dispersion near the edge:\begin{enumerate}
\item A very localized edge zone (hereafter edge zone) of width $\lambda\simeq c_b/n$ (i.e.\ comparable to the ring local height) where the ring surface density drops sharply to zero.
\item A zone perturbed by the shepherd satellite (hereafter satellite perturbed zone), of width $w\sim a (M_s/M_p)^{1/2}$, due to the action of the ring shepherds; the width estimate corresponds to the size of the region where resonantly forced test particle orbits start to cross (see section \ref{sec:wakecrit} for a justification of this expression).
\item The remainder of the ring, where neither the enhanced pressure nor the shepherd satellites play any role. This represents in fact most of the ring extent, for the configurations of narrow ring confinement found in the Uranian rings.  
\end{enumerate} 

Eq.~\eqref{dissexcess} below is derived for both ring and gap edges and allows us to estimate the velocity dispersion increase in the satellite perturbed zone. Note that for satellites orbiting close to the ring, the satellite induced distortion is characterized by large azimuthal wavenumbers ($|m_s|\gg 1$) so that the resonance relation $m_s(\Omega-\Omega_p)=\kappa$ leads to $|m_s|\sim a/|x|$ where $|x|=|a-a_s|$ is the distance between the satellite and the ring ($m_s$ is azimuthal wave number of the distortion of the edge due to the satellite, as opposed to the global ring azimuthal wavenumber $m$). Combining this with the radial extent estimate for the region perturbed by the satellite, the right-hand side of Eq.~\eqref{dissexcess} becomes $\sim |x|/w$. On the other hand, the energy dissipation is\footnote{Throughout this section, the simple hydrodynamic limit of Eqs.~\eqref{t2hydro}, \eqref{t1hydro} and \eqref{arphihydro} is used for simplicity, as the \cite{CG00} model is only semi-quantitative at its present stage of development.} $\propto 2q t_1 - 3 a_{r\theta}\sim \sigma_0 c^2$ [cf Eq.~\eqref{arphi}]. Noting $c$ the velocity dispersion in the ring interior (away from the satellite perturbation), the left-hand side of Eq.~\eqref{dissexcess} is $\sim c_b^2/c^2$ so that
\begin{equation}\label{precen}
c_b\sim c\left(\frac{|x|}{w}\right)^{1/2}.
\end{equation}
It is useful to have some orders of magnitude in mind. The minimal velocity dispersion $n d\sim 1$ mm/s (a similar order of magnitude holds for the ring particles mutual gravitational stirring, comparable to the ring particles escape velocity); also, recalling that $w\sim a (M_s/M_p)^{1/2}$, one gets $c_b/c\sim 30$ or 40 for $|x|\sim 10^3$ km and $w\sim 1$ km (relevant orders of magnitude for the Uranian rings). 

This substantial enhancement implies in turn a substantial increase of the pressure in the near-edge region. The resulting pressure-driven acceleration in the edge zone is $\boldsymbol{g}_p=-\nabla P/\sigma\sim \pm c_b^2 \boldsymbol{e}_r/\lambda\sim \pm c_b n\boldsymbol{e}_r$; this acceleration is directed outwards (inwards) at the outer (inner) edge. 

It is useful to compare the resulting apsidal shift rate $(dm\Delta/dt)_p\sim g_p/na\epsilon$ with the planet induced one Eq.~\eqref{syncons}. Distinguishing the usual eccentric mode $m=1$ from all other $m\neq 1$ modes, this yields:
\begin{eqnarray}
\left|\frac{(dm\Delta/dt)_p}{\omega'_{pl.}}\right|_{m=1} & \simeq & \frac{c_b/ae}{(21/8)J_2n(R_p/a)^2(\Delta a/a)},\label{enm1}\\
\left|\frac{(dm\Delta/dt)_p}{\omega'_{pl.}}\right|_{m\neq 1} & \simeq & \frac{c_b/ae}{(3/4)(m-1)n(\Delta a/a)},\label{[enmnot1}
\end{eqnarray}
where $\omega'_{pl.}=d(m\Omega-\kappa)/dt\times\Delta a/2$ is the planet-induced differential apsidal shift across half the total ring width $\Delta a$. As $J_2\simeq 3\times 10^{-3}$ for Uranus and $\simeq 16\times 10^{-3}$ for Saturn, one finds that this ratio varies from $\gtrsim 20$ (Saturn) to $\gtrsim 100$ (Uranus) for $m=1$, while it is $\lesssim$ or $\ll 1$ for $m\neq 1$. In other words, the pressure-induced precession of the edges largely exceeds the planetary one for $m=1$, but is sub-dominant or even negligible for $m\neq 1$. For this reason, the Chiang-Goldreich model is mostly relevant for elliptical ($m=1$) eccentric rings and we focus on this case in the remainder of this section, as the impact of the edge zone dynamics on $m\neq 1$ global modes is rather limited.

For $m=1$, the ring self-gravity needs to compensate this very large pressure contribution to the edge zone precession rate in order to ensure the rigid precession of the whole ring. If one represents the edge zone with a single streamline, the self-gravity contribution from the rest of the ring is dominated by the streamline that sits right next at a distance $\sim \lambda$ from this edge zone. The acceleration produced by this streamline, identified to a wire of linear mass density $\sigma_b\lambda$ (using again the asymptotic approximation of section \ref{sec:sg} on the perturbation driven by the ring self-gravity) is $\sim 2 G\sigma_b$,  where $\sigma_b$ is the density in this near-edge zone. Balancing this with the pressure-driven acceleration yields
\begin{equation}\label{sigedge}
\sigma_b \sim \frac{c_b^2}{2G\lambda}\sim \frac{c_b n}{2G}.
\end{equation}
From our previous order of magnitude estimates, such a surface density is $\sim 40$ times larger than the ones obtained from the standard self-gravity model.

\subsubsection{Model} 

It remains to ascertain that this surface density increase in the edge zone is reflected by a large enough mass enhancement requirement throughout the eccentric ring in order to resolve the issues discussed in section \ref{ssg} (standard self-gravity model). To this effect, Chiang and Goldreich have solved an $N$-streamline model under three simplifying assumptions

\begin{enumerate}
\item The hydrodynamic approximation of Eq.~\eqref{t2hydro} is ad-opted with a simple \textit{ad hoc} linear approximation of the pressure gradient over a region of width $\lambda$ at each edge; the pressure contribution is non-vanishing only in the edge zone.
\item Furthermore, $q$ is assumed constant throughout the ring.
\item Finally $\delta \epsilon \ll\epsilon$ is assumed as observed in most narrow eccentric rings.
\end{enumerate}

The resulting $N$-streamline model is then solved for the $N$ streamline masses $m_i$. It must be noted that this model is not expected to be very precise, but rather aims at deriving global features and scalings. 

With these approximations,  the pressure-driven contribution to the precession rate [$(dm\Delta/dt)_p=-(d\varpi/dt)_p$] reads

\begin{eqnarray}
\left(\frac{d\varpi}{dt}\right)_{p,i} & = & \ \  \frac{qH(q^2) c_b^2}{n a \epsilon\lambda}\left(1-\frac{\delta_e}{\lambda}\right) \qquad \mathrm{if}\ \delta_e=a_i - a_{ie} < \lambda,\label{pressin}\\
\left(\frac{d\varpi}{dt}\right)_{p,i} & = & -\frac{qH(q^2) c_b^2}{n a \epsilon\lambda}\left(1-\frac{\delta_o}{\lambda}\right) \qquad \mathrm{if}\ \delta_o=a_{oe} -a_i < \lambda,\label{pressout}
\end{eqnarray}
where $a_{ie}, a_{oe}$ are the ring inner and outer edge semi-major axes, respectively, and $a_i$ is the semimajor axis of the considered streamline. If one further assumes that $c_b= n\lambda$, this model has a single parameter, $c_b$; it is, however, useful to keep both $c_b$ and $\lambda$ as independent parameters for scaling purposes, keeping in mind the order of magnitude constraint $c_b\sim n\lambda$.

With the same assumption of constant $q$, the self-gravitational contribution becomes

\begin{equation}\label{sgprec}
\left(\frac{d\varpi}{dt}\right)_{sg,i} = \frac{naqH(q^2)}{\pi M_p\epsilon} \sum_{j\neq i} \frac{m_j}{a_i-a_j}.
\end{equation}
Collecting these relations together gives the relevant form of Eq.~\eqref{eigenbis} that is solved in this model:
\begin{equation}\label{pmsgm}
\left(\frac{d\varpi}{dt}\right)_{p,i} + \left(\frac{d\varpi}{dt}\right)_{sg,i} + \dot{\varpi}_{pl,i} - \Omega_p = 0,
\end{equation}
where $\dot{\varpi}_{pl,i}$ is the planet-induced precession rate for streamline $i$. In this relation, $\Omega_p$ is specified by Eq.~\eqref{patspeed2} for $m=1$ and is not an independent variable of the problem\footnote{This relation on $\Omega_p$ was not used by \cite{CG00}; instead they used Eq.~\eqref{pmsgm} in difference form with respect to a mid-ring streamline.}.

\begin{figure}[t!]
\centering
\includegraphics[width=\textwidth]{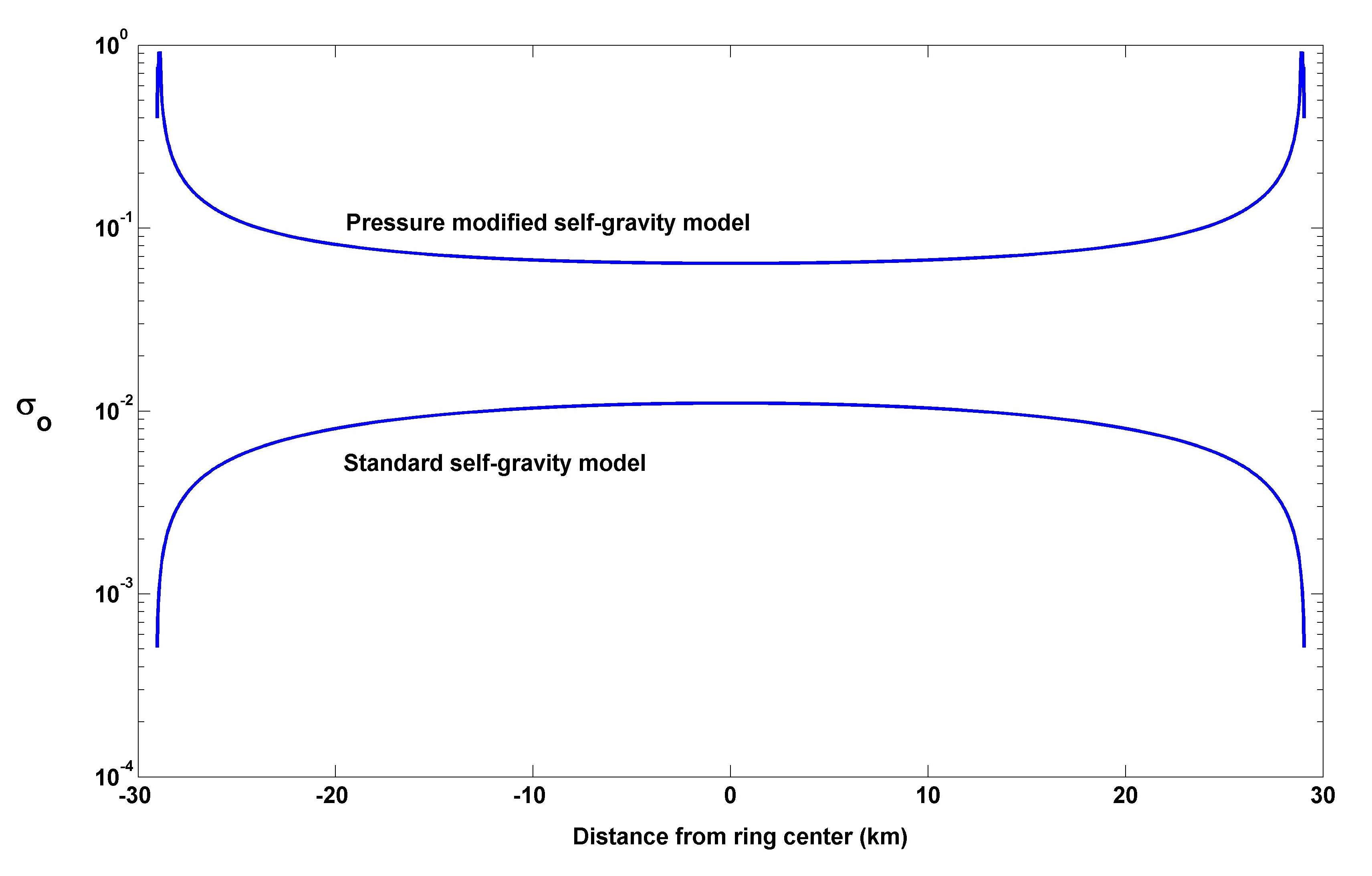}

\caption{Illustration of the pressure-modified self-gravity model (parameters of the $\epsilon$ ring of Uranus, and $c_b=4$ cm/s). The increase in mass with respect to the standard self-gravity model exceeds a factor of 10 in this case. Note that the two profiles are very dissimilar; in particular, the pressure-modified model produces a shape that is reminiscent of the characteristic double-peak profile observed in the $\epsilon$ ring; the effect would appear more pronounced in linear scale.}
\label{fig:pmsg}
\end{figure}

The resulting mass distribution is illustrated on Fig.~\ref{fig:pmsg}
; it is also compared with the standard self-gravity model for constant $q$. Besides the resulting substantial increase in total ring mass, note the aspect of the mass profile --- in particular, the characteristic double-peaked shape that is reminiscent of what is observed in a number of instances, most notably the $\epsilon$ ring.

%

\subsubsection{Scalings for negligible precession rate}

As argued in the introductory heuristic discussion and exemplified in Fig.~\ref{fig:pmsg}, for $m=1$ eccentric modes, the differential precession rate $ \dot{\varpi}_{pl.,i} - \Omega_p$ may be neglected in the problem, making the mass of the ring effectively independent of the ring geometry. It is possible to illustrate the resulting scaling with a simple four streamline model, which can be further reduced by assuming a symmetric mass distribution with respect to the mid-ring semi-major axis\footnote{Note that for a symmetric mass distribution, the resonance radius is the mid-ring radius.}: the boundary streamlines have equal mass ($m_1=m_4=m_b$) as well as the two inner streamlines $m_2=m_3=m_{in}$). The two boundary streamlines have width $\lambda$ while the two others have width\footnote{Unequal width streamlines raise issues of the convergence of the self-gravity integral, which is defined as a principal value; however, this four streamline model is indicative at best.} $(\Delta a - 2\lambda)/2\simeq \Delta a/2$ where $\Delta a$ is the full width of the ring. Under these assumptions, the fourth and first streamline equations are equivalent, as well as the second and third ones, reducing by half the effective dimension of the system, as expected from the symmetry assumption. 

With these approximations, ignoring the differential precession rate, the following expression for the ring mass $M_r$ can be derived:
\begin{equation}\label{mass1}
M_r=\frac{42\pi}{73}\frac{c_b\Delta a}{n a^2}M_p = \frac{42\pi}{73}\frac{c_b^2 a}{G}\frac{\Delta a}{\lambda},
\end{equation}
where $\lambda \ll \Delta a$ has been taken into account. On the other hand, Chiang and Goldreich quote the following scaling that was apparently found empirically
\begin{equation}\label{mass2}
M_r\sim \frac{c_b^2 a}{G}\left(\frac{\Delta a}{\lambda}\right)^{1/2}.
\end{equation}
The difference between the two is substantial as $\Delta a/\lambda \gg 1$, and is a measure of the crudeness of the four-streamline approximation. In any case, both expressions confirm that the ring mass is no longer related to the ring geometry, in contradistinction with the standard self-gravity model, but is instead related to the conditions prevailing in the edge zone.

\subsubsection{Discussion}

The Chiang and Goldreich model is promising, but still very crude in its present form. In particular, the crucial role played by the physical conditions in the edge zone call for a much more sophisticated modelling where the edge zone shepherding is precisely taken into account; no such complete model has been produced to date. \cite{ME02} have elaborated a somewhat different model; the main difference with \cite{CG00} lies in the presence of a pressure acceleration in the perturbed zone, which adds some measure of self-consistency to the model, without much affecting the results, while still neglecting the major issue just pointed out.

Note that even though the self-gravity model was put forward to explain how a narrow ring could counteract the differential precession of the planet, the shepherding requirement leads to the paradoxical result that the planet differential precession becomes secondary in the problem: the dynamics of the edges seems to control the whole ring equilibrium.

Finally, Chiang and Goldreich point out that the model does not require nor produce a positive eccentricity gradient; actually, solutions with negative eccentricity gradients have been produced in their paper. Thus, this pressure-modified self-gravity model loses one of the most successful features of the standard self-gravity one. This is not surprising, as the little relevance of the planet-induced differential precession removes the relation between the geometry of the ring and the required ring mass. However, this implies that an alternative explanation for the consistently observed positive eccentricity gradients must be found. 

Chiang and Goldreich make two suggestions in this respect. Firstly, negative eccentricity gradient solutions may be unstable. However, neither their preliminary analytic investigation (based on a four streamline model not reported in the article) nor the preliminary numerical test performed by \cite{ME02} showed any sign of instability. The question needs to be further examined, though; in particular, the ring shear stress may have an influence on this issue. Note that one also needs to show that positive eccentricity gradient solutions are stable.
Secondly, positive eccentricity gradients may reflect initial conditions, an option that still needs to be explored in some detail.

\subsection{Other suggestions to enforce uniform precession of narrow rings}\label{sec:otherprec}

A number of alternative suggestions have been made to account for the uniform precession of narrow eccentric rings. The remainder of this section provides some elements to assess the merits or limitations of these various ideas.

\subsubsection{Precessional pinch} 

This idea has been put forward by \cite{DM80}, who were motivated in particular by the fact that the dimensionless eccentricity gradient of narrow rings is always of order unity ($q \lesssim 1$), a feature that seems fortuitous in the standard self-gravity model or the pressure modified one. This suggestion has also been endorsed by \cite{PM05} and somewhat quantified by \cite{ME02}. The key idea is that, near periapse, ring particles can come in contact and be locked together; the differential precession would then be halted by the ring particle flow ``jamming" itself (presumably, ring particles are incompressible hard-spheres).

The precessional pinch idea has been around for a long time, but never really justified in dynamical terms; in fact, in the three papers mentioned above, it is simply assumed that such a process does take place in the ring. 

The issue is whether jamming or vertical splashing occurs in actual rings when the periapse pinch sets in. For jamming to take place, at least two conditions seem to be required: the vertical velocity dispersion must be very small, and the ring thickness must not adjust faster than the ring is radially compressed. There is little ground to believe that this holds true in eccentric rings. The level of vertical velocity dispersion is comparable to the radial one, due to mutual gravitational stirring of ring particles and also to coupling between their translational and rotational degrees of freedom (e.g., \citealt{S84,A91}). But more critically, the ring compression rate is always smaller than the collision rate, allowing the ring thickness to adjust to the compression before jamming can take place. 

Indeed, the fastest in-plane compression is due to the eccentricity gradient\footnote{For $m=1$ mode, the differential precession induces compression on a time-scale that is $J_2^{-1}$ longer, while for $m\neq 1$, the two processes have comparable time scales.}; the ring width undergoes large relative changes as particles move from apoapse to peripase, i.e., on a time-scale of order $n^{-1}$. First, even for low filling factors, the ring self-gravity enhances the collision rate by a factor of a few with respect to $n$ for rings of optical depth of order unity [see Eq.~\eqref{colfreq}]. But most importantly, when a close-packing condition starts to apply, the collision rate is increased by a factor $d/s$ where $s$ is the particle surface-to-surface distance and $d$ their size [see Eq.~\eqref{colfreq2}]. In other words, the very conditions needed for jamming to occur produce an increase in the collision frequency, allowing the vertical thickness to increase as or before jamming conditions start taking place. This argument makes a precessional pinch unlikely in dense narrow rings.

Note however that the initial motivation of \cite{DM80} remains: one needs to understand why dimensionless eccentricity gradients are of order unity in narrow elliptic rings.

\subsubsection{Large-scale pressure gradients (hot ring)}

The pressure term might be sufficiently large and nearly equal to the self-gravitational one everywhere, requiring a much larger ring mass to ensure uniform precession \citep{ME02}. The rationale of this suggestion can be understood from the two-streamline solution presented in Eqs.~\eqref{2streps} and \eqref{2straps}. It is assumed that $0<\Omega_{sg}-\lambda_2\ll \Omega_{sg}$ while $\lambda_2\gg\lambda_1$, i.e., the ring must be substantially hotter than usually believed. The first assumption allows us to increase the ring mass by a factor $\Omega_{sg}/(\Omega_{sg}-\lambda_2)$ for the same ring geometry while the second one is needed to maintain a small enough apsidal shift across the ring in the process.

In an $N$-streamline model, from a mathematical point of view, this model exploits the large degree of freedom offered by assuming that the velocity dispersion depends both on semi-major axis $a$ and mean longitude $\varphi$. A hot ring is by definition a low filling factor one, so that the stress tensor scales like $\sigma_0 c^2$. For a given surface density, and eccentricity and apsidal shift distributions, one can then always find a radial dependence of $c$ that achieves the required near cancellation of the self-gravitational contribution, while appropriate angular azimuthal profiles can always be found to make the ring viscous dissipation negligible compared to its pressure, due to the difference of azimuthal averaging in the definition of $t_1$ and $t_2$.

However, from a physical point of view, this proposal meets with two difficulties. First, in order to achieve such solutions in $N$-streamline models, the required velocity dispersion azimuthal profile is quite unrealistic, as pointed out by \cite{ME02} themselves in their study. Secondly, such a hot ring would require much more massive shepherd satellites to ensure their confinement (this second point was made by Peter Goldreich, as mentioned in \citealt{ME02}). 

These difficulties make this proposal unrealistic in actual rings.

\subsubsection{Shocks} 

The occurrence of shocks due to the ring compression at or near periapse can lead to impulses providing the required change in the periapse angle to balance the planetary differential precession \citep{GT79b}. Very small impulses (0.05 cm/s for the Uranian rings) are needed to prevent differential precession, and the formation of shocks is potentially able to produce such impulses. However, shocks in rings are difficult to model, and the process has never been studied. 

First, let us at least consider whether shocks are in principle a viable option from the well-known constraint that the compression velocity must exceed the sound speed for shocks to occur. Noting again that the fastest compression is due to the eccentricity gradient, this translates into $\delta v \sim a n\delta \epsilon \gtrsim c^* \simeq (d/s) c \simeq nd (H_0/d)^2$ or equivalently $H_0/d \lesssim (a\delta\epsilon/d)^{1/2}$. Close-packing has been assumed for $\tau \sim 1$, which maximizes the effective sound speed as information is transmitted instantaneously across a ring particle diameter during a collision (the sound speed inside the solid ring particle is much larger than the velocity dispersion of ring particles); this makes the shock formation condition more difficult to meet. For the $\epsilon$ ring, $(a\delta\epsilon/d)^{1/2}\sim 100$, so that the ring would need to be $\gtrsim 1$ km thick for shocks to be excluded on the basis of this argument, an unlikely constraint.

Note that this does not contradict the conclusion of the paragraph on precessional pinch, where it was argued that the ring thickness adjusts faster than the compression. This earlier argument was one of time scales while the present one bears on speeds; the vertical thickness adjustment has little or no effect on either the compression or sound speeds. One may question whether the usual hydrodynamical shock formation process through front steepening is relevant in rings. This is not at all obvious, considering the rather complex and peculiar ring rheology, but this point is rather difficult to explore any further, even in a heuristic way. On the other hand, the pasic premise according to which some sort of dynamical adjustement takes place when the compression speed exceeds the sound speed (which embodies the speed of transmission of information in the fluid) is physically motivated in all contexts, even if the details of this adjustement depend on the physics. The intriguing point of this discussion is that shocks are neither confirmed nor rejected as a potential cause of narrow ring rigid precession.

\subsubsection{Satellite-driven precession}

The shepherd satellites induce a precession rate of their own in the ring. If it is large enough, the ring may precess uniformly \citep{GT79b,BGT83c}. This alternative is the most straightforward to assess, as it involves only standard celestial mechanics. \cite{GT79b} made a quantitative study of this problem, both for locked-in and independent ring/satellite precession. They concluded that rather massive satellites are required for this process to be relevant. Such masses substantially exceed available observational constraints.

Indeed, in a Hamiltonian formulation of the problem for a close satellite\footnote{This formulation is based on the approximation Eq.~\eqref{sgacc} to evaluate the acceleration exerted on the ring by the satellite, with $\Delta_c$ evaluated by $J\Delta a$, where these quantities now involved differences between the ring, assimilated to a single ring wire, and the satellite.}, \cite{BGT83c} define a parameter $\Gamma$ that measures the relative importance of the planet and the shepherd satellites in the ring differential precession
\begin{equation}
\Gamma = \frac{21\pi M_p}{M_s} J_2\left(\frac{R_p}{a}\right)^2\left(\frac{x}{a}\right)^2,
\end{equation}
where $x=a-a_s$ measures the ring-satellite distance. This expression is appropriate for an $m=1$ narrow ring mode, but the problem is even more severe for $m\neq 1$. For illustrative purposes, it is interesting to compute the magnitude of $\Gamma$ for Ophelia and Cordelia, the two shepherds of the $\epsilon$ ring. Assuming masses $\sim 5\times 10^{16}$ kg and $x\sim 2\times 10^3$ km, this yields $\Gamma\sim 10^5$, i.e., the $\epsilon$ ring shepherds have a totally negligible role in its rigid precession.

In light of these arguments, it appears that the rigid precession of narrow rings is very unlikely to be enforced by small shepherd satellites.

\section{Trapped wave picture of sharp edges and narrow rings}\label{trapped}

The fact that narrow ring and edge modes can be viewed as a trapped wave was apparently first mentioned in the literature by \cite{BGT86} in the conclusion of their paper. It was also first pointed out there that the imaginary part of Eq.~\eqref{syncons} can be viewed as an eigenvalue problem for the ring eccentricity (eigenvector) and the pattern speed (eigenvalue); from this point of view, the standard and pressure-modified self-gravity models of rigid precession are special instances of this eigenvalue problem. Its most general formulation has been given in Eq.~\eqref{eigen}. We will discuss here some of the physical motivations for the trapped wave picture and its connection to the eigenvalue equation that constitutes the most relevant theoretical formulation of this question. The physical description provided here has been prompted by an unpublished preliminary numerical investigation by Peter Goldreich (private communication).

\begin{figure*}[ht!]
\centering
\includegraphics[width=\textwidth]{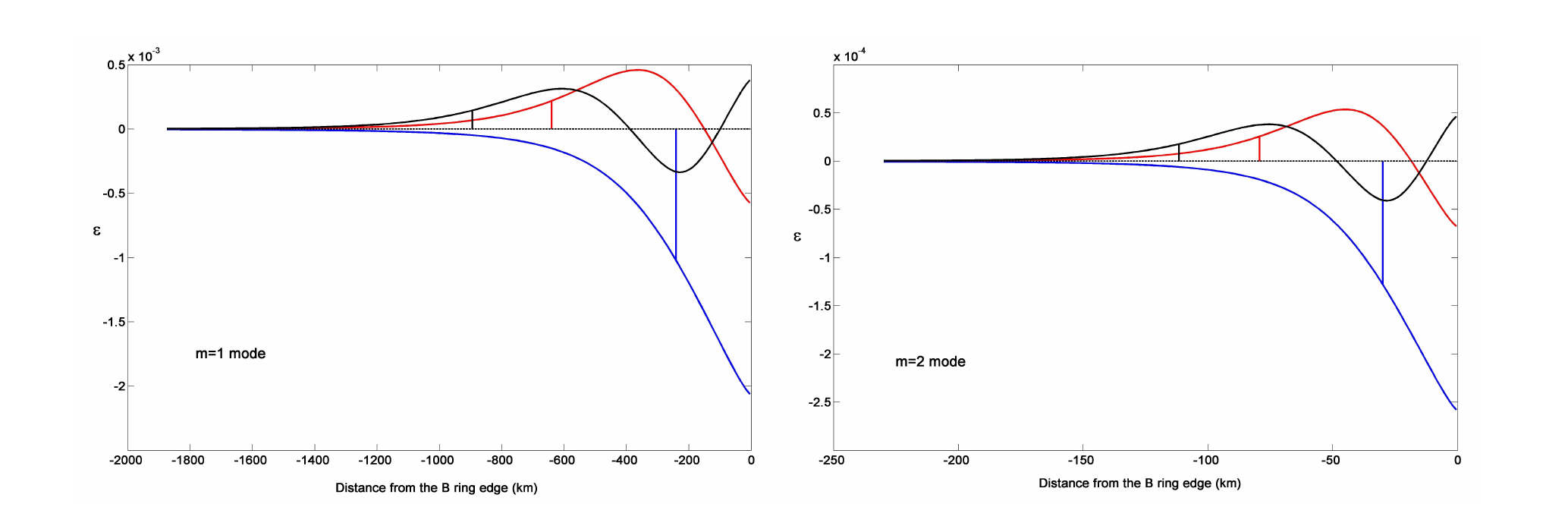}
\caption{First three free edge modes associated to $m=1$ and $m=2$ for parameters characterizing the B ring edge ($\sigma_0=150$ g/cm$^2$ assumed). The modes have very similar structures for both $m$ values, except for their radial extent due to the difference in magnitude of $m\Omega -\kappa$ (a factor $\sim J_2$). The vertical bars indicate the position of the mode resonance defined by $m\Omega -\kappa=m\Omega_p$. These are nonlinear solutions of the eigenvalue equations, but they differ only mildly from the linear ones. Note the similarity with truncated density waves.}
\label{fig:trapped}
\end{figure*}

Some elements of density wave theory are helpful to understand the basics of the trapped wave picture (for more details, the reader is referred either to the literature quoted, or to the chapter by Stewart in this volume, or to my own discussion of linear and nonlinear density waves in the framework of the streamline formalism, \citealt{L92}); the theory is essentially the same for externally forced waves or free internally unstable ones. The waves we are interested in are the long trailing and long leading waves, which propagate from and towards an inner and outer Lindblad resonance (ILR and OLR) implicitly defined by $m(\Omega - \Omega_p) = \kappa$, consistently with Eq.~\eqref{syncons}. Recall that both signs of $m$ are allowed in this relation, which defines either the resonance radius for a given pattern speed or conversely the admissible range of $\Omega_p$ for free waves. The ring self-gravity enforces the stationarity of density waves (as in the standard self-gravity model of narrow ring precession described above, except for the tight winding characterizing such waves) by providing the required contribution to the precession rate, leading to a simplified form of the WKB dispersion relation of density waves \citep{BGT85,SDLYC85} ($-$ sign for ILR, $+$ sign for OLR)
%
%
\begin{equation}
n [\kappa\mp |m|(\Omega-\Omega_p)]-\pi G\sigma_0|k|C(q)=0,\label{disprel}
\end{equation}
where $k=dm\Delta/da$ is the radial wavenumber and 

\begin{equation}
C(q)=\frac{4}{\pi}\int_0^\infty du\ \frac{\sin^2 u}
{u^2}H\left(q^2\frac{\sin^2 u}{u^2}\right).\label{cq}
\end{equation}
In the WKB limit, $\gamma$ is nearly equal to $\pi/2$ and $q=ka\epsilon$ [see Eqs.~\eqref{qcos} and \eqref{qsin}]. By taking the $q\rightarrow 0$ limit [$C(q)\rightarrow 1$], one can easily recover the well-known linear form of the dispersion relation (e.g., \citealt{LS68,GT78c}; see \citealt{Sh84} and references therein for a review). Two important pieces of information follow from this dispersion relation: i) both trailing $k >0$ and leading $k < 0$ waves have the same wavenumber, a property that remains valid if pressure and viscous dissipation are taken into account; ii) both leading and trailing waves are evanescent inside the ILR ($a < a_r$) or outside the OLR.

The second relevant piece of physics follows from the equation for the wave amplitude $\epsilon$. It is well-known that this equation can be derived either from a second-order WKB expansion or from the conservation of the wave action \citep{T69,Sh70,D72,GT78c}. Let us introduce the WKB form of the self-gravitational angular momentum luminosity\footnote{This quantity is second order in $\epsilon$ but is correctly derived from a first order theory.}, Eq.~\eqref{sgflux}, derived by \cite{BGT85}:
\begin{equation}
L_H^{sg}=-\mathrm{sgn}(k)\pi^2 G|m|\sigma_0^2 a_r^3\epsilon^2 B(q),\label{wkbsgflux}
\end{equation}
where $\mathrm{sgn}(k)$ is the sign of the wavenumber $k$ and
\begin{equation}
B(q)=\frac{4}{\pi}\int_0^{+\infty}du\ H\left(q^2\frac{\sin^2 u}{
u^2}\right)\frac{\sin 2u}{u}.\label{bq}
\end{equation}
The linear limit is again recovered when $q\rightarrow 0$, where $B(q)\rightarrow 1$. With these definitions, the amplitude equation reads, in the collisionless limit:
\begin{equation}
\frac{\partial L_H^{sg}}{\partial a}=\mathfrak{T}_s,\label{fluxcons}
\end{equation}
where $\mathfrak{T}_s$ is the torque density introduced in Eq.~\eqref{torque}. in other words, for free waves, the angular momentum luminosity is conserved; this relation expresses the conservation of the wave action. When the wave is viscously damped, the angular momentum luminosity is no longer conserved, as angular momentum is deposited in the ring while the wave propagates. The interesting point here is that free and forced leading and trailing waves must have the same amplitude $\epsilon$ as a function of $a$.

The last important piece of information comes from the known behavior of waves at boundaries and turning points \citep{GT78b}. For an outer edge associated to an ILR, a long trailing wave propagating outwards from the ILR will be reflected as a long leading wave once it reaches a boundary (the ring edge). The long leading wave will in turn be reflected again as a long trailing wave once it reaches the ILR. The reflection at the ILR is total because waves are evanescent inside the ILR; the reflection at the ring edge is total if the density drops to zero and the gap is wide enough (in order to prevent self-gravitational coupling through the gap). This behavior is exactly mirrored for an inner edge associated with an OLR, with the evanescent region now lying outside the OLR so that the leading wave propagates from the OLR while the trailing wave propagates to the OLR. All these features translate without modification to nonlinear waves.


We have now assembled all the required material to proceed towards a quantitative formulation of the problem. Consider an ILR for definiteness. First, the superposition of leading and trailing waves of identical apsidal shift and amplitude can be represented by a simple generalization of Eq.~\eqref{rmode} in the absence of dissipation:
\begin{eqnarray}\label{trapmode}
r(a_e,\varphi,t) & = & a_e \left( 1
 -\epsilon(a_e)\cos\left[m(\varphi-\Omega_p t)+ m\Delta(a_e)\right]\right.\nonumber\\
& & \qquad\ \left. -\epsilon(a_e)\cos\left[ m(\varphi-\Omega_p t)- m\Delta(a_e)\right]\right)\nonumber\\
& = & a_e \{ 1 -\epsilon\cos(m\Delta)\cos\left[ m(\varphi-\Omega_p t)\right]\},\nonumber\\
\end{eqnarray}
where $m>0$ is by definition the azimuthal wavenumber of the long trailing wave. In the presence of dissipation, one should add a residual apsidal shift in the second cosine factor. Such a mode can be represented by an effective (not strictly positive) eccentricity $\epsilon^*=\epsilon\cos(m\Delta)$. Consequently, in the dissipationless limit, the dynamics of such a mode is controlled by the rigid precession constraint Eq.~\eqref{eigenbis} as in the standard self-gravity model of narrow rings, except that $\epsilon^*$ is substituted for $\epsilon$ in this relation. This makes in effect the trapped wave dynamics an eigenvalue problem; this dynamics can be analyzed by standard means in the linear limit --- i.e., $q\rightarrow 0$ and $H(q)=1/2$ in Eq.~\eqref{eigenbis}, and the linear solution can be used as a guess for the nonlinear eigenvalue problem. For illustrative purposes, the first $m=1$ and $m=2$ modes are shown on Fig.~\ref{fig:trapped}.

Some generic features can nevertheless be deduced from the preceding remarks without actually solving the related eigenvalue problem:

\begin{enumerate}
\item The eccentricity $\epsilon^*$ decays to zero in the evanescent region ($a < a_r$ for an ILR and $a > a_r$ for an OLR).
\item The free modes are discretized due to the boundary condition at the edge. Because the edge moves freely, this boundary condition is that the fluid must be stress-free there, or equivalently (in the absence of collisional stress) that the self-gravity sum (or integral in the continuum limit) converges at the edge, i.e., $q \rightarrow 0$ as one moves towards the edge\footnote{Note that this boundary condition is appropriate for a single free edge mode, and ignores the physics involved in the maintainance of the sharpness of the edge. In a more complex setting, e.g., the B ring outer edge where both a forced and free mode components are present, the appropriate boundary condition is more complex, but the admissible solutions remain countable and characterized by the number of radial nodes.}. This suggests that the edge is associated with an extremum of the trapped wave amplitude\footnote{In the WKB limit, $d\epsilon^*/da$ is dominated by the derivative of $m\Delta$.}. Consequently, there can be many different edge modes for a given $m$ value, each with a different number of radial nodes; i.e., $m\Delta$ in Eq.~\eqref{trapmode} is equal to $(k+1)\pi$ at the edge, where $k$ is an integer representing the number of radial nodes. The wavelength close to the edge may differ somewhat from its free unbounded wave counterpart, however, due to the truncation of the self-gravity integral. An $N$-streamline model will display $N$ modes at most, from zero to $N-1$ radial nodes ($0 \le k \le N-1$).
\item As for narrow rings, the pattern speed in this model is effectively defined by Eq.~\eqref{patspeed}, which implicitly defines the resonance radius. Alternatively, the dispersion relation might be used to obtain a first idea of the resonance radius location and the associated pattern speeds. All these modes will have a different but closely related pattern speeds, corresponding to the discrete series of resonance radii associated to the series of nodes. This can create beatings between different eigenmodes with the same $m$ value if more than one such mode is present in the ring.
\item A narrow ring $m=1$ mode is also a nodeless solution to this eigenvalue problem, truncated at both edges by the shepherd satellites; in practice, the radial extension of narrow rings is small compared to the wavelength. In a similar fashion, forced edged modes (e.g., the forced Mimas $m=2$ mode) are truncated by the satellite where the torque balance is achieved. Note finally that the free $m=2$ mode observed at the B-ring can either be a free mode with a nonzero number of radial nodes or a viscously driven oscillation of the kind described in section \ref{sec:2str} (two-streamline model), as the beating of the forced and free modes' rotation direction is observed to be clockwise with the relevant definition of the slow angle\footnote{The choice of resonant angle in \cite{NFHMC14} and \cite{SP10} is the opposite of $m\Delta$, consistently with Eq.~\eqref{anom}. Note also that the direction of the slow motion is directly connected to the fact that self-gravity is the dominant perturbing force (e.g., it would be opposite if the ring pressure were dominant). Because of this, the free mode might also be a transient feature, but then the near equality of the two mode amplitudes would be even more coincidental.}; the near equality of the two mode amplitudes needs to be explained, though. This suggestion is further discussed in section \ref{sec:overvis} (overstability driving of ring eccentricities).
\end{enumerate}

\medskip

The dynamics of such edge modes remains to be explored in detail. Similarly, one needs to analyze multiple mode motions and couplings, considering the rather large number of modes seen at sharp edges and ringlet edges in the Saturnian and Uranian systems. Finally, a lingering question concerns the origin of these modes. This is the object of the next section, along with the question of the origin of narrow ring eccentricities.

\section{Origin of narrow ring eccentricities and edge mode amplitudes}\label{sec:ecc}

Two main processes have been proposed in the literature to explain the origin of the amplitudes of narrow ring and edge modes: external satellite forcing \citep{GT80,GT81,BGT83a,PM05} and an internal instability, in particular a viscous overstability \citep{BGT86,PL88,LR95}. In the first case, the steady-state eccentricity results from a balance between the satellite external forcing and the ring internal viscous damping; in the second case, the amplitude is self-limiting due to the necessary change of sign of the viscous coefficient $t_1$ (see section \ref{sec:press} on the stress tensor). For completeness, one must mention that once a mode is excited and maintained, other modes may in principle be produced through nonlinear coupling with itself or with other (even transient) modes, a process that has not been analyzed yet in the literature (see the discussion of section \ref{sec:gaps} on this point). Let us also mention right now that although such processes may explain why a given mode may reach a steady-state, no explanation has been put forward yet as to which modes are thus selected.

\subsection{Finite mode amplitude from viscous overstabilities}\label{sec:overvis}

Let us first ignore the effect of external satellites. 

It has been suggested by \cite{BGT85} and \cite{PL88} that eccentricities can be excited and maintained by a viscous overstability; such a process occurs when the stress perturbations correlate well enough with the velocity ones. \cite{LR95} subsequently examined the relevant dynamics in the context of narrow rings. Their analysis relies on the streamline formalism; it is not reproduced here, only the most important results will be quoted\footnote{The material assembled in section \ref{sec:pert} (perturbation equations) is sufficient for the interested reader to follow the details of the original analysis in case of need.}.

\cite{LR95} confirmed both the possibility of driving the ring mean eccentricity and internal librations as presented in the two-streamline model of section \ref{sec:2str} (two-streamline model) but also showed that the presence of internal librations substantially modifies the potential growth of the mean eccentricity through the existence of two regimes: a narrow ring global mode can be either in a large amplitude equilibrium state where $q$ is dominated by $\delta\epsilon^{eq}$ (the eccentricity difference between the two edges at equilibrium) and with very small libration amplitude, or conversely very small mean amplitude and large libration amplitude. Furthermore, they showed that the difference between the two was related to the initial mean eccentricity (amplitude) of the global mode: below some critical threshold, the ring is driven to the small amplitude regime, while above this threshold, it is driven to the large amplitude one. This dynamics results from the interaction between the ring libration and mean amplitude and from the change of sign of the viscous coefficient $t_1$ defined in section \ref{sec:press} (stress tensor) for large enough $q$; this change of sign is necessary due to the dissipative nature of collisions in rings, as shown in section \ref{sec:press}, so that these conclusions are model independent. Note that the critical role of librations on the outcome was ignored by \cite{BGT86} and \cite{PL88}, but the situation is somewhat different in a broad ring.

These findings have a number of implications for the origin of the amplitudes of modes in narrow rings (assuming that the ring is overstable):

\begin{enumerate}
\item A ring mode cannot spontaneously reach a sizable finite amplitude from a small one. If an overstable ring has a large mode amplitude, then either the initial amplitude was large enough for the overstability to drive it to  equilibrium, or some other agent (e.g., a satellite) helped in reaching the observed amplitude. 
\item Whether an overstable ring reaches an equilibrium amplitude by itself or with external forcing, the mean eccentricity will grow until $t_1(q)$ changes sign as a consequence of Eqs.~\eqref{2streq} and \eqref{edamp}. This might help explain why eccentric rings nearly always have $q\lesssim 1$, a feature that would otherwise be somewhat fortuitous.
\item The amplitude of librations will tend to be small in the large mean amplitude regime and behave as predicted by Eq.~\eqref{libsg3} (with $\lambda_1=0$), unless an external agent (e.g. a satellite) damps the mean amplitude. However, this seems contrived as one would need the ring not only to come into being with sufficient initial amplitude but the external agent to have a small enough damping effect as not to drive this amplitude below the critical threshold mentioned above that would drive the mean amplitude to zero. If on the contrary the satellite further excites the mean amplitude, the libration amplitude will be fully damped, consistently with the related change of sign of $t_1$.
\end{enumerate}

Such effects still need to be explored in detail to ascertain if they have any relevance for the modes observed in narrow rings. Furthermore, the dynamics of edge modes remains to be explored, as the numerical simulations performed in \cite{LR95} did not investigate the relevant physical conditions. However, one can expect that overstabilities may drive both the mean and libration amplitudes in this context; in particular, one expects that the libration amplitude is comparable to the equilibrium eccentricity gradient predicted by Eq.~\eqref{2streq}, so that either Eq.~\eqref{libsg2} or Eq.~\eqref{libsg3} with $|\Delta Z|\simeq \delta\epsilon^{eq}\simeq \epsilon^{eq}$ should describe the libration motion, if any, at least semi-quantitatively.

The B ring free $m=2$ mode raises an intriguing issue. It might result from a large viscously driven librating/circulating component of the forced mode, as the circulation is clockwise with our convention for the apsidal shift sign. However, as Mimas is exciting the mode amplitude in the first place, $q$ should be larger than the critical value $q_1$, making $\lambda_1$ effectively negative close to the edge, as a consequence of the viscous angular momentum luminosity reversal requirement for shepherding the edge (see section \ref{shflux}). The libration component of this mode (if any) would thus be damped right at the edge, whereas it should also be excited by coupling with the overstable part of the mode in the internal region where $q < q_1$. A detailed model is needed before a firm conclusion on the question of the libration amplitude at the edge can be reached for this mode; a related question bears on the near equality of the forced and free mode amplitudes.

\subsection{Satellite forcing}\label{sec:satamp}

Satellites are a conspicuous potential cause of the observed mode amplitudes. In particular, the external satellites that are responsible for the shepherding of the $\epsilon$ ring are thought to excite its eccentricity. External satellites modify narrow ring eccentricities through resonant interactions with the ring; non-resonant interactions have no or negligible effect on mode amplitudes and ring eccentricities. Also, because energy and angular momentum are conserved quantities, whatever is gained (lost) by the ring is lost (gained) by the satellite. As a consequence, there is a direct connection between the satellite and the ring eccentricity evolution as the ring or satellite specific energy and angular momentum depend only on their semimajor axis and eccentricity \citep{GT80,GT81}.

The question of origin of the eccentricities ($m=1$ mode) of narrow rings meets with an inherent difficulty: resonant energy and angular momentum exchanges between rings and satellites involve almost always an $m\neq 1$ satellite forcing, and not a direct forcing of the $m=1$ mode. Consequently, resonant driving of the $m=1$ mode requires second order mode coupling. Although it is often useful to focus on energy and angular momentum exchanges between rings and satellites (as these are sufficient to compute their semi-major axis and eccentricity evolutions), second order calculations such as those performed by \cite{GT80,GT81} or \cite{PM05} are needed due to this second order coupling issue and involve all osculating elements.

The two most relevant papers devoted to the origin of narrow rings' eccentricities ($m=1$ mode) are those of \cite{GT81} and \cite{PM05}. Both analyses involve resonances that \cite{GT81} and \cite{PG87} have called eccentric (a denomination also adopted by \citealt{PM05}). Therefore, the present discussion starts with a definition of these resonances and their basic properties.

In this section, and because the planet oblateness is not essential and neglected throughout, the usual elliptic elements ($a,e,\lambda,\varpi$) are used instead of the epicyclic ones used so far; this approximation also allows us to rely on different perturbation equations without further justification.

\subsubsection{Eccentric resonances}\label{sec:eccres}

Resonances in circular rings can be tagged as either corotation or Lindblad resonances. Lindblad resonances are the only resonances that have been considered so far in this review; they occur at semi-major axis $a_r$ implicitly defined by the resonance condition\footnote{One reverts to the usual $m >0$ convention in this section, for simplicity of comparison with the existing literature.} $m(\Omega-\Omega_p)=\pm \kappa$, with $\Omega_p =  \Omega_s + k\kappa_s/m$, and act primarily on the fluid particles' eccentricity and periapse angle; corotation resonances occur at $\Omega = \Omega_p$ and act primarily on its semi-major axis and mean longitude. 

In an eccentric ring, corotation resonances keep their identity while pure Lindblad resonances no longer exist. Instead, eccentric resonances must be introduced, which share characteristics with both the more usual Lindblad and corotation ones (in particular, the torque at such resonances is made up of a corotation-like and a Lindblad-like contribution). This follows because, in an eccentric ring, the background mean eccentricity cannot be ignored when it is larger than the amplitude of the perturbation forced by the satellite. The related resonance semi-major axis is implicitly defined by:
%
%
\begin{equation}
m(\Omega - \Omega_p) = q_e\kappa,\label{eccres}
\end{equation}
with $q_e$ a positive or negative integers while $m >0$; $q_e=0$ corresponds to  corotation resonances, and non-vanishing $q_e$ to eccentric resonances\footnote{$q_e$ is clearly not related to the streamline compression parameter $q$.}. The pattern speed definition ($\Omega_p = \Omega_s + k\kappa_s/m$) is the same as in circular rings. Such resonances are thus characterized by three integers: $m,q_e,k$. In the circular limit (circular rather than eccentric background motion), $|q_e|=1$ gives back the usual Lindblad resonances. One further defines inner and outer eccentric resonances (IER/OER) according to whether $\Omega_p > \Omega$ or $\Omega_p < \Omega$. 

Eccentric resonances are labelled by $m+k:m-q_e$, which gives the ratio of the ring to satellite rotation frequencies ($n/n_p$) in the limit of a spherical planet. Finally, $|k|+|q_e|$ defines the order of the resonance; this choice is motivated by the fact that the associated component of the disturbing function is proportional to $e_s^{|k|}e^{|q_e|}$. 

Note that this definition of eccentric resonances is consistent with the requirement of stationarity of the usual generic argument of the disturbing function for coplanar motions\footnote{The notation $\lambda$ is used here for the mean longitude instead of $\varphi$, for ease of comparison with the existing literature.} ($j_1\lambda + j_2 \lambda_s + j_3\varpi + j_4\varpi_s$) under the rotational invariance constraint $j_1+j_2+j_3+j_4=0$ (see e.g., \citealt{MD99} for an in-depth discussion of the disturbing function and its properties). Taking into account this rotational invariance, \cite{GT81} define the component of the disturbing function associated to Eq.~\eqref{eccres} as
\begin{equation}
\phi_{lmp}(a,e)\cos[p\lambda + (m-p)\varpi - (m+k)\lambda_s + k\varpi_s],\label{distgt}
\end{equation}
where $p=m-q_e$ and $l=m+k$, the subscript $s$ referring to the satellite. One sees that\footnote{The index $k$ here is different from the index $k$ in \cite{GT81}.} $m=j_1+j_2=-(j_3+j_4)$, $q_e=j_2$, $-(m+k)=j_3$ and $k=j_4$.

Note finally that the order of the disturbing function is not the order of the related torques. Indeed, in the circular limit, eccentric resonance torques are proportional to $e_s^{2|k|}e^{2(|q_e|-1)}$  while corotation torques are proportional to $e_s^{2|k|}$; this point is justified in Appendix \ref{app:torque}.

\subsubsection{Qualitative discussion} 

\cite{GT81} tackle the problem through an analysis of motions in eccentric resonances. In this approach the mode coupling issue is implicitly accounted for through the second order evolution of the ring particles' elliptic osculating elements ($a,e,\lambda,\varpi$) under the action of the satellite forcing and an \textit{ad hoc} dissipation term\footnote{The final results do not depend on dissipation, as a dissipationless limit is taken.} acting on all osculating elements (the planet oblateness is neglected). 

\cite{PM05} have analyzed this problem within the framework of their fluid description of narrow ring dynamics, to explicitly account for the coupling of $m_2\neq 1$ satellite forcing ($l=m_2$) with the $m_1=1$ mode. First order coupling between the satellite $m_2$ Lindblad forcing and the global ring eccentricity produces an $m_2 \pm 1$ mode, which corresponds to an $m_2\pm 2$ eccentric resonance ($|q_e|=2$) in terms of the associated disturbing function\footnote{Only one of these two modes is expected to fall within the ring, depending on the satellite orbiting inside or outside the ring.}, in the classification of \cite{GT81} and \cite{PG87}.
Then, second order coupling of this $m_2\pm 1$ ring mode with the satellite $m_2$ forcing feeds back on the $m_1=1$ mode. The whole secular free energy of this $|q_e|=2$ disturbing function component feeds the radial action (and therefore the eccentricity) of the global $m_1=1$ mode. 
The authors have applied this analysis to the effect of the 47:49 resonance of Cordelia on the $\epsilon$ ring of Uranus, for which it was designed. They conclude that the associated torque (taken from \citealt{GP87}) does excite the ring eccentricity, while a balance between this excitation and viscous damping determines in principle the long term equilibrium of the ring eccentricity. 

\begin{figure}[t!]
\centering
\includegraphics[width=0.7\textwidth]{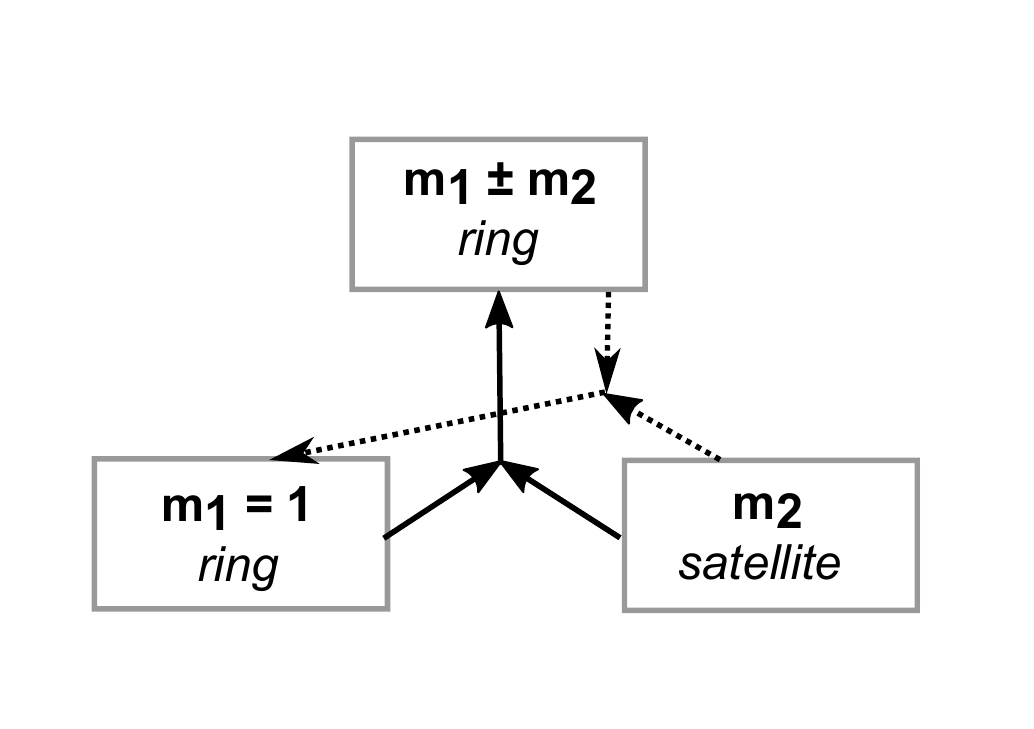}
\caption{Sketch of the mode coupling structure leading to eccentricity excitation by external satellites (adapted from \citealt{PM05}). First order coupling is represented by plain arrows, second order coupling by dotted arrows.}
\label{fig:coup}
\end{figure}

The mode coupling picture just described is sketched on Fig.~\ref{fig:coup}, adapted from \cite{PM05}. 

\medskip
\noindent{\textit{Number and type of resonances}\\
The \cite{PM05} type of approach is definitely more consistent with a fluid treatment of the ring than the analysis of \cite{GT81}. However, the most important point brought forward by the analysis of \cite{GT81} is that, in a more general context, the number and type of resonances involved is critically relevant to the result. This approach also allows us to compute eccentric resonance satellite torques as a side bonus. Therefore, we follows the approach of \cite{GT81} in the remainder of this review.

\cite{GT81} have shown that eccentric resonances share characteristics with the usual corotation and Lindblad resonances: the Lindblad part of an eccentric resonance acts primarily on ($a,e$) and excites the ring eccentricity while the corotation part acts primarily on ($a,\lambda$) and damps the eccentricity. The situation they have investigated assumes that a satellite on a circular orbit creates many eccentric resonances in a narrow ring. In the linear (unsaturated) regime and keeping the two leading orders in the disturbing function (both $|q_e|=1$ and 2 for any given $m$ while $k=0$), \cite{GT81} find that the Lindblad $|q_e|=1$ contribution vanishes so that the corotation $|q_e|=1$ contribution wins by a small margin ($\sim 5$\%) over the Lindblad $|q_e|=2$ contribution (the corotation $|q_e|=2$ contribution being negligible) and the eccentricity damps. Therefore, in this situation, the authors argue that the fate of the ring eccentricity critically depends on the saturation of the corotation terms; gap opening could in principle also reduce the Lindblad contribution, but this does not seem to be relevant for actual narrow rings. However, the reinvestigation of this problem detailed in Appendix~\ref{app:discrep} reveals that, surprisingly, the $|q_e|=1$ Lindblad contribution does in fact not vanish, so that external satellites always excite narrow ring eccentricities, contrarily to the conclusion stated in the abstract of their paper.

Note that, in the context analyzed by \cite{GT81}, \textit{all} eccentric resonances do feed back on the global $m_1=1$ mode, as is apparent from their second order calculation (summarized below). This formally mirrors the \cite{PM05} argument: the first order response to the satellite $m_2=p$ forcing produces a $m_2 + q_em_1$ response to first order in the osculating elements from the considered ($l,m,p$) potential component, resulting in $m_2 + q_em_1$ first order-responses in the position vector, as can be seen by reinserting the perturbed osculating elements into Eqs.~\eqref{epir} and \eqref{epiphi}. Then the resonant second order coupling with the satellite feeds back on the original $m_1 = 1$ mode, as, by construction of the osculating elements, the secular second order variation of the eccentricity contributes only to $m_1= 1$ in the position vector\footnote{One also sees in the same way that contributions to other $m\neq 1$ modes come from the non-secular response; these responses are eventually dissipated. Note that the product of first-order terms also produce constant corrections to the osculating elements, but these are unessential, only the second order resonant contribution leads to a secular increase of the ring eccentricity.}. 

However, the situation assumed by \cite{GT81} probably does not apply to an actual narrow ring, and definitely not the $\epsilon$ ring for which it was designed. This became obvious only after the discovery of this ring's shepherd satellites (there is only a small number of resonances with external satellites in this ring, see \citealt{PG87}). In any case, the fate of the eccentricity depends on which resonances are present in the ring; note that other satellites than shepherds may contribute to the eccentricity evolution. Conclusions for individual rings therefore need to be drawn on a case by case basis, a point not discussed in any detail in this review.

\medskip
\noindent\textit{Response to resonant forcing}\\
Finally, the physics of the ring response to the resonant satellite forcing is of some importance as well, although it is known that  \textit{linear} global net torques and energy exchanges between rings and satellites at resonances do not depend on the physics, at least for circular background motions (probably the most general proof of this statement has been provided by \citealt{MVS87}). The generalization of this result to eccentric background motions does not seem to have been investigated in the literature, but seems likely.

\cite{GT80} make use of the results of density wave theory for the computation of the evolution of a satellite eccentricity in multiple resonant interaction with a circular ring, while, as we have seen, \cite{GT81} rely on simpler models of the ring response for an eccentric ring. For a circular ring, favoring one approach over another will not matter as long as the response is linear in amplitude; however, the use of density wave theory will more closely reflect the nature of the actual ring physical response, at least when resonances are present within the ring and not only at the ring edge. For an eccentric ring, one must necessarily use simpler physical models of the disk response, as no density wave theory in an eccentric background motion has been developed to date. 

Now, although the linear global exchanges may not depend on the nature of the response, the situation is less clear in the nonlinear case; note in this respect that \cite{GP87} have shown that the torques confining the $\epsilon$ ring must be nonlinear [see their Eq.~(15)]. \cite{BGT84} discuss nonlinear density wave torque saturation in a circular ring; \cite{GT81} provide a different way to estimate nonlinear saturation, most notably of corotation torques, by heuristically quantifying the fraction of material trapped in librating motion where no net torque is applied; in a now classic paper, \cite{OL03} provide a much more quantitative fluid discussion of the same topic, and the saturation parameter is the same in both the heuristic and quantitative approaches (ratio of the satellite strength to the viscous diffusion efficiency). Nonlinear torque saturation estimates have been further elaborated by \cite{GP87}. It is nevertheless fair to stress that such nonlinear estimates are only semi-quantitative, and that in the nonlinear regime, the physical nature of the ring response to the resonant forcing may have some influence on the saturated torque amplitude and on the related energy exchange.

\subsubsection{Eccentricity evolution in the Goldreich \& Tremaine picture}\label{sec:gtpic}

We will give here results for individual first and second order resonances. We follow the logic of \cite{GT81}, from which one can get a broader view of the problem. The calculation will be presented in some detail, as the results differ from the original publication in important respects that will be discussed and justified; also, this will provide a physical and technical guideline for this rather arduous paper\footnote{\cite{GT81} also evaluate the evolution of the satellite eccentricity, which proceeds in much the same way as the calculation of the ring eccentricity evolution.}. Finally, expressions for all resonances up to second order ($|q_e|+|k| \le 2$) will be given for both the eccentricity evolution and associated torques (see Appendix \ref{app:torque} on this second point), as these have never been published in a systematic way up to now; the torque expressions are required in the analysis of narrow ring confinement by external satellites in section \ref{sec:satmass} (minimal satellite masses for confinement).

The physics is simplified in the following way. The perturbation equations below relate to individual fluid particles' osculating elements. However, as the global $m=1$ mode is dominant and as one is only interested in secular evolutions, they also bear directly on the behavior of this global mode as pointed out in the qualitative discussion above. The physics of the mode rigid precession is ignored as it is not essential in the problem; in particular the planet oblateness is neglected. 

Under these assumptions, it is more appropriate to revert to Laplace equations of perturbed elliptic motions, instead of Eqs.~\eqref{aepi} to \eqref{meanepi}. The perturbation equations then take the following form, valid to leading order in eccentricity\footnote{This does not preclude the validity of the second order perturbation expansion to identify resonant contributions. Also, the subscript $r$ has been dropped from the notation.} and easily seen to be equivalent to the Laplace equations of perturbed elliptic motion (see, e.g., \citealt{MD99}, chapter 6):
\begin{eqnarray}
\frac{da}{dt} & = & -\frac{2}{na}\frac{\partial \Phi}{\partial \lambda_0},\label{agt}\\
\frac{d\lambda_0}{dt} & = &\quad\!\! \frac{2}{na}\frac{\partial \Phi}{\partial a},\label{lambgt}\\
\frac{dI}{dt} & = & \quad\!\! \frac{\partial \Phi}{\partial \varpi},\label{egt}\\
\frac{d\varpi}{dt} & = & -\frac{\partial \Phi}{\partial I},\label{pomgt}
\end{eqnarray}
%
%
%
where $I=na^2[1-(1-e^2)^{1/2}] \approx na^2 e^2/2$ is radial action, $\lambda = nt + \lambda_0$ so that $\lambda_0$ is the mean longitude at epoch\footnote{A technical comment is in order here. For once, it is simpler to avoid the usual modification of the mean longitude at epoch, $\lambda = \int_0^t n dt + \lambda'_0$, as the secular terms can be calculated without explicitly evaluating the derivative with respect to $a$ (see, e.g., section 7.7.1 of \citealt{R88} for details on the two forms of the perturbation equations); this procedure also gives back the usual expression of the corotation torques in the most direct way.} and $\Phi$ the potential, given by Eq.~\eqref{distgt}. This is a quasi-Hamiltonian form of the equations. Defining $\mathbf{X}=(a,I,\varpi,\lambda)$ and denoting by $\mathbf{F}(\mathbf{X})$ the collective right-hand sides of the previous equations, the two first perturbation orders can be summarized as 
\begin{eqnarray}
\frac{d\mathbf{X}_1}{dt} & = & \mathbf{F}(\mathbf{X}_0) -\gamma \mathbf{X}_1\label{X1},\\
\frac{d\mathbf{X}_2}{dt} & = & \nabla_{\mathbf{X}_0}\mathbf{F}(\mathbf{X}_0)\cdot\mathbf{X}_1.\label{X2}
\end{eqnarray}
where the subscripts 0,1, and 2 refer to zeroth, first and second order quantities, respectively (the small parameter being the usual one, e.g., the ratio of the perturber's to the planet's mass).

A heuristic dissipation term $\gamma > 0$ has been introduced in the first order perturbation equation, in order to regularize the resonant response. The $\gamma\rightarrow 0$ limit is taken at the end of the calculation so that the physical basis for this procedure is not essential, at least in first approximation; its role is to keep the response linear. The physical reason of this regularization is that collective effects tend to make the circulating orbits the only relevant ones, an assumption always performed in ring fluid analyses\footnote{This assumes that collective effects are sufficiently strong to make librating orbits irrelevant; such an assumption is implicit, e.g., in the choice of the mode shapes, Eqs.~\eqref{rmode} and \eqref{phimode}. This needs to be checked, in particular at corotation resonances; see the discussion of \cite{GT81} on the magnitude of $\gamma$ for this to be true, in relation to their Eq.~(25).}. 

Eq.~\eqref{meanevec} is chosen as the definition of the mean eccentricity, recast in terms of the complex eccentricity, Eq.~\eqref{compe}:
\begin{equation}
e_r = |\langle Z\rangle| = |\langle e\cos\varpi\rangle - \mi \langle e\sin\varpi\rangle|,\label{vececc}
\end{equation}
%
%
where $\langle X\rangle$ is the mass average of $X$. Note that this averaging involves two different operations, as we are no longer dealing with streamlines but individual fluid particles: first, an average over all fluid particles at the same $a$, which equivalently can be expressed as an average over the particle's mean longitude at epoch\footnote{Because of this, no short-period average is needed to isolate the secular contribution.} $\lambda_0$; second, an average over $a$ weighted by $\sigma_0$. To take this into account, one uses two notations below:
\begin{itemize}
\item $\langle X\rangle$ indicates that the averaging must be performed on both variables $a$ and $\lambda$.
\item $\langle X\rangle_\lambda$ indicates that the averaging with respect to $\lambda_0$ has been performed, and only the radial mass average remains to be done.
\end{itemize}

The evolution of $e_r$ involves a second order expansion in the perturbation potential $e_r =  e_{r,0} + e_{r,1} + e_{r,2}$ with $e_{r,0}\gg e_{r,1} \gg e_{r,2}$ and $\varpi_r = \varpi_0+\delta\varpi_r$ with $\delta\varpi_r = \varpi_1 + \varpi_2 \ll 1$. Finally, it is assumed, as observed in most eccentric narrow rings, that the difference of eccentricities and periapse angle of ring streamlines is small and therefore negligible across the ring; correlatively, the planet-induced precession is ignored, so that $\varpi_0$ is constant and can be set to zero by an appropriate choice of the origin of angles. With these remarks, one finds
\begin{equation}
e_r^2 = |\langle Z\rangle|^2 = \langle e\rangle^2 + \langle e\delta\varpi\rangle^2 - \langle e\rangle\langle e\delta\varpi^2\rangle,\label{ering}
\end{equation}
up to second order in $\delta \varpi$.

The second order perturbation equation for $e_r$ is potentially less straightforward than might be expected, due to Eq.~\eqref{ering}. Remembering that $e_r\approx e_{r,0}$ (a type of substitution commonly performed in secular perturbation theory), one finds
\begin{eqnarray}\label{meaneccevol}
\frac{d\left\langle e_{r}\right\rangle}{dt} & = & \left\langle\frac{de_{r,2}}{dt}\right\rangle\nonumber\\
& =  &
\left\langle\frac{de_{2}}{dt}\right\rangle
- e_r \left\langle\varpi_1\frac{d\varpi_{1}}{dt}\right\rangle  + \frac{1}{e_r}\left\langle e_{1}\right\rangle
\left\langle\frac{de_{1}}{dt}\right\rangle 
+ e_r {\left\langle \varpi_{1}\right\rangle
\left\langle\frac{d\varpi_{1}}{dt}\right\rangle}\nonumber\\
& = & \left\langle\frac{de_{2}}{dt}\right\rangle
- e_r \left\langle\varpi_1\frac{d\varpi_{1}}{dt}\right\rangle.\label{dI2}
\end{eqnarray}
The last equality follows because $e_1,\varpi_1$ are harmonic in time and their average over $\lambda_0$ (as well as that of their derivatives) cancels out; for the same reason, $\langle X_1 dX_1/dt\rangle = 0$, leading to the intuitive result that the evolution of $\langle e_r\rangle$ is identical to the evolution of $e_2$. Furthermore,
\begin{equation}
\left\langle\frac{de_{2}}{dt}\right\rangle = \frac{1}{na^2 e_r}\left\langle\frac{dI_{2}}{dt}\right\rangle
- \frac{1}{n^2a^4 e_r^3}\left\langle I_1\frac{dI_{1}}{dt}\right\rangle.\label{de2}
\end{equation}
In these relations, the order of the time derivative and spatial averaging has been inverted. This is justified to leading order in eccentricity\footnote{The contribution of $da/dt$ is negligible compared to the contribution of $de_r/dt$ in the mass averaging integral;  this is also true for the connection between $de/dt$ and $dI/dt$.}.

The evaluation of the different second order contributions is straightforward (see Appendix \ref{app:discrep}) and yields:
\begin{eqnarray}
\left\langle\frac{de_{r,2}}{dt}\right\rangle & = & \frac{1}{na^2 e_r}\left( \left\langle\frac{dI_{2}}{dt}\right\rangle_L + \left\langle\frac{dI_{2}}{dt}\right\rangle_C\right),\label{e2}\\
\left\langle\frac{dI_{2}}{dt}\right\rangle_L & = &
\left\langle \frac{q_e^2\pi}{2e_r n a^2}\delta(pn-ln_s)
\left(\frac{\partial \phi^2_{lmp}}{\partial e} \right)_{e_r} \right\rangle_\lambda,\label{e2l}\\
\left\langle\frac{dI_{2}}{dt}\right\rangle_C & = &
 - \left\langle \frac{\pi q_e p}{n a} \frac{\partial}{\partial a}\left[ \phi^2_{lmp}\delta(pn-ln_s)\right]_{e_r}\right\rangle_\lambda,\label{e2c}
\end{eqnarray}
where the $L$ an $C$ index refer to the Lindblad- and corotation-like contribution to the eccentricity evolution. For corotation resonances ($q_e=0$), only the corotation term arises; note that, consequently, corotation resonances do not contribute to the ring eccentricity evolution as the corotation term is proportional to $q_e$. 

The last step differs from \cite{GT81} as we investigate the effect of isolated resonances, whereas they focus on a quasi-continuum of $m$ values. The mass average is greatly simplified by the fact that, for $m\gg 1$, the semi-major axis dependence is weak, except in the combination $n-n_s$, so that one readily obtains the following final expressions for the Lindblad and corotation contributions:
\begin{eqnarray}
\left\langle\frac{de_r^2}{dt}\right\rangle_L & = & 
\frac{2\pi}{3(m-q_e) n^3 a^3\Delta a}
\left[\frac{q_e^2}{e}\frac{\partial \phi^2_{lmp}}{\partial e} \right]_{e_r},\label{phil}\\
\left\langle\frac{de_{r}^2}{dt}\right\rangle_C & = &
\frac{4\pi q_e}{3n^2 a^2 \Delta a \sigma_0} \phi^2_{lmp} \frac{d}{da} \left(\frac{\sigma_0}{n}\right)\nonumber\\
& \simeq & \frac{2\pi q_e}{n^3 a^3 \Delta a} \phi^2_{lmp}.\label{phic}
\end{eqnarray}

The last expression ignores the contribution of the derivative of the surface density, an approximation often made in the evaluation of corotation resonance torques. Note that for eccentric resonances (the only ones relevant for the ring eccentricity evolution), the corotation contribution is negligible (it is smaller by $\sim e_r^2$ compared to the Lindblad one). As a consequence, \textit{all} eccentric resonances contribute to increase the ring eccentricity. This result is at odds with the conclusion drawn by \cite{GT81} who found instead a vanishing leading $|q_e|=1$ Lindblad contribution; considering the importance of this point, this outcome is scrutinized and justified in Appendix~\ref{app:discrep}. 

Our definition of the ring mean eccentricity implicitly assumes that the amplitude gained in small scale ($m \gg 1$) disturbances is transferred to the mean eccentricity in the resonance region, and then to the overall ring; otherwise, the ring average of a localized disturbance would not reflect the observed geometry of narrow rings, whose eccentricity is nearly uniform throughout their width. The first part of this redistribution depends on the ability of high $m$ disturbances to transfer the free energy gained from the satellite to the mean ring eccentricity (apart from the dissipated part), as discussed in the second order mode coupling theory of \cite{PM05}; this does not pose any particular problem, and the qualitative discussion above justifies the assertion that all eccentric resonances contribute to the secular variation of the eccentricity. The second part of the transfer, however, is described neither in \cite{PM05} nor here; it is implicitly assumed that the ring collective effects (self-gravity) will make such a global transfer to the whole ring effective before the local gain in eccentricity is locally damped. Such an assumption is reasonable but would need to be checked more carefully in the future\footnote{The argument developed by \cite{GT81} in the first section of their paper implies that the mean vector eccentricity is only very slowly damped directly by collisions; however, this ignores the type of damping evaluated in Eq.~\eqref{edamp}, which results from a coupling between the ring self-gravity and the stress tensor. However, the two-streamline analysis performed earlier does suggest that the required transfer time-scale is indeed shorter than the dissipation one.}; in particular, it is quite possible that this averaging does not faithfully capture the end result in terms of ring mean eccentricity evolution.

This second transfer issue is irrelevant in the \cite{GT81} analysis, where all ring fluid particles undergo the same evolution, due to a quasi-continuum of $m$ values, which all yield essentially the same eccentricity evolution for all streamlines in the asymptotic limit described in Appendix \ref{app:asymp} for the disturbing function; however, the assumption of resonance quasi-continuum is only valid if the distance between consecutive resonances (for given $q_e,k$) is smaller than the resonance width, a situation that does not arise in the narrow rings whose shepherd satellites have been identified.

At this point, further progress is accomplished and more specific expressions are obtained through a number of simplifying assumptions: $m\gg 1$, $\eta_r\equiv e_r a_s/x \ll 1$ and $\eta_s\equiv e_s a_s/x \ll 1$ where $x\equiv a - a_s$ is the satellite-ring distance and $e_r,e_s$ are the ring and satellite eccentricities. As usual $\Delta a \ll |x| \ll a$ where $\Delta a$ is the ring width. In the $\epsilon$ ring, the parameters $\eta_r, \eta_s$ are smaller than unity, but not very small; the approximation concerning $\eta$ is therefore not strictly satisfied, although not really violated. These asymptotic approximations ($m\gg 1$, $\eta_r,\eta_s\ll 1$) bring simplifications in the form of the disturbing function. The corresponding evaluations are performed in Appendix \ref{app:asymp} for the leading eccentric resonances. This leads to\footnote{The first expression might seem surprising, as it is independent of the ring eccentricity, suggesting that a circular ring can become eccentric under the action of an external satellite. This seems to violate symmetry considerations: how can a specific direction of periapse angle be selected in an average theory where no direction is preferred? A similar issue can be raised about the selection of the mode whose amplitude is amplified by the satellite. Remember however that the ordering $e_{r,0}\gg e_{r,1} \gg e_{r,2}$ is needed in the derivation of Eqs.~\eqref{ering} and \eqref{dI2}, and the fate of an initially circular ring is outside the scope of the present analysis; similarly, the mode coupling argument of \cite{PM05} requires a non-vanishing initial eccentricity. I am indebted to Scott Tremaine for raising this point and for related discussions; the point of view expressed here is mine, though, as remaining possible subtle errors in this line of argument.}:

\bigskip

\begin{description}
\item[$\mathbf{k=0, |q_e|=1}$] 
\begin{eqnarray}
\frac{d\langle e_r^2\rangle}{dt} & = & \frac{8F_1^2}{9\pi}\left(\frac{M_s}{M_p}\right)^2\frac{a}{\Delta a}\left|\frac{a}{x}\right| n,\nonumber\\
& \simeq & 1.80 \left(\frac{M_s}{M_p}\right)^2\frac{a}{\Delta a}\left|\frac{a}{x}\right| n\label{k0q1}
\end{eqnarray}
\item[$\mathbf{k=0, |q_e|=2}$] 
\begin{eqnarray}
\frac{d\langle e_r^2\rangle}{dt}& = & \frac{8F_2^2}{81\pi}\left(\frac{M_s}{M_p}\right)^2\frac{a}{\Delta a}\left|\frac{a}{x}\right|^3 n e_r^2\nonumber\\
& \simeq & 4.57 \left(\frac{M_s}{M_p}\right)^2\frac{a}{\Delta a}\left|\frac{a}{x}\right|^3 n e_r^2,\label{k0q2}
\end{eqnarray}
\item[$\mathbf{k=q_e=\pm 1}$] 
\begin{eqnarray}
\frac{d\langle e_r^2\rangle}{dt} & = & \frac{16F_2^2}{81\pi}\left(\frac{M_s}{M_p}\right)^2\frac{a}{\Delta a}\left|\frac{a}{x}\right|^3 n e_s^2\nonumber\\
 & \simeq & 9.14 \left(\frac{M_s}{M_p}\right)^2\frac{a}{\Delta a}\left|\frac{a}{x}\right|^3 n e_s^2,\label{k1q1}
\end{eqnarray}
\end{description}
In these expressions, $F_1$ and $F_2$ are dimensionless numerical factors given in Eqs.~\eqref{F1} and \eqref{F2}.

Although less relevant to actual rings, as pointed out earlier, let us conclude this discussion with an expression for the resonance quasi-continuum limit of \cite{GT81}. In this limit, many resonances are present at any given location in the ring. Their individual contributions are all identical for given\footnote{Resonances with $q_e+k=0$ correspond to $a=a_s$ and are of no interest here.} $q_e,k$ ($q_e+k \neq 0$) in the asymptotic regime explored here ($|x| \ll a$) and one needs only to focus on the dominant type of eccentric resonance ($k=0, |q_e|=1$). In this case, the ring eccentricity evolution is given by Eq.~\eqref{k0q1}, scaled by the number of resonances in the ring width\footnote{This simple prescription would not be appropriate for recovering the (negligible) corotation-like contribution.}, $\delta N\simeq (2/3)|a/x|^2(\Delta a/a)$ (obtained from $m\approx 2|k+q_e|a/|x|$ due to the resonance relation), leading to
\begin{eqnarray}
\left(\frac{d\langle e_r^2\rangle}{dt}\right)_{cont.} & = & \frac{16F_1^2}{27\pi}\left(\frac{M_s}{M_p}\right)^2\left|\frac{a}{x}\right|^3 n\nonumber\\
& \simeq & 1.20 \left(\frac{M_s}{M_p}\right)^2\left|\frac{a}{x}\right|^3 n,\label{k0q1gt}
\end{eqnarray}
which corrects the corresponding expression of \cite{GT81}, Eq.~(80).

\section{Sharp edges and shepherding}\label{sec:shep}

The formation of sharp edges and the confinement (shepherding) of narrow rings have a common physical origin in the various physical analyses put forward over the years by BGT \citep{GT79a,BGT82,BGT84,BGT89,GRS95}; the confinement of sharp edges and narrow rings relies crucially on the reversal of the flux of angular momentum pointed out in section \ref{sec:press} (stress tensor) as will be discussed in the first subsection. 

Very few papers have been devoted to this problem besides the BGT ones; they are reviewed in section \ref{sec:othersh}. Also, the absence of obvious confining satellite in quite a number of instances challenges the BGT picture; this is further discussed in section \ref{sec:conf}.

\subsection{Confinement in the BGT picture: broad outline}\label{shflux}

Two limit cases must be distinguished in the discussion of the BGT picture. The first one corresponds to an isolated satellite resonance \citep{BGT82}. In this context, the ring streamlines are described by an $m$-lobe pattern consistently with the kinematics presented in section \ref{sec:kin} for integer $m$; shepherding through isolated resonances is relevant for sharp edges of broad rings and for narrow rings. The second one corresponds to the wake-like response of a ring edge perturbed by a close satellite \citep{BGT89}; this occurs when resonances overlap. An integer $m$-lobe pattern cannot form around the whole azimuthal extent of the ring in this context. Instead, the satellite perturbation is more aptly described in the impulse approximation where the effect of the satellite is assimilated to a jump in osculating elements at closest encounter, producing a finite eccentricity at constant semi-major axis. The ensuing wake pattern is then partially or fully damped between two encounters\footnote{The wake model of \cite{BGT89} assumes full damping.}. The related kinematics can still be captured by the formalism of section \ref{sec:kin} with a real (positive or negative) $m$ instead of an integer one; $m$ then depends on the distance between the satellite and the streamline. The ring particles' eccentricity and apsidal shift become azimuthally dependent. These two types of responses are successively discussed, as they differ in some important aspects. In particular, viscous angular momentum flux reversal occurs in different ways in these two limiting cases.


The initial shepherding picture put forward by \cite{GT79a} fits in neither of these two categories, although in principle it bears a more direct relationship to the wake confinement mechanism. This picture is qualitatively correct in its broad outline but is flawed in a very major way, as first pointed out by \cite{BGT84}. This flaw is quite severe: a narrow ring in which the wakes produced by the two shepherd satellites extend throughout the ring can only be confined by the shepherds if viscous angular momentum flux reversal applies \textit{everywhere} in the ring. Such an extensive reversal requirement in this context has apparently not been explicitly discussed in the literature yet. Such a discussion is provided below, as well as the related implication for the shepherd satellite mass limits.  
%
%
Finally, the possibility that angular momentum flux reversal holds throughout a narrow ring suggests that a related mechanism of confinement may be relevant: a single satellite may suffice to confine the ring \citep{GRS95}. 
%
%

These various contexts are discussed below, after providing a criterion to distinguish the occurrence of wake or resonant responses in the ring.

\subsection{Satellite wakes and isolated vs ovelapping resonance response}\label{sec:wakecrit}

Following \cite{BGT89}, it is useful to provide a criterion distinguishing the two limit cases described above from a relevant wake kinematic modelling; this criterion is also a resonance overlap condition. At the same time, some dynamical aspects of the wake response to satellite forcing will be given for later use.

\subsubsection{Wake kinematics}

Let us slightly alter the kinematics of section \ref{sec:kin} to capture a wake response to the forcing of a close satellite on a circular orbit. Let us introduce the azimuthal phase in the rotating frame\footnote{The planet oblateness is unessential in this discussion and is ignored.}
\begin{equation}\label{synphase}
\phi=\varphi-n_s t.
\end{equation}
With this definition, downstream corresponds to $\phi >0$ for streamlines interior to the satellite orbit, and  to $\phi < 0 $ otherwise (that is, the stream leads the satellite on its inward side and trails on its outward side, per Kepler's third law).
The azimuthal wavenumber is now a continuous function of $x\equiv a - a_s$ implicitly defined by the resonance condition $m(n-n_s)=n$:
\begin{equation}\label{wakem}
m=-\frac{2 a_s}{3x}.
\end{equation}
The shape of streamlines is now described by\footnote{A more precise representation assumes that $a$ depends on $\overline{x}=\overline{a}-a_s$ and $\phi$, where $\overline{a}$ is the average of $a$ over $\phi$ for any given fluid particle.}:
\begin{equation}
r(x,\phi) = a \left[ 1-\epsilon(x,\phi) \cos[ m(x)\phi + \Delta(x,\phi)] \right],\label{rwake}
\end{equation}
Note that the wake azimuthal wavelength corresponds to $L= 2\pi a/m = 3\pi x$, a well-known result; the Encke gap edges provide a prototypical example of this feature \citep{CS85,SCME86}.

In the impulse approximation (i.e., at $\phi=0$) and to first order in the ratio of the satellite mass to the planet's, the eccentricity and apsidal shift are given by [cf \citealt{JT66}, Eq.~(37); see also \citealt{Der84}, section II.A]
\begin{eqnarray}
\epsilon_0 & = & g \frac{M_s}{M_p}\left(\frac{a_s}{x}\right)^2,\label{epsimpulse}\\
\Delta & = & - \mathrm{sgn}(x)\frac{\pi}{2},\label{deltimpulse}
\end{eqnarray}
with
\begin{equation}\label{impulsefactor}
g=\frac{8}{9}[2K_0(2/3)+K_1(2/3)]=2.24
\end{equation}
This result can be heuristically understood by considering that a radial acceleration $\dot{v}\sim GM_s/x^2$ is exerted at closest distance on the ring particles during a time $\sim v_{syn}/x\sim 1/\Omega$ ($v_{syn}$ is the relative velocity of the ring and the satellite). The ensuing relative velocity $v\sim \dot{v}/\Omega\sim ae_0$ gives the scaling factors of Eq.~\eqref{epsimpulse}.

With these results, and neglecting streamline interactions for the time being, the definition of the streamline compression parameter $q$ [Eqs.~\eqref{qcos} and \eqref{qsin}] yields\footnote{The streamline compression parameter characterizes the variation of $r$ with $a$ so that Eq.~\eqref{qsin} must be generalized to $q\sin \gamma =ae \partial(m\phi +\Delta)/\partial a$.}
\begin{equation}
q^2 = (3m\epsilon_0)^2\left[1+ \left(\frac{m\phi}{2}\right)^2\right],\label{qwake}
\end{equation}
which shows that streamline compression increases downstream from the impulse by the satellite. 

\subsubsection{Isolated vs overlapping resonance response}

In the absence of interactions, streamline crossing would take place when $q =1$, i.e., at the critical phase $\phi_c$:
\begin{equation}\label{phic1}
\phi_c =  - \mathrm{sgn}(x) \frac{3M_p}{2gM_s}\left(\frac{x}{a_s}\right)^4
\end{equation}
If $\phi_c \gg 2\pi$ the satellite response will of the resonant type, as successive satellite encounters occur before kinematic crossing occurs and as dissipation will produce a resonant streamline shape in the end through damping of the free mean motion;  self-gravity will then prevent the relative apsidal drift of streamlines and streamline crossing. Otherwise the response will be a wake and streamline crossing is prevented by viscous damping. 

As a rule of thumb, \cite{PG87} simplify Eq.~\eqref{phic1} into
\begin{equation}
\phi_c\sim \frac{M_p}{M_s}\times \left(\frac{1}{m^4}\right)\lesssim 1\label{phic2}
\end{equation}
for a wake response to occur in the ring. This criterion is essentially identical to the requirement that consecutive leading order resonances do overlap when the resonance width is estimated from test particle's periodic orbits forced by the satellite, the amplitude $a\epsilon$ of which is $\sim a^2(M_s/M_p)/(a-a_r)$ \citep{BGT82,GT82}; indeed, such orbits cross when $da\epsilon/da > 1$, which sets the onset of dominant collective effects (precisely to prevent this crossing). Identifying  the width of the perturbed region with the size of the region of crossing test particle orbits gives an estimate of the resonance width $w$:
\begin{equation}
w \simeq (M_s/M_p)^{1/2} a.\label{reswidth}
\end{equation}
Now, for leading order resonances [$m(\Omega - \Omega_p)=\pm\Omega$] consecutive resonances are separated by $\delta a_r\sim a_r (x/a_r)^2$ where $a_r$ is the resonance radius for any given $m\gg 1$. The constraint $w > \delta a_r$ (overlapping resonances) is identical to the rule of thumb mentioned above. 

\subsubsection{Some elements of wake dynamics} 

Torques play an important role in confinement dynamics, and it is useful to provide here an expression for the torque density (per unit semi-major axis) associated to such a wake response, for later use. The simplest argument makes use of Eq.~\eqref{epsimpulse} and of the Jacobi constant for a circular satellite orbit $\Delta E - n_s\Delta H =0$ by assuming that the impulse in eccentricity is fully damped between two encounters; this leads to a change in semi-major axis \citep{LP79,Der84}. The resulting torque density per unit semi-major axis is readily obtained from this procedure
\begin{equation}\label{torquedens}
\mathfrak{T}_s = \mp \frac{g^2}{2}\left(\frac{M_s}{M_p}\right)^2 \frac{a^4}{x^4}\sigma_0 n^2 a^3+ \mathcal{O}(e_s^2),
\end{equation}
with positive (negative) sign for inner (outer) satellites and where the numerical coefficient $g$ is given in Eq.~\eqref{impulsefactor}.

The second element of wake dynamics that will be needed in the following discussions is the wake damping equation [Eq.~(48) of \cite{BGT89}], which reads 
\begin{equation}\label{wakedamp}
\frac{d\epsilon^2}{d\phi} = -\frac{4aqt_1}{3x\sigma_0n^2a^2}.
\end{equation}
This equation follows from Eq.~\eqref{t1} in the following way. The short time scale of the problem is the orbital one. As $d\phi/dt = (n-n_s)= n/m$, the phase change corresponding to one orbit is $\delta\phi =2\pi/m$. The averaging performed in Eq.~\eqref{t1} is in fact an average over this short phase in the problem at hand. Furthermore, in the limit $\phi_c\ll 1$, wake damping occurs in the tight winding limit where the derivative of the phase $m\phi+\Delta$ dominates over the derivative of the eccentricity in the contribution to the streamline compression parameter $q$; thus $q = ae\partial (m\phi +\Delta)/\partial a$. In this regime $\gamma=\pi/2$ and the dominant term in Eq.~\eqref{t1} is the $t_1\partial (m\phi +\Delta)/\partial a$ term (where the modification of the phase derivative has been taken into account to properly generalize this equation). Eq.~\eqref{t1} then yields Eq.~\eqref{wakedamp} in a straightforward manner from these considerations; this derivation is more straightforward than the one adopted in \cite{BGT89}.

\subsection{Sharp edge confinement at an isolated resonance}\label{sec:resconf}

The most cogent analysis of the problem raised by the existence of sharp edges has been provided by \cite{BGT82}. Their discussion is mostly heuristic; the authors also mention $N$-streamline numerical calculations in support of this analysis, but these are not reported in the paper and have never been published. However, the heuristic argument captures the most critical elements of the problem and is presented below in a slightly more general form for both dilute and dense rings\footnote{The original discussion focused on dilute rings.}. For definiteness, the discussion focuses on outer edges, but is exactly mirrored for inner edges, with an appropriate sign change of the torque density.

The heuristic argument of \cite{BGT82} points out the existence of a significant increase of the velocity dispersion (for dilute rings) or of the ring thickness (for dense ones) to absorb the  energy dissipation required in the perturbed part of the edge for stationary confinement to occur. To understand the physical origin of this requirement, it is convenient to divide the edge zone into two regions: an inner one, with $a \le a_i$ where the flow is essentially circular, and an outer one $a_i \le a \le a_o$ where the flow is perturbed, $a_o$ being the ring edge.

The energy and angular momentum viscously transferred from the unperturbed to the perturbed region are $L_E^{vis}(a_i) = L_E^{vis}(q=0) = n L_H^{vis}(a_i)$ from Eq.~\eqref{visflux}. Conservation of angular momentum implies that $|T^s|= L_H^{vis}(a_i)$ where $T^s$ is the total satellite torque exerted in the perturbed region. From Jacobi's constant, the total exchange of energy with the satellite is $\Omega_p |T^s| = \Omega_p L_h^{vis}(a_i)$. Consequently, the total energy that must be dissipated in the perturbed region is $(n-\Omega_p)L_h^{vis}(a_i)$, with a related average dissipation rate per unit mass
\begin{equation}\label{edgediss}
\left\langle\frac{dE}{dt}\right\rangle_{a>a_i}=\frac{(n-\Omega_p)L_H^{vis}(a_i)}{2\pi a(a_o-a_i)\sigma_0}.
\end{equation}
This is substantially larger than the dissipation per unit mass in the unperturbed region, which reads [from Eq.~\eqref{visflux}]
\begin{equation}
\left\langle\frac{dE}{dt}\right\rangle_{a < a_i} = \frac{3n L_H^{vis}(a_i)}{4\pi a^2\sigma_0}.\label{standarddiss}
\end{equation}
To estimate this enhancement, one needs to estimate the width of the perturbed region. This follows from Eq.~\eqref{reswidth} which gives $(a_o - a_i)/a\sim (M_s/M_p)^{1/2}$, leading to
\begin{equation}\label{dissexcess}
\frac{\langle dE/dt\rangle_{a>a_i}}{\langle dE/dt\rangle_{a<a_i}}\sim  \frac{1}{m}\left(\frac{M_p}{M_s}\right)^{1/2}\sim 6\times 10^2,
\end{equation}
where the last estimate is relevant for the B-ring edge. The arguments of section \ref{sec:press}  (stress tensor) imply that this enhancement translates into an increase of the velocity dispersion (dilute rings) or of the unperturbed ring thickness (dense rings) near the edge\footnote{Self-gravitational wakes are ignored in this discussion for simplicity; see the remarks on this point in section \ref{sec:sgw} on self-gravitational wake transport.}.

The viscous angular momentum luminosity must cancel at the ring edge, so that the conditions for angular momentum reversal must be prevalent there. This follows from two complementary lines of argument that are presented in turn right below: from a mathematical point of view, the constraint of conservation of angular momentum requires angular momentum flux reversal at the edge; from a physical point of view, the enhanced dissipation in the perturbed edge region implies that without this reversal, the shepherding of the edge could not be achieved as the viscous flux would exceed the torque that is responsible for the perturbation in the first place. 

The constraint of conservation of angular momentum reads\footnote{The self-gravity luminosity is negligible in this context but this would only simplify the present argument. Also a Eulerian approach is implicitly assumed here; then, an advective angular momentum term $2\pi a\sigma_0 H (da/dt)$ should also be present in this relation [see Eq.~\eqref{dX2} of Appendix \ref{app:torque} and section 6.4 of \cite{L92} on this point]. However, this term is generically negligible because the steady-state mass conservation implies that $\sigma_0(da/dt)$ is equal to its unperturbed value [as a consequence of Eq.~\eqref{mass} averaged over $\varphi$]; the enhancement just discussed in relation to Eq.~\eqref{dissexcess} then makes the advective term much smaller than the viscous flux term [see Eqs.~\eqref{arphi}, \eqref{visflux} and the discussion after Eq.~\eqref{localtorque}].} $\partial(L_H^{vis}+L_H^{sg})/\partial a$ $=\mathfrak{T}^s$ in steady state, leading to the well-known global angular momentum balance constraint:
\begin{equation}
L_H^{vis}(a_i)=\left|\int_{a_i}^{a_o} da \mathfrak{T}^s\right|=|T^s|.\label{torquebalance}
\end{equation}
Now, assuming that there is no ring material beyond the edge, one necessarily has $L_H^{sg}(a_o^+)=0$ from Eq.~\eqref{sgflux}; similarly $a_{r\theta}(a_o^+)=0$ so that $L_H^{sg}(a)+L_H^{vis}(a)=|\int_a^{a_o} da \mathfrak{T}^s|=|T^s(a)|$. Using Eq.~\eqref{sgflux}, one has $L_H^{sg}(a)\rightarrow 0$ when $a \rightarrow a_0$. Consequently, as the satellite torque density is finite, one must also have $L_H^{vis}(a)\rightarrow 0$ when $a \rightarrow a_0$. Without this property, the edge could not reach steady state\footnote{This assumes that the edge is sharp, i.e., that the drop of surface density at the edge occurs on a very small length scale compared to the others; in any case, the surface density must drop to zero at least on length scales comparable to the ring thickness at the edge, of the order of a few tens or hundreds of meters.}. For later use, and taking into account that the self-gravity contribution is negligible, one defines the local angular momentum balance condition as
\begin{equation}
L_H^{vis}(a)=\left|\int_{a}^{a_o} da \mathfrak{T}^s\right|,\label{localtorque}
\end{equation}
of which Eq.~\eqref{torquebalance} is a particular case.

The enhanced dissipation in the perturbed edge region makes angular momentum luminosity reversal even more necessary. Indeed, in the perturbed edge region ($a_o > a > a_i$, without some cancellation of some sort, the viscous angular momentum luminosity $L_H^{vis}(a)$ would greatly exceed $L_H^{vis}(a_i)$ due to the natural scaling of $a_{r\theta}$ with the ring velocity dispersion and/or surface density (see sections \ref{sec:dilute} and \ref{sec:compact}), while the integrated satellite torque from $a$ to the edge is by necessity smaller than the total torque exerted on the perturbed region. In other words, one would then expect $L_H^{vis}(a) \gg L_H^{vis}(a_i) = |T^s| > |\int_a^{a_o} da \mathfrak{T}^s|$, i.e., the satellite could never confine the edge in such conditions. 

This discussion has an interesting consequence. Combining Eq.~\eqref{edgediss}, the dissipation contribution to Eq.~\eqref{arphi}, and Eq.~\eqref{specen} leads to
\begin{equation}\label{excesst1}
|\langle t_1 \rangle_{a>a_i}| \simeq \frac{2a}{qm(a_o-a_i)}a_{r\theta}(a_i)\gg a_{r\theta}(a_i).
\end{equation}
As $a_{r\theta}(a)\sim a_{r\theta}(a_i)$ through most of the perturbed region due to the local angular momentum balance constraint Eq.~\eqref{localtorque}, this implies that the conditions for angular momentum reversal must be met in most of the edge region, and not only close to the edge. Otherwise $t_1$ and $a_{r\theta}$ would be of the same order of magnitude from the scalings given in section \ref{sec:press} (stress tensor). 
 
\subsection{Overlapping resonances confinement through wake response: broad ring}\label{sec:wakeconfbroad} 

\cite{BGT84} were the first to point out that angular momentum flux reversal could be mediated through the wake response of a ring to satellites' perturbations. In this context, however, a somewhat different shepherding mechanism is required, as proposed and analyzed in detail by \cite{BGT89}. It involves angular momentum flux reversal in the downstream wake produced at the edge by the satellite in the impulse approximation, and requires the ring to be large enough so that the wake amplitude can be approximated to zero far from the edge. This second condition clearly applies to a broad ring edge, and is the context analyzed in this discussion of this process. The dynamics is more sophisticated than for a resonant edge confinement, and will only be briefly outlined; however, the material provided in this review is sufficient to follow the details of the \cite{BGT89} analysis in case of need. The edge position is noted $x_e \equiv a_s - a_e$ (positive for an inner edge, negative for an outer one).

The wake evolves kinematically downstream until a condition of near streamline crossing holds; then the wake damps at nearly constant $q$. Self-gravity plays little role in the problem, and streamline crossing is prevented by viscous damping. Qualitatively, viscous flux reversal ($a_{r\theta}< 0$) takes place for $0 < \phi \lesssim 3 \phi_c$ (a range identified numerically by \citealt{BGT89}), while on average over $\phi$ the luminosity must cancel at the edge, for reasons already discussed for the resonant shepherding process.

It is useful to examine angular momentum and energy constraints in this context. The global angular momentum constraint Eq.~\eqref{torquebalance} is unchanged, which from Eqs.~\eqref{torquedens} and \eqref{visflux} gives
\begin{equation}\label{edgearphi}
a_{r\theta}(a_i) =  \frac{g^2}{12\pi}\left(\frac{M_s}{M_p}\right)^2 \left(\frac{a}{|x_e|}\right)^3\sigma_0 n^2 a^2.
\end{equation}
Instead of Eq.~\eqref{excesst1} it is preferable to estimate $|\langle t_1\rangle|$ from Eqs.~\eqref{wakedamp} and \eqref{epsimpulse}:
\begin{eqnarray}
|\langle t_1(x)\rangle_{\phi}| & = & \frac{1}{2\pi}\left|\int_0^{2\pi}d\phi\ t_1\right|\nonumber\\
& = & \frac{3g^2}{8\pi q}\left(\frac{M_s}{M_p}\right)^2 \left(\frac{a}{x}\right)^4\sigma_0 n^2 a^2 \label{dissip1}\\
& = & \frac{2}{q}\left(\frac{|x_e|}{x}\right)^3\frac{a}{x}\times a_{r\theta}(a_i)\gg a_{r\theta}(a_i),\label{dissip2}
\end{eqnarray}
where $\langle X \rangle_\phi$ stands for the azimuthal average\footnote{This notation differs from the convention of \cite{BGT89}, where the bracket average stands for an average over their intermediate time-scale, whereas it is used here for their long (synodic) time scale. Note also that $\sigma_0$ is factored out of $t_1$ and $a_{r\theta}$ in their paper.} of any quantity $X$. A similar conclusion is obtained from Eq.~\eqref{excesst1} with $a_o - a_i\sim x$. 

Note finally that the perturbed flux coefficient is of the same magnitude as the dissipation one in the perturbed (wake) angular sector of the streamline [i.e., for $0 < \phi \lesssim 3 \phi_c$, cf.~\cite{BGT89}, Eq.~(60)], but $|\langle a_{r\theta}(x)\rangle_{\phi}| \ll |\langle t_1(x)\rangle_{\phi}|$ in most of the perturbed region, due to the local angular momentum balance condition. 

These results imply again that the conditions for viscous angular momentum flux reversal apply in most of the perturbed region. Indeed, the dimensional scaling is the same for all of the stress tensor coefficients, so that the only way $t_1$ can largely dominate $a_{r\theta}$ is by having the dimensionless $q$ dependence of $a_{r\theta}$ satisfy a condition of near cancellation. This has important implications for the original \cite{GT79a} picture that it is interesting to revisit, although, as pointed out in section \ref{sec:satamp} (excitation of eccentricities by external satellites), the actual relevance of the underlying physical context (overlapping resonances throughout a narrow ring radial extent) is unclear.

\subsection{Overlapping resonances confinement through wake response: narrow ring}\label{sec:wakeconfnarrow}

The points made above imply that the initial \cite{GT79a} shepherding model simply cannot work.  This was first pointed out by \cite{BGT84}, and it is useful to examine again and update this argument in  light of the understanding provided by this work's analysis of wake edge confinement. The satellites are assumed to have the same mass and to orbit at equal distance $|x|$ from the ring, so that the sum of their torques cancels at the ring mid-semimajor axis $a_m$. The ordering $\Delta a \ll |x| \ll a$ is assumed ($\Delta a$ is the ring width); the first inequality ensures that the configuration does exist for some time (satellites closer than $\Delta a$ would be quickly repelled by the rings) while the second is needed for the wake criterion Eq.~\eqref{phic1} to be satisfied.

From Eq.~\eqref{torquedens}, each satellite exerts a torque on the ring of magnitude (the sign depending on the semi-major axis of the satellite with respect to the ring's)
\begin{equation}\label{torqueshep}
T^s_0 = \mp \frac{g^2}{2}\left(\frac{M_s}{M_p}\right)^2 \frac{a^3\Delta a}{x^4}\sigma_0 n^2 a^4.
\end{equation}
However, the torque densities exactly cancel in the mid-ring streamline, and nearly cancel elsewhere, so that only the differential torque is in fact exerted on each ring half; accordingly, the angular momentum balance equation Eq.~\eqref{torquebalance} applied to each ring half yields
\begin{equation}
|T^s_0| = |xL_H^{vis}(a_m)|/2\Delta a,\label{gttorque}
\end{equation}
where $a_m$ is the mid-ring semi-major axis. Although the compounded satellite torques nearly cancel, the compounded dissipation associated with the satellite disturbances does not. This is most transparent in the $\phi_c \ll 2\pi$ limit; in this case, each satellite disturbance is very localized in azimuth on any streamline; these disturbances do not superpose most of the time, so that they damp independently, although the changes in semi-major axis due to both torques do nearly cancel one another once averaged along the whole streamline. As a consequence, Eq.~\eqref{dissip1} holds throughout the ring\footnote{This calculation assumes a circular ring, where $\epsilon=0$ and the background compression parameter (ignoring the wake) cancels for $\phi \gtrsim 3\phi_c$. We are dealing with eccentric rings here; however, the range over which the wake contribution to $\epsilon$ damps should be independent of this, so that the expression given in Eq.~\eqref{dissip1} should apply as well, as it relates only to the wake contribution (similarly $q$ refers only to the wake contribution to the streamline compression parameter in this expression).}, except for a factor of two [which arises because $\epsilon_0$ from Eq.~\eqref{epsimpulse} is damped twice along any given streamline, once for each shepherd satellite]:
\begin{eqnarray}
|\langle t_1(x)\rangle_{\phi}| & = & \frac{1}{2\pi}\left|\int_0^{2\pi}d\phi\ t_1\right|\nonumber\\
& = & \frac{3g^2}{4\pi q}\left(\frac{M_s}{M_p}\right)^2 \left(\frac{a}{x}\right)^4\sigma_0 n^2 a^2 \label{dissip4}
\end{eqnarray}
The energy transfer associated to the torque is $\Omega_s T_0 \simeq n T_0$ for each satellite. On the other hand, from the dissipation term of Eq.~\eqref{arphi}, Eq.~\eqref{dissip1} implies  that the dissipated part of the transfer is of the same order of magnitude as the total energy transfer:
\begin{equation}
2nT_0\sim \left(\frac{dE}{dt}\right)_{dissip}\sim n a \delta a |\langle t_1\rangle_{\phi}|.\label{dissip5}
\end{equation}
Now, $|\langle a_{r\theta}(x)\rangle_{\phi}| \sim |\langle a_{r\theta}(a_m)\rangle_{\phi}|$ for most of the ring due to the  local angular momentum balance constraint Eq.~\eqref{localtorque}. Then combining Eq.~\eqref{dissip5} with $L_H^{vis}(x)=2\pi a^2 |\langle a_{r\theta}(x)\rangle_{\phi}|$, Eq.~\eqref{gttorque} leads to [note the difference with Eq.~\eqref{dissip2}]
\begin{equation}
|\langle t_1(x)\rangle_{\phi}| \sim \left(\frac{ax}{\delta a}\right)^2 |\langle a_{r\theta}(x)\rangle_{\phi}| \gg  |\langle a_{r\theta}(x)\rangle_{\phi}|,\label{gtbalance}
\end{equation}
i.e. angular momentum flux reversal must hold throughout the ring for the shepherds to be able to enforce confinement in a regime of overlapping resonances. 

Note that, as a consequence, the torque balance is no longer a constraint on the satellite mass. Indeed, angular momentum luminosity reversal makes the magnitude of $L_H^{vis}(a_m)$ undetermined. Instead, the global torque balance translates into the required level of luminosity cancellation, as just pointed out. Either the ring adjusts so that this required level is achieved, or the ring is not confined. The satellite mass needed to achieve confinement is constrained in a different way, as discussed in section \ref{sec:satmass}. This is fundamentally different from the satellite mass constraint deduced from the analysis of \cite{GT79a}.

\subsection{Single-sided shepherding} \label{sec:singleshep}

An extreme situation of angular momentum flux reversal is reached when the viscous luminosity not only cancels, but is negative throughout a ringlet. In this case ($L_H^{vis} < 0$ everywhere), the ringlet spontaneously contracts due to the resulting effective negative viscosity. This possibility has been explored by \cite{GRS95}, prompted by the numerical results of \cite{HS94,HS95}. They  have pointed out that a single shepherd satellite can confine a narrow ring in such conditions. On the other hand, because of the resulting torque imbalance (with respect to a situation with a shepherd satellite on each side of the ring), such a ringlet will undergo fast radial migration, the time-scale being controlled by the rate of angular momentum flux between the ringlet and the satellite; see section \ref{sec:subtimes} and \cite{GRS95} for more details on these time scale issues. A more recent simulation \citep{LSLW11} explores the dynamics of the contraction in more detail; in particular, the role of the damping of the eccentricity of the wake created by the confining satellite, concentrated at the most highly compressed phases of the wake, is pointed out.

\subsection{Why is a sharp edge sharp}\label{sec:edgesharpness} 

In observed edges, the ring density is generally constant or nearly so right to the edge. This has not been explained yet. Angular momentum viscous luminosity reversal is a necessary but not sufficient condition for this. In fact, there does not appear to be any generic reason for this to be the case; rather this seems to emerge as a chance property of the stress tensor physics explored so far.

Somewhat surprisingly, no detailed analysis of this question has been published for sharp edges confined at an isolated resonance in the framework of the streamline formalism (the alternative analysis of \citealt{HSP09} is discussed in section \ref{sec:othersh}). However, in the context of wake shepherding, \cite{BGT89} have produced detailed solutions of the equivalent problem, as discussed above. They explicitly show that solutions with reasonably constant surface densities can be found, and they point out that a generic condition for this to be possible is satisfied by all the pressure tensor models they have explored [see their equation (65) and the accompanying discussion].

In order to check that the density remains more or less constant right to the edge in resonant shepherding, one would need to find both the complex amplitude $Z$ and surface density $\sigma_0$ profiles when a quasi-stationary state is obtained (ignoring the very long term diffusion of the whole ring), i.e., the full set of Eqs.~\eqref{completez} and \eqref{completea} must be solved [or equivalently, their semi-Eulerian counterparts in ($a,\varphi$) coordinates, which follow from Eq.~\eqref{dX2} of Appendix \ref{app:torque}]. This is sufficient in the dense limit, but for dilute rings, the velocity dispersion $c$ must be added as a variable to be determined, and consequently, the internal energy balance constraint Eq.~\eqref{dispvel} must be added to the problem\footnote{Or rather, its exact equivalent, though \cite{BGT83b} have shown from their numerical results that this relation fits very well the numerical data for the $\varepsilon(q,\tau)$ relation with $\beta = 0.3$. See \cite{SDLYC85}, however.}. However, this requires the knowledge of the $\varepsilon(c,\tau)$ relationship, which is still highly uncertain. Once again, the potential role of self-gravitational wakes on transport and dissipation is ignored, in the absence of relevant information on this issue in perturbed ring regions (see the discussion of section \ref{sec:sgw}).

To summarize, in light of these arguments (still partly to be checked in a detailed quantitative way), a sharp edge can be sharp as long as the macroscopic equations used here are valid, i.e., everywhere except within a distance to the edge comparable to the ring thickness at the edge $c/n$.

\subsection{Minimum satellite masses}\label{sec:satmass}

It is of some interest to quantify the minimum satellite mass that is needed to achieve confinement. The isolated resonance and wake response contexts are discussed in turn.

The minimum satellite mass follows from the requirement that the maximum satellite torque $T_M$ exceeds the minimum unperturbed viscous angular momentum luminosity: $T_M > L_H^{vis}$. Indeed, a maximum torque for a given mass is equivalent to assuming a minimum mass at given torque. In the same logic, the minimum satellite mass will be obtained by balancing the minimum viscous angular momentum luminosity with the maximum satellite torque. 

If self-gravitational wake contributions are negligible, the minimum luminosity is reached in the close-packing limit. From Eq.~\eqref{visflux} and the discussion of close-packing in section \ref{sec:compact}, the minimum unperturbed luminosity reads (see also \citealt{GP87})
\begin{equation}
L_{min}^{vis} = 2\pi n a^2a_{r\theta} = 3\pi\sigma_0\nu_m n a^2,\label{minflux}
\end{equation}
where $\nu_m = n (\sigma_0/\rho)^2$; see Eq.~\eqref{numin}. If self-gravitational wakes dominate the transport behavior of the ring [i.e., if the criterion Eq.~\eqref{sigc} is satisfied], the use of $\nu_{sg}$, Eq.~\eqref{nusg}, will give more realistic mass estimates in dense rings. However the discussion at the end of section \ref{sec:sgw} leaves the fate of wakes uncertain in perturbed regions. 

To summarize, the minimum viscosity will definitely yield the lowest possible satellite mass that is required to confine ring edges, but one should keep in mind that the resulting mass may be underestimated by about one order of magnitude if self-gravitational wakes are present.   

\subsubsection{Isolated resonance: broad ring}

Let us first look for the maximum torque in the case of isolated resonance shepherding. At the ring edge, the (Lindblad) resonant interaction creates an edge mode (see section \ref{trapped} on the trapped wave picture of ring modes); without dissipation, all ring streamlines would be aligned with the satellite-imposed pattern speed and no torque would arise. The torque exerted there results from the small apsidal shift due to viscous dissipation, somewhat like what occurs in a planetary tide. It is customary to estimate the maximum torque exerted on an edge from the torque that a density wave would produce if the ring edge were not present. This follows for two reasons. First, if the ring were not truncated at the edge, a density wave would indeed propagate, as discussed in the trapped wave model of edge modes \ref{trapped}. The second reason is the one discussed above in section \ref{sec:satamp} (eccentricity excitation by external satellites): the torque does not depend on the exact physical mechanism that actually takes place in response to satellite forcing, as long as it is linear\footnote{This condition is not necessarily valid, but linear torques are always larger than nonlinear ones \citep{BGT84,GP87} and the linear torque therefore provides an upper limit to the actual torque, consistently with the strategy adopted to obtain a minimum satellite mass.}. The main overall effect of the edge is therefore to limit the extent of the perturbation exerted by the satellite and the resulting torque below its value estimated through the theory of density waves. 

These arguments show that the maximum torque should be computed from the linear theory of density waves. The torque magnitude has been derived in a number of papers (see e.g.\ \citealt{GT78c,GT79c,SYL85}; a derivation adapted from \citealt{SYL85} in the framework of the streamline formalism can be found in \citealt{L92}). The leading order torques are Lindblad resonance torques that can be expressed as:
\begin{equation}\label{restorque}
T_s = \mp f e_s^{2|k|}\left(\frac{M_s}{M_p}\right)^2 \sigma_0 n^2 a_r^4.
\end{equation}
where $f$ is a dimensionless factor; the minus (plus) sign applies to an ILR (OLR). In the limit $m\gg 1$, $f=8.6 m^2$ for a first order resonance ($k=0$) and $f=3.0 m^4$ for a second order one ($|k|=1$) (see \citealt{GT80} and \citealt{GP87}; the  $|q_e| = 1$ eccentric resonance torques of the next section also give back these expressions by noting that $m = 2a/3|x_e|$, as eccentric resonances reduce to Lindblad ones in the limit of circular background motion). The ($m,k$) labels are those of the satellite perturbing strength, Eq.~\eqref{psimk}.

The resulting constraint on the satellite mass thus reads
\begin{equation}\label{msreson}
\frac{M_s}{M_p} \ge  e_s^{-|k|}\left(\frac{3\pi}{f}\right)^{1/2}\frac{\nu_m^{1/2}}{a_r n^{1/2}}.
\end{equation}
Corotation torques are discussed in the following section.

\subsubsection{Isolated resonance: eccentric narrow ring}\label{sec:narrow}

Torque expressions in this context have been derived in Appendix \ref{app:torque} and are evaluated in the relevant asymptotic regime $\Delta a \ll |x_e| = |a_s - a_e| \ll a$ with the help of the results of Appendix \ref{app:asymp} ($a_e$ is the ring edge semimajor axis). This context is very similar to the broad ring one of the preceding section, as leading order torques (eccentric $|q_e|=1,k=0$ and corotation $q_e=0, |k|=1$) are identical to the corresponding Lindblad and corotation torques in the limit of a circular ring; specific next-to-leading order torques arise due to the non-vanishing global eccentricity of the ring.

Torques are quoted for the dominant corotation and eccentric resonances (defined in section \ref{sec:eccres}) for forcing potentials of the form \eqref{distgt} (excluding $q_e+k=0$, which corresponds to $a = a_s$, i.e., satellites orbiting within the rings). Note that the results of Appendix \ref{app:torque} show that torques scale like $e_r^{2(|q_e|-1)}e_s^{2|k|}$ for eccentric torques and $e_s^{2|k|}e_r^{2|q_e|}$ for corotation torques ($e_r$ being the ring eccentricity). Corotation torques are always negative if the surface density gradient can be neglected, so that they are only effective for ring confinement at outer edges; they act to prevent confinement at inner edges. Conversely, if the density gradient contribution is dominant (a likely occurrence at a ring edge), a corotation torque always counteracts ring confinement, but one should keep in mind that a leading order corotation torque ($|k|=1, q_e=0$) is smaller than the related leading order eccentric torque ($|q_e|=1, k=0$) by a factor $e_s^2$. These torques read (taking into account that the corotation-like contribution to eccentric resonance torques is negligible):

\bigskip

\begin{description}
\item[$\mathbf{k=0, |q_e|=1}$] 
\begin{eqnarray}
T_E & = & \mathrm{sgn}(x) \frac{16F_1^2}{27} \left(\frac{M_s}{M_p}\right)^2\left|\frac{a}{x_e}\right|^2 \sigma_0 n^2 a^4,\nonumber\\
& \simeq & 3.76\ \mathrm{sgn}(x)\left(\frac{M_s}{M_p}\right)^2 \left|\frac{a}{x_e}\right|^2 \sigma_0 n^2 a^4. \label{Tk0q1}
\end{eqnarray}
\item[$\mathbf{k=0, |q_e|=2}$] 
\begin{eqnarray}
T_E & = & \mathrm{sgn}(x) \frac{16F_2^2}{243}\left(\frac{M_s}{M_p}\right)^2
\left|\frac{a}{x}\right|^4 \sigma_0 n^2 a^4 e_r^2\nonumber\\
& \simeq & 9.57\ \mathrm{sgn}(x_e) \left(\frac{M_s}{M_p}\right)^2
\left|\frac{a}{x_e}\right|^4 \sigma_0 n^2 a^4 e_r^2. \label{Tk0q2}
\end{eqnarray}
\item[$\mathbf{k=q_e=\pm 1}$] 
\begin{eqnarray}
T _E & = & \mathrm{sgn}(x) \frac{32F_2^2}{243}\left(\frac{M_s}{M_p}\right)^2
\left|\frac{a}{x_e}\right|^4 \sigma_0 n^2 a^4 e_s^2\nonumber\\
& \simeq & 19.13\ \mathrm{sgn}(x)\left(\frac{M_s}{M_p}\right)^2 \left|\frac{a}{x_e}\right|^4 \sigma_0 n^2 a^4 e_s^2.
 \label{Tk1q1}
\end{eqnarray}
\item[$\mathbf{|k|=1, q_e=0}$] 
\begin{eqnarray}
T_C & = & -\frac{16F_1^2}{81\pi} \left(\frac{M_s}{M_p}\right)^2\left|\frac{a}{x_e}\right|^3 \sigma_0 n^2 a^4 e_s^2 \times \left(1 + \frac{2a}{3\sigma_0} \frac{d\sigma_0}{da}\right),\nonumber\\
& \simeq &  - 0.40\ \left(\frac{M_s}{M_p}\right)^2 \left|\frac{a}{x_e}\right|^3  \sigma_0 n^2 a^4 e_s^2 \times \left(1 + \frac{2a}{3\sigma_0} \frac{d\sigma_0}{da}\right).
\label{Tk1q0}
\end{eqnarray}
\end{description}
In these relations, the subscripts $E,C$ refer to eccentric and corotation resonances, respectively, $F_1$ and $F_2$ are dimensionless numerical factors specified in Eqs.~\eqref{F1} and \eqref{F2}, and $|x_e|$ is the distance from the satellite to the edge.

We are interested in the smallest possible satellite mass; this will be provided by the largest satellite torque at given mass, i.e., the $k=0, |q_e|=1$ eccentric torque. The minimum satellite mass follows immediately from Eq.~\eqref{Tk0q1}:
\begin{equation}\label{mseccres}
\frac{M_s}{M_p} \ge \left(\frac{81\pi}{16 F_1^2}\right)^{1/2} \left|\frac{x_e}{a}\right|\frac{\nu_m^{1/2}}{a_r n^{1/2}},
\end{equation} 
and is identical to Eq.~\eqref{msreson} for a first order Lindblad resonance. The corresponding expressions for other resonances are readily obtained from the relevant torques.

\subsubsection{Overlapping resonances: broad ring edge}

Let us consider wake confinement next. This is relevant for a broad ring and for the most well-known gaps. For a broad ring, the wake pattern is by necessity of negligible amplitude far from the edge.  The total torque is then obtained by integration of Eq.~\eqref{torquedens} from the edge ($x_e$) to infinity to read
\begin{equation}\label{edgetorque}
T_s = \mp \frac{g^2}{6}\left(\frac{M_s}{M_p}\right)^2 \left|\frac{a}{x_e}\right|^3\sigma_0 n^2 a^4+ \mathcal{O}(e_s^2),
\end{equation}
from which the related minimum satellite mass is readily obtained
\begin{equation}\label{msimpulse}
\frac{M_s}{M_p} \ge \left(\frac{18\pi}{g^2}\right)^{1/2} \left|\frac{x_e}{a}\right|^{3/2}\frac{\nu_m^{1/2}}{a_r n^{1/2}},
\end{equation}
where $x_e$ is the value of $x$ at the edge. Note that for $m\gg 1$, for $k=0$, the minimum mass scales as $|x_e|/a$ in the isolated resonance confinement [cf Eq.~\eqref{mseccres}] instead of $(|x_e|/a)^{3/2}$ here.

\subsubsection{Overlapping resonances: narrow ring}\label{sec:shepnarrow}

The last question to be addressed here concerns the mass of shepherd satellites confining a narrow ring in the resonance overlap regime (i.e., when wake responses are present throughout the ring). The constraint of angular momentum luminosity cancellation, Eq.~\eqref{gtbalance}, sets the magnitude of the satellite masses, instead of the global torque balance.

Consider first a hypothetical situation with a single satellite. The streamline-averaged value of $a_{r\phi}$ is the sum of two contributions, as discussed in section \ref{sec:wakeconfnarrow}: a large (negative) value $\sim 3a^2 a_{r\theta}^*\phi_c$ in the range $0\le \phi \lesssim 3\phi_c$ and an unperturbed (positive) contribution through $3\phi_c \lesssim \phi \le 2\pi$. These two contributions nearly cancel; starred quantities are characteristic values in the wake region. The first contribution can be obtained from Eq.~\eqref{dissip1} by noting that $a_{r\theta}^*\sim t_1^*$. More precisely, from Eq.~(60) of \cite{BGT89}, one has
\begin{equation}
a_{r\theta}^* = \frac{2q^* t_1^*}{9}.\label{arphi*}
\end{equation}
This relation is relevant in the large $q$ regime for the generic stress tensor models of \cite{BGT86,BGT89}; this relation is also needed for confinement of the edge at nearly constant surface density \citep{BGT89}. As most of the wake damping occurs at constant $q$ in this limit, one expects this relation to hold for the whole damping range.

In the more relevant case of two satellites, the same reasoning follows, except that there are two wake-like perturbed zones in azimuth instead of one. Thus, from Eq.~\eqref{dissip4}, the flux contribution of the wake region ($0\le \phi \lesssim 3\phi_c$) reads:
\begin{eqnarray}
F^* & \simeq & - 2a^2\int_0^{3\phi_c} \frac{2q}{9} t_1 d\phi \simeq 2a^2\int_0^{2\pi} \frac{2q}{9} t_1 d\phi\nonumber\\
& = & \frac{2g^2}{3}\left(\frac{M_s}{M_p}\right)^2 \left(\frac{a}{x}\right)^4 \sigma_0 n^2 a^4.\label{wakeflux}
\end{eqnarray}
The second contribution ($3\phi_c \lesssim \phi \le 2\pi$) is given by Eqs.~\eqref{minflux} in the limit $\phi_c\ll 2\pi$. Then, balancing $F^*$ with $L_{min}^{vis}$ leads to
\begin{equation}
\frac{M_s}{M_p} \gtrsim \left(\frac{9\pi}{2g^2}\right)^{1/2} \left(\frac{x}{a}\right)^{2}\frac{\nu_m^{1/2}}{a_r n^{1/2}}.\label{msnarrow}
\end{equation}
This mass constraint is the least demanding of all, due to the $(x/a)^2$ scaling [compare with Eqs.~\eqref{mseccres} and \eqref{msimpulse}].

This analysis has a side benefit: from Eq.~\eqref{wakeflux}, \eqref{torqueshep} and \eqref{gttorque}, one can relate the actual mid-ring viscous angular momentum luminosity $L_H^{vis}(a_m)$ to the unperturbed one $L_{unpert}^{vis}\simeq |F^*|$:
\begin{equation}
L_H^{vis}(a_m) \simeq \frac{3}{2} \frac{\delta a^2}{ax} L_{unpert}^{vis}.\label{fluxreduc}
\end{equation}
This shows that the viscous angular momentum flux reversal constraint must be quite substantial in narrow rings for shepherding to take place when resonances are overlapping throughout the ringlet radial extent.

\subsubsection{Implication}

With these mass limits, it appears that most sharp edges and ringlets in the Saturnian system cannot be confined by satellites once the observational limits imposed by the Cassini data are taken into account (see the chapter by Nicholson \textit{et al.} for details). Some other process must be at work for these objects. Alternative suggestions are briefly discussed in the concluding section.

\subsection{Other studies of sharp edge shepherding}\label{sec:othersh}
 
A very detailed investigation of sharp edges at an isolated resonance has been performed by \cite{HSP09}; this analysis is tailored for the B-ring edge, but the framework is general. The authors have developed a fluid formalism similar to the approaches adopted in, e.g., \cite{SYL85,SDLYC85} or \cite{PM05}. The stress tensor is modeled in the hydrodynamic limit, with constant kinematic and bulk viscosities. They find that the near-alignment of the ring streamlines with Mimas reduces the satellite torque so much as to make confinement impossible, unless the bulk viscosity exceeds the kinematic viscosity by a very large factor, $\gtrsim 10^4$. 

This unrealistic bulk-to-kinematic viscosity ratio follows from the fact that angular momentum luminosity reversal is never reached in this model. This contradicts the analysis of section \ref{sec:resconf} (shepherding at isolated resonances) which implies that viscous angular momentum luminosity reversal must occur in a substantial part of the perturbed region. Note that this is a necessary, model-independent constraint on the ring rheology for confinement to be possible. Consequently, the constant viscosity assumption must be abandoned to the benefit of more sophisticated stress tensor models, at least for the analysis of this specific process.

\cite{SPSS09} have also investigated the problem of the B-ring edge with a simplified $N$-body dynamical model. They find that the confinement problem disappears due to the phase-locking of all particles with Mimas forcing, treated in the impulse approximation; collisions are reduced to the point of disappearance due to this phase-locking, and with them collision-induced spreading. This, however, is clearly an artifact of their very simplistic collisional dynamics. No amount of phase synchronization can suppress the velocity dispersion and ensuing collisional spreading continuously driven by the ring particles' mutual gravitational stirring if nothing else.

Some papers have also been devoted to the theoretical analysis of a collisionless edge dynamics, e.g., \cite{Ste91} and \cite{SSS10}. A moonlet can open a gap by collisionless gravitational interaction alone, but through a different process as the one studied here, i.e., through clearing of the chaotic zone in the moonlet vicinity \citep{W80,DQT89}; the resulting gap width scales as $(M_m/M_p)^{2/7}$ according to the famous result of \cite{W80} ($M_m$ is the moonlet mass). In this regime, the collisionless pressure tensor limits the nonlinear compression parameter $q$ due to phase-mixing \citep{Ste91}. Although the model is able to given some interesting insight into the role played by the moonlet eccentricity on the wake shape \citep{SSS10}, the relevance of the collisionless regime is most likely very limited  due to the collision frequency in ring regions of interest here.

\section{Theory and observations: a miscellany of issues}\label{sec:misc}

Most of the physics discussed in this review has been quantified within the framework of the streamline formalism, although incursions in other types of  fluid or $N$-body dynamical analyses have been made where and when relevant. In any case, other fluid analyses of ring dynamics do share with the streamline formalism a number of distinctive characteristics, most of them being rooted in or connected to the work of GT/BGT. Before turning to a discussion of unsolved issues, it might therefore be useful to point out the strengths and weaknesses of this framework.

\subsection{Formalism: strengths and weaknesses}\label{sec:form}

Probably the most powerful aspect of the formalism stems from its ability to deal with a large range of time scales, from the orbital one, which also characterizes the relaxation time of the stress tensor (typically a few hours) to the large scale mass and angular momentum redistribution one, relating to the viscous component of the stress tensor (up to hundreds or thousand of years). Self-gravity and the synodic period associated to close satellites define the most important intermediate time scales (from one year to tens of years). Dealing with such a wide range of time-scales is made possible by taking advantage of their well-defined hierarchy and by effectively reducing the dynamics to one spatial dimension. 

This is definitely an advantage over either fluid or $N$-body brute force simulation, where the time-span of the simulation is set by the smallest time-scale in the problem. Note however that by solving 2D instead of 3D problems, modern numerical means may give access to time-spans comparable to the intermediate time-scale at reasonable spatial resolution in such approaches. For the same reason, present-day numerics would give us access to an unprecedented level of details in resolving the streamline formalism dynamical equations for a wide range of problems, a step that has not yet been taken but is increasingly needed.

The limitations of the formalism are directly rooted in what makes it so powerful. Up to know, little effort has been devoted to formalize multi-mode motions, whereas these are ubiquitous in ring dynamics: for example, the shepherding of narrow eccentric rings at isolated resonances or the driving of their eccentricities involve several modes; similarly, nearly all ring edges host quite a number of different modes. In order to investigate the related physics within the streamline formalism, a serious formalization effort is required to encompass multi-mode motion. This would still maintain the numerical demand within reason, unless the number of modes treated becomes comparable to the number of streamlines, in which case the model becomes effectively 2D. Conversely, 2D numerical simulations are naturally equipped to deal with such problems.

On a related front, specific yet-to-come formalizations are needed to capture the effects on the large scale dynamics of small-scale, non-axisymmetric, traveling or transient wave motions, such as small scale-viscous overstabilities and self-gravitational wakes. A similar remark holds for the potential role of embedded objects that are too small to create gaps. 

\subsection{Physics: needs and shortcomings}\label{sec:phys}

On the physics side, it is fair to say that the results and insights obtained so far are more proofs of concept than actual detailed analyses of the most salient problems. This in itself is quite an achievement, considering the subtlety of the physics involved, an inherent characteristic of weak dynamical effects. But in the post-Voyager and soon post-Cassini era, more is needed to account for the wealth of data collected by these missions. 

We are facing two different kinds of challenges:
\begin{itemize}
\item The sophistication of existing analyses must be increased on issues where the basic physics seems to be captured, if only to characterize the extent to which the problem is truly understood.
\item New physical ideas and detailed theoretical analyses must be put forward where present ideas appear to fail.
\end{itemize}

\medskip

These challenges are outlined for the various issues that have been discussed in this review.

\subsubsection{Rheology}\label{sec:rheo}

The ring rheology remains the weak point in the fundamental physical characterization of the dynamics, prior to tackling any specific large scale process\footnote{This point makes planetary rings distinctively different from most other disk systems. In fluid disks, for example, turbulence and turbulent transport are often the most difficult to characterize. Planetary rings cannot develop three-dimensional turbulence, as the granularity scale is comparable to the scale-height. And their collisional dynamics is quite specific, even in comparison with closer analogs such as planetesimal disks or debris disks. See the chapter by Latter \textit{et al.} for more details on these issues.}. It was pointed out in section \ref{sec:press} (stress tensor) that the role of small scale unstable structures (self-gravitational wakes) on the stress tensor in perturbed regions is not yet understood; this is a strong limitation considering the  transport amplification such physics produces in unperturbed regions, in light of the key role played by the stress tensor in some of the outstanding issues discussed in this concluding section. Also, if the velocity dispersion amplification discussed in section \ref{sec:resconf} does indeed prevent wake formation, this might nelp explaining the discrepancy in the results on the ring surface density obtained by \cite{HN16} and \cite{RSLCS10}, although other issues might be at work here (e.g., the role of a size distribution in the analysis of gravitational wakes).

In principle, two strategies can be adopted in ring rheology theoretical studies: \textit{ab initio} analyzes of the stress tensor behavior, and constraints on the rheology through \textit{ad hoc} models of the stress tensor incorporating all known generic constraints, such as the model proposed by \cite{BGT86}, coupled to more or less sophisticated heuristic rules making direct contact with the model parameters and the ring particles collisional properties. Both strategies have been used in the past; quite possibly, the next generation of detailed models of the problems listed below will bring further constraining information on the rheology of ring systems. At present, only very basic constraints have been provided in this way, e.g., estimates of the ring velocity dispersion from density wave damping lengths, based on heuristic expressions for the ring viscosity [e.g., Eq.~\eqref{kinvis}].

\subsubsection{Confinement}\label{sec:conf}

This problem is probably the most critical large-scale dynamical issue, both in terms of the lack of detailed analyses and the need of new theoretical ideas. The exposition of section \ref{sec:shep} (shepherding) shows that the contexts in which the confinement mechanism is understood or at least promisingly identified are the exception rather than the rule. 

In particular, the shepherding of narrow rings by external satellites is convincingly argued in only one case, the $\epsilon$ ring of Uranus; one may positively argue in the same direction for the F ring of Saturn, although the issue here is much more complex (see chapter by Murray et al.).

Similarly, for gap edges, the Encke and Keeler gaps are the only known instances where a satellite orbiting inside and confining the gap has been conclusively identified. The outer edges of the A and B ring have also long been associated with resonances with external satellites, Janus and Mimas.

\paragraph{Limitations of available analyses.}

Even when sharp outer edges are convincingly associated with known satellites, problems remain in the detailed application of the available theories. To this date, the most detailed theoretical analysis of edge shepherding are those of \cite{BGT89} (close satellite) and \cite{HSP09} (isolated resonance). The first one assumes that satellite wakes are totally damped between two encounters, which is not the case in the Encke gap. The relevance of this point for the confinement dynamics is unclear. Theoretical extensions to multiple wake excitation before damping are possible but have not been produced so far. The question of the confinement of sharp edges at isolated resonances is more critical. The analysis of \cite{HSP09} has been discussed in section \ref{sec:othersh}, where we argue that its conclusions are model-dependent as this analysis does not produce the necessary angular momentum flux reversal for confinement to be achieved. Detailed analyses based on a more complex ring rheology still need to be produced before any conclusion can be drawn for the A and B ring edges.

The most critical problem, though, is the absence of shepherd satellites for most of the known gaps and ringlets in the Saturnian system. The minimum satellite masses given in section \ref{sec:satmass} are much larger than what is allowed by the Cassini constraints on small satellite sizes (see chapter by Philip Nicholson, Richard French and Joseph Spitale), even taking into account the required angular momentum flux reversal throughout narrow ringlets confined by close satellites. Whether this is also a problem for the Uranian rings remains to be seen, as the observational constraint is substantially less tight in this system; one might feel somewhat uncomfortable, though, to rely on a host of as yet unseen moons to achieve confinement in the Uranian system. For a while (prior to the Voyager encounter with Uranus), it was believed that the distribution of narrow rings in the Uranian system might be explained by resonances with satellites (the most extensive version of this idea has been investigated by \citealt{FG99}); however, this proposal never gained wide acceptance after the encounter due to the lack of relevant resonances for a majority of these rings (e.g., \citealt{PG87}).

\paragraph{An unexplored process: mutual edge confinement.}

Although in a large number of instances, the source of confinement has not yet been identified, the physical constraints discussed in section \ref{sec:shep} (shepherding), in particular torque balance and angular momentum flux reversal, must apply whatever the mechanism at work. One possible such mechanism that has apparently never been discussed in the ring literature is the mutual confinement of the two gap edges, or equivalently, mutual confinement between a gap and an embedded ringlet's edges\footnote{Apparently, a similar process has been studied by \cite{SK91} in the galactic dynamics context; I am indebted to an anonymous referee for pointing this out to me.}. 

Indeed, any non-circular potential of large enough magnitude may act as a source or sink of gravitational torque, under the right conditions. In particular, the same azimuthal wavenumber $m$ disturbances with the same pattern speed must be present at both edges for mutual edge confinement to arise; this is required for secular torques (and angular momentum transfer) to exist. Also, there must be enough mass in these non-axisymmetric features for the maximum possible torque to be large enough to maintain the gap open, while ring dissipation would produce the apsidal shift that would control the actual magnitude of the torque. 

The required direction of angular momentum exchanges brings some interesting constraints. If the mass involved is small, the expected mode amplitudes and couplings will be large only if both edges are resonantly forced. Furthermore, the sign of the torque will be related to the sign of the apsidal shifts of the two edge modes; this sign is deduced by standard tidal reasoning, where the lag in time is always positive and the resulting relative phase lag (opposite of the apsidal shift) is positive or negative depending on the relative rotation frequencies of the two bodies. In the present case, the resonances between the two edge modes must be Lindblad resonances, so that the pattern speed should be $\Omega_p=(m\pm 1)n/m$. Quite clearly, the phase lag of the inner (outer) edge will be positive (negative) since $n > \Omega_p$ ($n < \Omega_p$), and the angular momentum will then flow from the inner to the outer edge of the gap, as required. In other words, the inner edge is an ILR and the outer one an OLR with the same pattern speed and wavenumber $m$. This constraint implies $m \simeq 2 n/\delta n = 4a/3\delta a \gg 1$, where $\delta a$ is the gap width. Peter Goldreich (private communication) has however pointed out that the fine tuning required by this process makes it unlikely.

Alternatively, an external resonance at the inner edge with an external satellite might provide a large enough amplitude at both edges even though the outer edge amplitude would be substantially smaller than the inner one; furthermore, the relative lag should have the correct sign if the physical conditions at both edges are similar enough, simply because fluid particles revolve faster at the inner edge than at the outer one, although both lags would have the same sign in this case. This might possibly explain the confinement of the Cassini division gap edges from the Mimas 2:1 forcing, assuming that the gaps themselves are related to consecutive resonances with a low frequency arising from the B ring outer edge dynamics, which would produce some of the related angular momentum transfer (see right below).

\subsubsection{Gaps and narrow rings characteristics}\label{sec:gaps}

\paragraph{Cassini Division gaps.}

The preceding suggestion leaves open the question of the origin of the gaps and ringlets even if they are able confine one another. A somewhat related issue concerns the distribution of gap locations and widths in Saturn's rings. In particular, the apparent equal spacing of the series of gaps observed in the Cassini division is intriguing, especially in light of the existence of a low frequency at the B ring edge. \cite{HMBBS10} have analyzed the possibility that low frequencies at the B ring edge might produce the observed gap structure, but this investigation was not totally conclusive (the beating of the forced and free $m=2$ modes, whose magnitude is in the required range to produce resonances of the appropriate spacing falls off the mark). 

However, it is probably worth revisiting this problem. Note in this respect that two physical processes are potentially responsible for the existence of low frequencies in edge modes and ringlet modes: viscous overstabilities (sections \ref{sec:2str} on the two-streamline model and \ref{sec:overvis} on the driving of ring mode amplitudes by a viscous overstability) and the beating between edge modes of different number of radial nodes (section \ref{trapped} on the trapped wave picture of ring modes). It is unclear that all relevant combinations of high and low frequency modes have been explored so far, and a more detailed modeling of edge modes and of their possible overstable libration motions is probably required on the theoretical front to make progress in a systematic manner on this question.

\paragraph{Narrow ring eccentricities and inclinations.}

The origin of narrow ring eccentricities might still be problematic, either in a quantitative or a qualitative way. Forcing by satellites has been explored in landmark papers, but there is still some uncharted territory in this specific mechanism from a fluid point of view (see section \ref{sec:satamp}). Another intriguing issue is the peculiar trend observed in the Uranian rings' eccentricities and inclinations, which (ignoring the $\epsilon$ ring), tend to decrease outwards. A similar question concerns the origin of the inclinations of narrow rings; more generally, much less attention has been paid to the questions posed by ring inclinations than ring eccentricities, but it would be surprising if some of the issues in eccentric mode dynamics were not mirrored in their inclined counterparts.

\paragraph{Multiple mode excitation.}

The most unexpected feature, maybe, concerns the large number of different modes observed at gap and narrow ring edges. The vast majority of these modes does not correspond to any clearly identified external forcing. In the absence of such forcing, only two options remain: either these modes are transient features (but one still needs to identify their dynamical excitation mechanism in the first place) or they are internally driven. Viscous overstabilities are an obvious source of internal driving, and the problem for an edge mode is somewhat different from a narrow ring global eccentricity. Indeed, in this latter case, a narrow ring global mode with initially very small amplitude cannot reach a large ($e \gg\delta e$) finite amplitude through this process, as viscous overstabilities may \textit{maintain} existing eccentricities, but not \textit{amplify} them below some amplitude (section \ref{sec:overvis}). However, the boundary conditions of an edge mode combined with the intrinsic nonlinearity of the problem make the mean and librating components comparable in amplitude in this context; modes may then be both created and maintained at an edge by viscous overstabilities, but with possibly rather complex characteristics, especially once the confinement dynamics is taken into account. 

As to which specific azimuthal wavenumbers $m$ are excited, either viscous overstable modes trace back to initial conditions, or a combination of transient excitation and viscous overstability explains the observed mode amplitudes if they are not simply transient features. Note finally that mode coupling may also create new modes, but this also requires the seeding modes to be present in the first place; for example, assuming that many different modes are produced by a transient excitation (e.g., a small object crossing the ring plane), mode coupling might help to understand which modes might survive or decay, and under which physical conditions.

It seems fair to say at this point that we have no reliable explanation for the presence of the observed modes, neither in type (azimuthal wavenumber) nor in amplitude, although a viscous overstability naturally tends to saturate amplitudes\footnote{In this respect, the low $q$ value of the B ring edge $m=1$ mode is surprising.} at $q \lesssim 1$. At the very least, the mode coupling problem needs to be formalized beyond existing attempts as a start on these questions. There is a vast literature on mode coupling in other fields, in particular in relation with weak (or wave) turbulence, a problem that has been extensively studied in the 1970s, mostly by Russian authors (see, e.g., \citealt{SUZ88} and references therein). The usual resonant coupling condition ($m_1+m_2+m_3=0$ with appropriate signs for the azimuthal wavenumbers) underlies the resonant energy exchanges that define the resonantly driven mode amplitudes. In a rotating medium, another condition bearing on angular momentum exchanges needs to be met as well\footnote{All couplings produced by the usual condition do not result in secular angular momentum exchanges between modes; in this case, the modes created will not be truly secular modes but their amplitudes will slowly oscillate in time.}; this second condition makes mode coupling at a given edge unlikely (Peter Goldreich, private communication), although it can generate mode coupling between two different narrow ring edges through the evanescent region inside the ring\footnote{A similar process has been studied in the physics of accretion tori \citep{GGN86}.}, or at two different gap edges (the suggestion for gaps has been discussed in section \ref{sec:gaps} above). The existence of an evanescent region makes the coupling of edge modes quite specific compared to the coupling of Fourier components in homogeneous media with periodic boundary conditions (the usual setting for most wave turbulence studies).

%
\paragraph{Anomalous apsidal shifts.}

Finally, some narrow ring modes seem to display large apsidal shifts. This is inconsistent with our basic understanding of these structures, unless a librating or circulating pattern is driven by a viscous overstability (see section \ref{sec:2str} on the two-streamline model of ring modes). More observational work as well as more detailed modeling of edge and ring modes under dynamically sustained librating or circulating conditions are required to prove or disprove such a possibility.

\subsubsection{Rigid precession}\label{sec:prec}

The last outstanding issue is also one of the oldest. It is now known the initial proposal of \cite{GT79b} whereby the ring self-gravity alone acts in a narrow ring to explain its rigid precession is incorrect in a number of instances, if not always (see section \ref{ssg}). The pressure-modified self-gravity model (section \ref{sec:cg}) is more promising; in particular, the qualitative shape of the surface density profile is more reminiscent of actual observed profiles. Detailed models are again needed to ascertain this possibility. 

The problem is more daunting than it seems, however, as such theoretical analyses require an explicit treatment of the confinement of the edge in a globally eccentric background. This in turn requires a comprehensive theory including all the processes that have been studied independently up to now: mode coupling, edge confinement, eccentricity excitation or damping by the satellites shepherding or by a viscous overstability, and rigid precession. A notable first step in this direction was taken by \cite{PM05}, but this still leaves quite a few questions unanswered. Also, such models need to be produced not only in a generic way, but for specific rings.

\subsection{Ring arcs}\label{sec:arcs}

Ring arcs are slightly off-topic with respect to the main theme of this review. They are discussed in detail in the chapters by de~Pater et al. and Hedman et al. elsewhere in this volume, and only a few remarks bearing on the theory of these systems will be made here.

Rings arcs are similar to narrow rings except that the ring material has only a limited azimuthal extent. The Neptunian ring arcs are thought to be constituted by material trapped in libration motion in corotation resonance sites, in order to prevent the fast azimuthal spreading that would otherwise be induced by differential rotation. Interparticle collisions will make ring arc material lose energy and therefore eventually leave the corotation resonance sites, which are maxima of the effective gravitational potential they orbit in. Thus it was very quickly realized that one not only needed (a) satellite(s) to produce the required corotation sites, but also the required Lindblad resonance that would inject orbital energy in the arcs and counteract the effect of collisional dissipation. The first such model was proposed by \cite{L85} and involved two satellites, one providing the corotation resonance and a second one providing the Lindblad one. An alternative theory was later proposed by BGT \citep{GTB86}, whereby a single satellite on an inclined orbit can provide both resonances at once, making the joint occurrence of both resonances less coincidental. This second theory makes use of a clever adaptation of the streamline formalism to model libration motions inside the corotation resonance separatrix.

The satellite Galatea is the most obvious candidate for such a process in the Neptunian ring arc context, as one of its relevant corotation resonances and associated Lindblad resonances falls in the immediate vicinity of the ring arcs. The model of \cite{NP02}, based on a revised arc mean motion, favors the 43:42 corotation eccentric resonance with Galatea, which is made to coincide with the ring arcs location by taking into account the effect of the ring arc mass on the precession rate of Galatea. This model specifies the mass of the Fraternit\'e arc as a function of Galatea's mass and orbital eccentricity.

This model has an unwanted feature that does not seem to have been pointed out yet in the literature: the required arc mass is so large as to make neighboring corotation sites unstable, due to the depth and steepness of the potential well produced by Fraternit\'e; in fact, the corotation sites located as far as $\sim 70\degree$ away from Fraternit\'e\footnote{Calculation by the author.} are no longer potential maxima. This drawback 
is in line with the very dynamic nature of the other arcs surrounding Fraternit\'e \citep{dP05}, discovered after the inception of Namouni and Porco's model. Some more detailed theoretical analysis is needed, however, to characterize the dynamics in a context where most of Galatea's corotation sites are no longer permanent traps for ring arc material, and to confirm the viability of the model of \cite{NP02} with respect to all available observational constraints.

\subsection{Concluding words: dynamics and history}\label{sec:conc}



An intriguing perspective has emerged in the course of this review: in a number of problems, initial conditions may not be irrelevant, contrary to what was widely believed for awhile (or maybe just hoped). For example, specific initial conditions are required for viscous overstabilities to explain observed ring eccentricities; similarly, the pressure-modified self-gravitational process for ring rigid precession most probably requires specific initial conditions to account for the systematic positive sign of narrow ring eccentricity gradients. Opening such a window on the rings' history, and not only through the large scale diffusion dynamics, may turn out to be a particularly rewarding enterprise, as rings appear to be more dynamic objects than expected at first.

\section*{Acknowledgments}
\label{sec:ack}
\addcontentsline{toc}{section}{\nameref{sec:ack}}
\markboth{Acknowledgments}{}

I wish to thank Peter Goldreich and Scott Tremaine for insightful comments on some of the issues discussed in this work. I am indebted to Henrik Latter, Phil Nicholson and the referee for their careful reading and numerous suggestions that helped improve the content of this review. It goes without saying that I take responsibility for any error that may still be present in this material.

\clearpage

\section*{List of symbols}
\label{sec:symb}
\addcontentsline{toc}{section}{\nameref{sec:symb}}
\markboth{List of Symbols}{}

\begin{center}

\setlength{\tabcolsep}{0pt}
\setlength\LTleft{-1.7cm}
 
\begin{ThreePartTable}
\label{tab:symb}
 
\begin{TableNotes}
  \footnotesize
  \item[a] The meaning (elliptic or epyciclic) should be clear from the context.
  \item[b] The planet contribution is not included.
  \item[c] Inner Lindblad Resonance.
  \item[d] Measured across half the ringlet width.
\end{TableNotes}

\begin{longtable}{lll}
\toprule
Symbol && Definition\\
\midrule

$a$ && Elliptic semimajor axis\\
$a_e,a$ && Epicyclic semimajor axis\tnote{a}\\
$a_i,a_j$ && Streamlines $i,j$ semimajor axes\\
$a_i, a_o$ && Inner limit of an ILR\tnote{c}\hspace{1mm} edge perturbed region, edge outer limit\\
$a$ && Ringlet mean semimajor axis\\
$a_r$ && Resonance radius\\
$a_s$ && Satellite semimajor axis\\
$a_{r\theta},a^i_{r\theta}$ && Azimuthal average of the $P_{r\theta}$ pressure tensor component\\
$B(q)$ && Nonlinear correction to density-waves self-gravitational angular moment luminosity\\
$c$ && Velocity dispersion\\
$c_b$ && Velocity dispersion at ring edge\\
$c^*$ && Effective ring sound speed in close-packing configuration\\
$C(q)$ && Nonlinear correction to the density-waves dispersion relation\\
$d$ && Ring particle size\\
$d/dt$,$D/Dt$ && Lagrangian derivative\\
$\delta a, \delta\epsilon, \delta m\Delta$ && Variation across\tnote{d}\hspace{1mm} a narrow ring of $a$, $\epsilon$, $m\Delta$\\
$\delta^{\pm}(a_i)$ && Width of streamline $i$\\
$\Delta$ && Apsidal shift for $m=0$ modes\\
$\Delta_g$ && Gap width\\
$\Delta a$ && Total ring(let) width\\
$\Delta a_{ij}$ && Streamlines ($i,j$) semimajor axis difference $a_i - a_j$\\ 
$\Delta^{\pm}(X)$ && Variation of $X$ across a streamline\\
$e$ && Elliptic eccentricity\\
$e_{r}$ && Mean (mass average) ring elliptic eccentricity\\
$e_s$ && Satellite elliptic eccentricity\\
$\bm{e}_\alpha,\bm{e}_\beta$ && Unit orthonormal vectors\\
$\epsilon,\epsilon_i$ && Epicyclic eccentricity\\
$\bar{\epsilon}$ && Mean (mass average) ring epicyclic eccentricity\\
$\epsilon^*$ && Effective epicyclic eccentricity of trapped edge modes\\
$\epsilon_0$ && Satellite-induced epicyclic eccentricity jump (impulse approximation)\\
$\varepsilon_c$ && Ring particle normal coefficient of restitution\\
$E, E_i$ && Specific energy\\
$F$ && Self-gravitational enhancement factor of the vertical epicyclic frequency\\
$f,F_1,F_2$ && Dimensionless factors characterizing satellite torques\\
$FF$ && Ring filling factor\\
$g$ && Satellite torque dimensionless factor (impulse approximation)\\
$G$ && Gravitational constant\\
$G(q)$ && Dimensionless velocity shear\\
$\bm{g}_{sg}, g_r, g_\theta$ && Self-gravitational acceleration\\
$\gamma$ && See $q\cos\gamma,q\sin\gamma$\\
$\Gamma$ && Planet to shepherd satellites ratio of contributions to the ring differential precession\\
$H, H_i$ && Specific angular momentum\\
$H_0$ && Unperturbed ring thickness\\
$H(q^2)$ && Dimensionless factor characterizing streamline self-gravitational interactions\\
$\eta,\eta_{nl}$ && Dynamic viscosity (non-local)\\
$l$ && Ring particle mean free path\\
$\mi$ && $\sqrt{-1}$\\
$i,j$ && Streamline indices (discrete streamline formulation)\\
$I$ && Radial action\\
$j_1,j2,j_3,j_4$ && Integers characterizing the satellite disturbing function\\
$J$ && Jacobian of the change of variable $r,\theta\rightarrow a_e,\varphi_e$\\
$J_2$ && First gravitational harmonic coefficient of the planet potential\\
$k$ && Density-wave radial wavenumber\\
$k$ && Number of trapped edge-modes radial nodes\\
$k$ && Integer characterizing the satellite disturbing function\\
$\kappa$ && Epicyclic (horizontal) frequency\\
$\kappa_s$ && Satellite epicyclic frequency\\
$l$ && Particle mean free path\\
$L$ && Satellite wake azimuthal wavelength\\
$l$ && Integer characterizing the satellite disturbing function\\
$L_E^{sg},L_H^{sg}$ && Self-gravitational energy and angular momentum luminosities\\
$L_E^{vis},L_H^{vis}$ && Stress tensor energy and angular momentum luminosities\\
$\lambda$ && Elliptic mean longitude\\
$\lambda_0$ && Elliptic mean longitude at epoch\\
$\lambda_s$ && Satellite elliptic mean longitude\\
$\lambda$ && Width of ring edges (surface density drop to zero)\\
$\lambda, \lambda_c$ && Length of self-gravitational instability\\
$\lambda_i$ && Streamline $i$ linear mass density\\
$\lambda_1,\lambda_2$ && Effective viscous-like and pressure-like frequencies\\
$m$ && Azimuthal wavenumber, number of lobes\\
$m$ && integer characterizing the satellite disturbing function\\
$m_i$ && Mass of streamline $i$\\
$m_p$ && Ring particle mass\\
$m\Delta$ && Streamline apsidal shift\\
$m\Delta_0$ && Streamline apsidal shift at $t=0$\\
$M_e,M$ && Epicyclic mean anomaly\tnote{a}\\
$M_p$ && Mass of the planet\\
$M_r$ && Ring(let) mass\\
$M_s$ && Satellite mass\\
$n, n_i$ && Ring mean motion\\
$n_s$ && Satellite mean motion\\
$N$ && Number of streamlines\\
$\nu,\nu_l$ && Kinematic viscosity (local, non-local)\\
$\nu_m$ && Ring minimum viscosity (close-packing configuration)\\
$\nu_{sg}$ && Effective kinematic viscosity due to self-gravitational wakes\\
$\omega_c$ && Collision frequency\\
$\omega_1,\omega_2,\omega_i$ && Auxiliary frequencies (two-streamline model)\\
$\omega'_{pl}$ && Planet-induced apsidal shift across a ringlet mid-width\\
$\Omega$ && Epicyclic angular velocity\\
$\Omega_p$ && Pattern speed\\
$\Omega_s$ && Satellite epicyclic angular velocity\\
$\Omega_{sg}$ && Self-gravitational frequency\\
$p$ && Integer characterizing the satellite disturbing function\\
$P_{\alpha\beta}, P_{rr}, P_{r\theta}$ && Pressure tensor components in orthonormal frames\\
$\varphi_e,\varphi$ && Epicyclic mean longitude\tnote{a}\\
$\varphi_0$ && Epicyclic mean longitude at $t=0$\\
$\phi$ && Azimuthal phase in the satellite rotating frame\\
$\phi_c$ && Critical azimuthal phase for satellite wake streamline crossing\\
$\phi_s$ && Satellite potential\\
$\Phi_{mk}$ && $m,k$ component of the satellite potential\\
$\Phi,phi_{lmp}$ && Satellite potential component (opposite of the disturbing function)\\
$\Phi_p$ && Gravitational potential of the planet\\
$\Psi_{mk}$ && Effective potential energy associated with $m,k$ component of the satellite potential\\
$\varpi_e,\varpi$ && Epicyclic periapse angle\tnote{a}\\
$\varpi_0$ && Epicyclic periapse angle at $t=0$\\
$\varpi_s$ && Satellite periapse angle\\
$\dot\varpi_{pert}$ && Streamline precession rate due to perturbations\tnote{b}\\
$\dot\varpi_{plan}$ && Planet-induced streamline precession rate\\
$q, q_i$ && Streamline compression parameter\\
$q_e$ && Integer characterizing eccentric resonances\\
$q_a$ && Critical value of $q$ where $L^{vis}_H$ changes sign\\
$q_1$ && Critical value of $q$ where $t_1$ changes sign\\
$q\cos\gamma,q\sin\gamma$ && Eccentricity and apsidal shift contributions to the compression parameter\\
$q_{ij}\cos\gamma_{ij},q_{ij}\sin\gamma_{ij}$ && Discrete form of $q\cos\gamma,q\sin\gamma$ for non consecutive streamlines\\
$\mathbf{r}$ && Position vecteur\\
$r,\theta$ && Polar coordinates\\
$R,S$ && Radial and tangential components of perturbing accelerations\\
$R_p$ && Radius of the planet\\
$\rho$ && Ring volume density\\
$s$ && Interparticle distance (surface to surface)\\
$\sigma$ && Ring surface density\\
$\sigma_0,\sigma^i_0$ && Unperturbed ring surface density\\
$\sigma_b$ && Surface density at ring edge\\
$\sigma_c$ && Critical surface density for the onset of self-gravitational wakes\\
$t$ && Time\\
$t_1,t_2,t^i_1,t^i_2$ && Effective viscous-like and pressure-like pressure tensor coefficients\\
$T_s, T^s_i, T^s_0$ && Satellite torque\\
$\mathfrak{T}_s$ && Satellite torque density\\
$\tau$ && Optical depth\\
$\tau_{sp}, \tau_{sat}$ && Ringlet spreading time-scale (free, shepherded)\\
$\tau_{gap}$ && Gap closing time-scale\\
$u_\alpha,u_\beta$ && Velocity components in an orthonormal frame\\
$w$ && Width or ring perturbed region\\
$x_\alpha,x_\beta$ && Orthonormal frame coordinates\\
$x,x_e$ && Ringlet-satellite, ring edge-satellite distance (signed quantity)\\
$Z$ && Complex eccentricity\\
$\bar{Z}$ && Mean (mass average) complex eccentricity\\

\bottomrule

\insertTableNotes

\end{longtable}

\end{ThreePartTable}

\end{center}

\restoregeometry

\newpage\null\thispagestyle{empty}
\clearpage

\appendix
\section*{Appendix: ring-satellite interactions revisited}
\label{sec:app}
\addcontentsline{toc}{section}{\nameref{sec:app}}
\markboth{Appendix}{}
\renewcommand{\thesubsection}{\Alph{subsection}}
\setcounter{subsection}{0}

It appears that an algebraic error has been made by \cite{GT81}, an exceptional fact in their production. This has important consequences as it changes some of the physical conclusions drawn from their investigation. The present appendix therefore provides all the required details to support this claim, as well as the additional information needed in section \ref{sec:satamp} (external satellite forcing of ring eccentricities). It also serves as \textit{corrigendum} for this important paper.

\subsection{Satellite excitation of ring eccentricities: tracking discrepancies}\label{app:discrep}
\renewcommand{\theequation}{\thesubsection\arabic{equation}} 
\setcounter{equation}{0}

The results obtained in Eqs.~\eqref{phil} and \eqref{phic} differ from \cite{GT81} in several important respects, a surprising outcome that requires appropriate justification. To this effect, additional mathematical and physical information is provided and discussed in this Appendix.

Let us start from Eqs.~\eqref{X1} and \eqref{X2} and their original form, Eqs.~\eqref{agt} to \eqref{pomgt}. The solution of Eq.~\eqref{X1} is readily obtained for the potential component specified by Eq.~\eqref{distgt} ($\varpi_s=0$ by choice of the origin of angles):
\begin{eqnarray}
a_1 & = & \frac{2p\phi_{lmp}}{na}\left[-\frac{\omega_r}{\omega^2}C + \frac{\gamma}{\omega^2}S\right],\label{a1}\\
\lambda_{0,1} & = & \frac{2}{na}\frac{\partial}{\partial a}\left[\phi_{lmp}\left(\frac{\gamma}{\omega^2}C + \frac{\omega_r}{\omega^2}S\right)\right],\label{lamb1}\\
I_1 & = & - q\phi_{lmp}\left[-\frac{\omega_r}{\omega^2}C + \frac{\gamma}{\omega^2}S\right],\label{I1}\\
\varpi_1 & = & -\frac{\partial\phi_{lmp}}{\partial I}\left[\frac{\gamma}{\omega^2}C + \frac{\omega_r}{\omega^2}S\right],\label{pom1}
\end{eqnarray}
where 
\begin{eqnarray}
C & = & \cos[(pn - ln_s)t + q \varpi_o + p\lambda_{0,o}],\label{coslmp}\\
S & = & \sin[(pn - ln_s)t + q \varpi_o + p\lambda_{0,o}],\label{sinlmp}\\
\omega_r & = & p n - l n_s,\label{omr}\\
\omega^2 & = & \omega_r^2 + \gamma^2.\label{om}
\end{eqnarray}
In these relations, $\varpi_o$ and $\lambda_{0,o}$ are the unperturbed periapse angle and mean longitude at epoch. Only three terms are needed in the evaluation of Eq.~\eqref{e2}, $dI_2/dt$, $I_1dI_1/dt$ and $\varpi_1 d\varpi_1/dt$. The first of these follows from Eq.~\eqref{X2}, which takes the following expanded form
\begin{equation}
\frac{dI_2}{dt} = \frac{\partial^2 \Phi}{\partial\varpi\partial a}a_1 +
\frac{\partial^2 \Phi}{\partial\varpi\partial \lambda_0}\lambda_{0,1} +
\frac{\partial^2 \Phi}{\partial\varpi\partial I}I_1 +
\frac{\partial^2 \Phi}{\partial\varpi^2}\varpi_1.\label{I2}
\end{equation}
Remember that the derivative with respect to $a$ involves the angular part of the potential as well. Eqs.~\eqref{I1} and \eqref{pom1} imply
\begin{equation}
\left\langle I_1\frac{d I_1}{dt} \right\rangle_{\lambda} = \left\langle \varpi_1\frac{d \varpi_1}{dt} \right\rangle_{\lambda} = 0,\label{quadav}
\end{equation}
where $\langle X \rangle_{\lambda}$ stands for the average over the mean longitude at epoch $\lambda_0$. Combining Eqs.~\eqref{dI2}, \eqref{de2}, \eqref{I2} and \eqref{quadav} finally yields Eqs.~\eqref{e2l} and \eqref{e2c}.

Note that the secular variation of the amplitude of the global $m=1$ mode is given by $ae_2 + e a_2$; both second order contributions are of order $e^{2|q_e|-2}e_s^{2|k|}$ (for $|q_e|\ge 1$), so that the contribution of $a_2$ is negligible. Consequently, the secular second-order change of $e$ controls by itself the change of amplitude of the eccentric mode. For the sake of completeness on this question, it would also be of interest to reformulate the \cite{PM05} approach to capture the leading $|q_e|=1$ eccentric resonance in the framework of their fluid formalism; however, this task will not be undertaken here.

The second-order calculations performed above are straightforward but the evaluation of the two terms of the form $\langle X_1 dX_1/dt\rangle$ ($X_1=I_1, \varpi_1$) leads to an unexpected conclusion. As $X_1$ is harmonic, such terms vanish due to the averaging over $\lambda_0$ as $X_1$ and $dX_1/dt$ are out of phase by $\pi/2$.  However, Eqs.~(55e,f) of \cite{GT81} give instead a non vanishing contribution in the $\gamma\rightarrow 0$ limit:
\begin{eqnarray}
\left\langle I_1\frac{d I_1}{dt} \right\rangle_{\lambda_0, GT} & = & \frac{\pi}{2}q_e^2 \phi_{lmp}^2\delta(pn - l n_s),\label{Iavgt}\\
 \left\langle \varpi_1\frac{d \varpi_1}{dt} \right\rangle_{\lambda_0, GT} & = & \frac{\pi}{2}\left(\frac{\partial \phi_{lmp}}{\partial I}\right)^2\delta(pn - l n_s).\label{pomavgt}
\end{eqnarray}
This point has important physical implications: the cancellation of these terms changes the sign of the dominant ($|q|=1$) resonant eccentricity driving (the dominant Lindblad contribution is now strictly positive for this resonance, while it is vanishing in \citealt{GT81}), and therefore makes a narrow ring eccentricity always excited by an external satellite forcing, a conclusion that differs from the \cite{GT81}. This therefore requires careful consideration. 

In fact, the \cite{GT81} result can only be recovered by using Eq.~\eqref{X1} and taking the limit $\gamma\rightarrow 0$ in the expression of both $dX_1/dt$ and $X_1$ \textit{before} averaging $X_1 dX_1/dt$ over $\lambda_0$. With this procedure, $\langle X_1 dX_1/dt\rangle_{\lambda} = \langle X_1 F(\mathbf{X}_0)\rangle_{\lambda} $ in the notations of Eq.~\eqref{X1}, which yields Eqs.~\eqref{Iavgt} and \eqref{pomavgt}. However, $\langle X_1 dX_1/dt\rangle_{\lambda} = 0$ is the correct result, for the following set of reasons:

\begin{itemize}
\item First, actual disk physical responses to the forcing by the satellite are regularized by collective effects, dissipative or other; thus from a physical point of view, the correct result is the one obtained on the regularized expression of $X_1$. The $\gamma\rightarrow 0$ limit is physically unessential; it is just a way to obtain a generic expression independently of the collective effect at work. Ignoring this limit, $\langle X_1 dX_1/dt\rangle_{\lambda}$ is clearly zero.
\item  Second, taking the limit before other operations is incorrect from a mathematical point of view:  $\gamma\rightarrow 0$ does not ensure that $\gamma X_1, \gamma X_1^2 \rightarrow 0$ as $X_1 \rightarrow \infty$; actually, $\gamma X_1 \rightarrow 0$ but $\gamma X_1^2$ remains finite and exactly cancels the $X_1 F(\mathbf{X}_0)$ term; taking the limit prior to other operations amounts to ignoring this cancellation and leads to the \cite{GT81} result. 
\item Finally, the eccentricity evolution can also be computed from energy and angular momentum budgets, as discussed in Appendix~\ref{app:budget} below, and this alternate route gives back the result obtained in Eq.~\eqref{phil}, not the \cite{GT81} one. 
\end{itemize}

\smallskip

Let us also discuss Eq.~\eqref{e2c}. In the quasi-continuum limit, the corotation-like contribution always damps the eccentricity [see Eq.~(75) of \cite{GT81}], whereas here the sign of this contribution depends on the sign of $q_e$ [see Eq.~\eqref{phic}]. However, in the quasi-continuum limit, the integration by parts performed to obtain Eq.~\eqref{phic} from Eq.~\eqref{e2c} would not be valid as the boundary contributions would not vanish. Conversely, in this quasi-continuum limit, the replacement of the sum over $m$ by an integral, needed to derive Eq.~(75) of \cite{GT81} from their Eq.~(56), is legitimate as long as the distance between successive resonances (at given $q_e,k$) is much smaller than the resonance width set by $\gamma$. The difference in sign behavior for the corotation-like term is therefore rooted in the difference of physical contexts and not in an artifact of the mode of calculation. Both Eqs.~\eqref{phic} and Eq.~(75) of \cite{GT81} are correct in their respective domain of application.

\subsection{Resonant angular momentum and energy budget}\label{app:torque}
\renewcommand{\theequation}{\thesubsection\arabic{equation}} 
\setcounter{equation}{0}
The time evolution of the ring total energy and angular momentum is best evaluated in semi-Eulerian coordinates instead of the semi-Lagrangian ones used in the remainder of this chapter (see also \citealt{L92}, section 6.4). To this effect, ($a,\varphi$) are used as coordinates instead of Lagrangian labels. Note that $\varphi = \lambda$ for any fluid particle.

Let us introduce the total energy $\mathcal{E}$ and angular momentum $\mathcal{H}$ of the ring 
\begin{eqnarray}
\mathcal{E} & = & \int_{0}^{\infty} da\int_0^{2\pi} d\varphi\ a \sigma_0 E, \label{en}\\
\mathcal{H} & = & \int_{0}^{\infty} da\int_0^{2\pi} d\varphi\ a \sigma_0 H,
\label{angmom}
\end{eqnarray}

For any quantity $X=X(a,\varphi,t)$, one can always write
\begin{equation}
\frac{dX}{dt}=\frac{\partial X}{\partial t}+ 
\frac{da}{dt}\frac{\partial X}{\partial a}+
\frac{d\varphi}{dt}\frac{\partial X}{\partial \varphi},\label{dX}
\end{equation}
where $d/dt$ is the Lagrangian derivative used so far. The equation of conservation of mass  reads: 
\begin{equation}
\frac{\partial\sigma_0}{\partial t}+
\frac{1}{a}\frac{\partial}{\partial a} \left(a\sigma_0\frac{da}{dt}\right)
+\frac{1}{a}\frac{\partial}{\partial \varphi} \left(a\sigma_0\frac{d\varphi}{dt}\right) =0.\label{mass}
\end{equation}
Thus, for any quantity $X$, using the mass conservation constraint Eq.~\eqref{mass}, one has
\begin{equation}
a\sigma_0\frac{dX}{dt} =\frac{\partial a\sigma_0 X}{\partial t}+ 
\frac{\partial}{\partial a}\left(a\sigma_0 X\frac{da}{dt}\right)
+\frac{\partial}{\partial \varphi} \left(a\sigma_0 X\frac{d\varphi}{dt}\right),\label{dX2}
\end{equation}
so that after integration over ($a,\varphi$) one obtains
\begin{eqnarray}
\frac{d\mathcal{E}}{dt} = \frac{\partial \mathcal{E}}{\partial t} & = &\int_{a_i}^{a_o}da\ \frac{\partial}{\partial t}\left(2\pi a\sigma_0 E\right)\nonumber\\
& = & \int_{a_i}^{a_o}da\ 2\pi a\sigma_0\left\langle\frac{dE}{dt} \right\rangle_{\lambda},\label{Ebudget}\\
\frac{d\mathcal{H}}{dt} = \frac{\partial \mathcal{H}}{\partial t} & = &\int_{a_i}^{a_o}da\ \frac{\partial}{\partial t}\left(2\pi a\sigma_0 H\right)\nonumber\\
& = & \int_{a_i}^{a_o}da\ 2\pi a\sigma_0\left\langle\frac{dH}{dt} \right\rangle_{\lambda},\label{Hbudget}
\end{eqnarray}
where $a_i, a_o$ are the ring boundaries, and where $\varphi = \lambda$ has been used to express the azimuthal integral in terms of the mean longitude at epoch average of the Lagrangian derivative of $E,H$.

We have now assembled all the needed material to compute resonant secular changes of energy and angular momentum (i.e., torques). Let us consider torques first. From Eq.~\eqref{specang}, \eqref{agt} and \eqref{egt},
\begin{equation}
\frac{dH}{dt} = \frac{na}{2}\frac{da}{dt} - \frac{dI}{dt}= -\frac{\partial\Phi}{\partial\lambda_0} -\frac{\partial\Phi}{\partial\varpi}.\label{dH} 
\end{equation}
The second order time variation of $H$ is readily obtained from an expansion similar to Eq.~\eqref{I2} and leads to results similar to Eqs.~\eqref{e2l} and \eqref{e2c}:
\begin{eqnarray}
\left\langle\frac{dH}{dt}\right\rangle_{\lambda} & = & \left\langle \frac{dH}{dt}\right\rangle_{\lambda,L} + \left\langle \frac{dH}{dt}\right\rangle_{\lambda,C},\label{HH}\\
\left\langle \frac{dH}{dt}\right\rangle_{\lambda,L}  & = & -\frac{\pi m q_e}{2}\frac{\partial \Phi_{lmp}^2}{\partial I}\delta(pn - l n_s),\label{HHl}\\
\left\langle \frac{dH}{dt}\right\rangle_{\lambda,C} & = & \frac{\pi m p}{na}\frac{\partial}{\partial a}\left[\Phi_{lmp}^2\delta(pn - ln_s)\right],\label{HHc}
\end{eqnarray}
from which the final result follows:
\begin{eqnarray}
\frac{\partial \mathcal{H}}{\partial t} & = &
\left( \frac{\partial \mathcal{H}}{\partial t}\right)_L +
\left( \frac{\partial \mathcal{H}}{\partial t}\right)_C,
\end{eqnarray}
\begin{eqnarray}
\left( \frac{\partial \mathcal{H}}{\partial t}\right)_L & = &
-\frac{2\pi^2 m \sigma_0}{3(m-q) n^2 }
\left[\frac{q_e}{e}\frac{\partial \phi^2_{lmp}}{\partial e} \right]_{e_r},\label{H2l}\\
\left( \frac{\partial \mathcal{H}}{\partial t}\right)_C & = &
- \frac{4\pi^2 m a}{3n} \phi^2_{lmp} \frac{d}{da} \left(\frac{\sigma_0}{n}\right)\nonumber\\
& = &  -\frac{2\pi^2 m\sigma_0}{n^2}\phi^2_{lmp}\left(1 +\frac{2a}{3\sigma_0}\frac{d\sigma_0}{da}\right).\label{H2c}
\end{eqnarray}
For a corotation resonance, $q_e=0$ and only the corotation part of the torque arises. For an eccentric resonance, on the contrary, $q_e\neq 0$, the corotation part of the torque is  $\sim e_r^2$ smaller than the Lindblad one and can be neglected. These equations are identical\footnote{\cite{GT81} derive what appears to be Lagrangian torque expressions, but in fact these are the semi-Eulerian ones Eq.~\eqref{Hbudget}, in agreement with density wave theory (see the torque expressions given in, e.g., \citealt{GT80}).} to Eqs.~(32), (44) and (47) of \cite{GT81}.

To evaluate changes in energy, instead of following the same route, it is easier to use a constant of motion that is closely related to Jacobi's. Indeed, one can check that Eqs.~\eqref{agt} and \eqref{egt} allow us to relate changes in $a$ and $I$ for potential components of the form \eqref{distgt}:
\begin{equation}
q_e \frac{n a}{2}\frac{da}{dt} + p\frac{dI}{dt} = 0.\label{constant}
\end{equation}
This in turn implies a relation between the rate of changes of the specific energy and angular momentum with the help of Eqs.~\eqref{specen} and \eqref{specang}:
\begin{equation}
\frac{dE}{dt} = \frac{pn}{m}\frac{dH}{dt} = \frac{ln_s}{m}\frac{dH}{dt}.\label{newjacobi}
\end{equation}
The last equality holds at resonance only ($pn-ln_s=0$), but the limit $\gamma \rightarrow 0$ makes it the relevant location in relating changes in energy and angular momentum; the same limit makes relation \eqref{constant} exact to leading order in eccentricity. Consequently,
\begin{equation}
\frac{\partial \mathcal{E}}{\partial t} = \frac{ln_s}{m} \frac{\partial \mathcal{H}}{\partial t}.\label{jacobi2}
\end{equation}

\subsection{Eccentricity evolution from energy and angular momentum budgets}\label{app:budget}
\renewcommand{\theequation}{\thesubsection\arabic{equation}} 
\setcounter{equation}{0}
Only eccentric resonances contribute to the ring evolution, and for these resonances, the Lindblad contribution is the dominant one. This allows us to derive the ring eccentricity evolution from energy and angular momentum budgets in a simple way.

From Eqs.~\eqref{ering}, \eqref{de2} and \eqref{quadav}, and remembering the $e_r \simeq e_0$,
\begin{equation}
\frac{d\langle e_r^2\rangle}{dt} = \frac{1}{M_r} \int da\ 2 \pi \sigma_0 a \left\langle \frac{d e^2}{dt}\right\rangle_{\lambda,L}.\label{ealt}
\end{equation}
On the other hand, to leading order in eccentricity,
\begin{equation}
\frac{dH}{dt} = \frac{na}{2}\frac{da}{dt} - \frac{na^2}{2}\frac{de^2}{dt},\label{dH2}
\end{equation}
so that the satellite torque can also be expressed as
\begin{eqnarray}
\frac{\partial \mathcal{H}}{\partial t} & = & \int da\ 2\pi a\sigma_0 \left[ 
\frac{na}{2}\left\langle\frac{da}{dt}\right\rangle_{\lambda,L} - \frac{na^2}{2}\left\langle\frac{de^2}{dt}\right\rangle_{\lambda,L} \right]\nonumber\\
& = & \frac{1}{n}\frac{\partial \mathcal{E}}{\partial t} - \frac{1}{2}na^2 M_r\frac{d\langle e_r^2\rangle}{dt}.\label{Halt} 
\end{eqnarray}
In the last expression, the fact that the Lindblad mean longitude at epoch averages are proportional to $\delta(pn - l n_s)$ has been used [see Eqs.~\eqref{e2l}, \eqref{e2c}, \eqref{HHl} and \eqref{HHl}]; this allowed us to move relevant factors in and out of the integrals. Neglecting the corotation part is self-consistently justified, as Eqs.~\eqref{H2l} and \eqref{H2c} imply that this contribution is $\sim e_r^2$ smaller than the Lindblad part.

Combined with Eq.~\eqref{jacobi2}, this finally leads to the desired result, once the resonance relation is used ($pn = l n_s$):
\begin{equation}
\frac{1}{2}na^2 M_r\frac{d\langle e_r^2\rangle}{dt} = - \frac{q_e}{m}\frac{\partial \mathcal{H}}{\partial t}.\label{e2alt}
\end{equation}
This is consistent with Eq.~\eqref{phil}, as can be seen from Eq.~\eqref{H2l}.

\subsection{Asymptotic evaluation of the potential}\label{app:asymp}
\renewcommand{\theequation}{\thesubsection\arabic{equation}} 
\setcounter{equation}{0}
Further progress can be made with the help of the asymptotic approximations $m\gg 1$, $\eta_r,\eta_s\ll 1$ ($\eta_r =a_s e_r/x, \eta_s =a_s e_s/x$=, which leads to simplifications in the form of the disturbing function. Recalling that $x=a-a_s$, in the limit $m\gg 1$ ($|x|/a_s\ll 1$), Eq.~(66) and (67) of \cite{GT81} gives the relevant generic form of $\phi_{lmp}$ [defined in Eq.~\eqref{distgt}]:
\begin{eqnarray}
\phi_{lmp} & = & - \frac{GM_s}{2\pi^3 a_s}\int_{0}^{2\pi}dM  \int_{0}^{2\pi}dM_s \ (1 + e_s \cos M_s) \times \nonumber\\
& & \ \ \ K_0\left[\frac{m|x|}{a_s}(1 - \eta_r\cos M + \eta_s\cos M_s)\right]\times\nonumber\\
& & \ \ \ \cos\left[qM+kM_s -\frac{2 mx}{a_s}(\eta_s\sin M_s - \eta_r\sin M)\right]. \label{philmp}
\end{eqnarray}
where $K_0$ is the modified Bessel function of the second kind. This expression comes from a first order expansion in the satellite and ring eccentricities of the potential component of Eq.~\eqref{satpot} expressed in terms of Laplace coefficients, and from an asymptotic approximation of the Laplace coefficient $b^{(m)}_{1/2}$ itself in terms of $K_0$. 
%
%
The resonance relation Eq.~\eqref{eccres} yields a useful generic relation\footnote{Resonances with $q_e+k=0$ correspond to $a=a_s$ and are not relevant in the problem.}: 
\begin{equation}
\frac{m|x|}{a_s} \approx \frac{2|q_e+k|}{3}.\label{mx}
\end{equation}
which combined to Eq.~\eqref{philmp} leads to the various  components of interest of the satellite potential:
\begin{eqnarray}
\phi_{m,m,m\pm 1} & = & - \frac{2 G M_s}{\pi a_s}\frac{\eta_r}{3} F_1 \quad (k=0, |q_e|=1),\label{phiq1}\\
\phi_{m\pm 1,m,m} & = & + \frac{2 G M_s}{\pi a_s}\frac{\eta_s}{3} F_1\quad (q_e=0, |k|=1),\label{phik1}\\
\phi_{m,m,m\pm 2} & = & - \frac{2 G M_s}{\pi a_s}\frac{\eta_r^2}{18}F_2 \quad (k=0, |q_e|=2),\label{phiq2}\\
\phi_{m\pm 2,m,m} & = & - \frac{2 G M_s}{\pi a_s}\frac{\eta_s^2}{18}F_2 \quad (q_e=0, |k|=2),\label{phik2}\\
\phi_{m\pm 1,m,m\mp 1} & = & + \frac{2 G M_s}{\pi a_s}\frac{\eta_r\eta_s}{9} F_2  \quad (k=q_e=\pm 1),\label{phikq}
\end{eqnarray}
where
\begin{eqnarray}
F_1 & = & 2K_0(2/3)+K_1(2/3)\simeq 2.52\label{F1}\\
F_2 & = & 20K_0(4/3)+19K_1(4/3)\simeq 12.05\label{F2}
\end{eqnarray}
The $q_e=0$ terms do not contribute to the eccentricity evolution, but have been given as they are needed for the torque evaluations of section \ref{sec:satmass} (minimal satellite mass for edge confinement).

\clearpage 
\newpage\null\thispagestyle{empty}
\clearpage

\addcontentsline{toc}{section}{References}
\bibliography{ringbiblio}
\bibliographystyle{plainnat}

\clearpage
\newpage\null\thispagestyle{empty}
\vfill
\begin{figure}[h]
    \centering
    \includegraphics[scale=0.4]{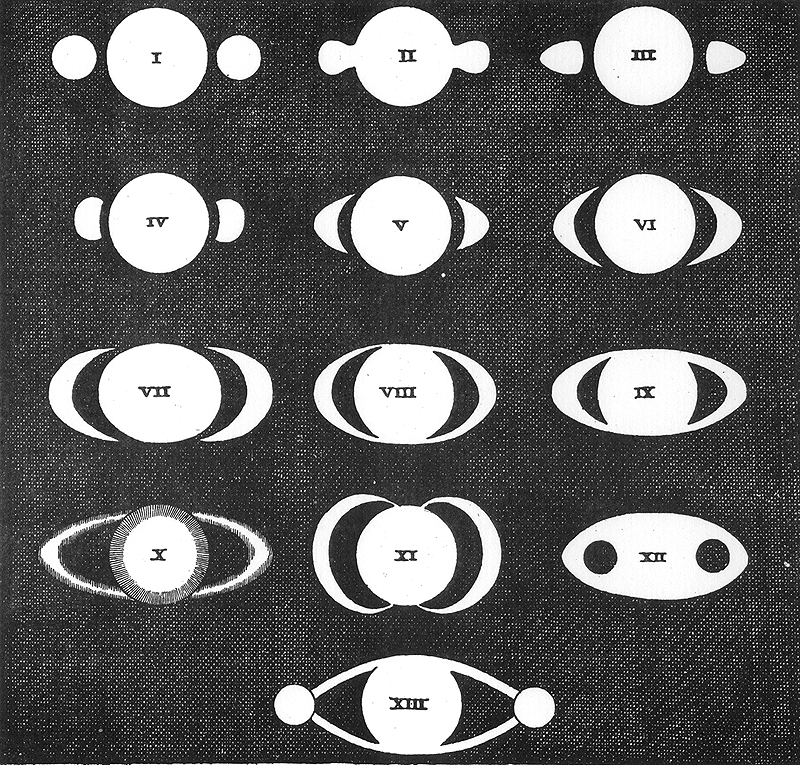}
\end{figure}
\vskip 3truecm
\begin{figure}[h]
    \centering
    \includegraphics[scale=0.4]{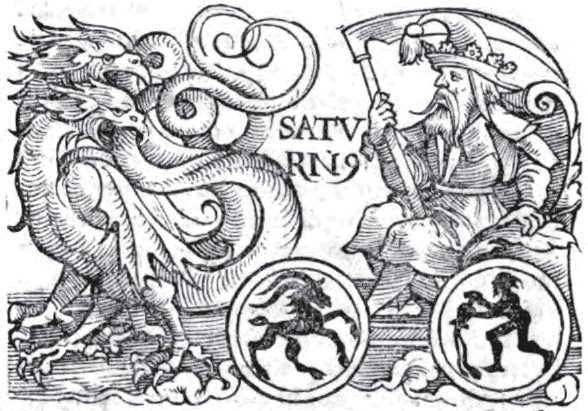}
\end{figure}
\vfill
\end{document}